\documentclass[preprint2]{emulateapj}
\usepackage{ulem}
\usepackage{color}
\newcommand{\mbh}{\ensuremath{M_{\rm{BH}}}\,}

\shorttitle{Spatial Clustering of RASS-AGN: V. -- $z=0.44-0.64$}
\shortauthors{Krumpe et al.}

\usepackage{amsfonts}
\usepackage{graphics,graphicx}
\usepackage{ulem}
\usepackage{appendix}
\usepackage{lineno}

\begin{document}
\def\mpch {$h^{-1}$ Mpc} 
\def\kpch {$h^{-1}$ kpc} 
\def\kms {km s$^{-1}$} 
\def\lcdm {$\Lambda$CDM } 
\def\xir {$\xi(r)$}
\def\wprp {$w_p(r_p)$}
\def\xisp {$\xi(r_p,\pi)$}
\def\xis {$\xi(s)$}
\def\rr {$r_0$}
\def\etal {et al.}

\title{The Spatial Clustering of ROSAT All-Sky Survey Active Galactic Nuclei\\ V. 
The Evolution of Broad-line AGN Clustering Properties in the Last 6 Gyr.}

\author{Mirko Krumpe\altaffilmark{1},  
Takamitsu Miyaji\altaffilmark{2,1,11}, 
Antonis Georgakakis\altaffilmark{3,4},
Andreas Schulze\altaffilmark{5,6}, 
Alison L. Coil\altaffilmark{8}, 
Tom Dwelly\altaffilmark{4}, 
Damien Coffey\altaffilmark{4}, 
Johan Comparat\altaffilmark{4},
H\'ector Aceves\altaffilmark{2},
Mara Salvato\altaffilmark{4}, 
Andrea Merloni\altaffilmark{4}, 
Claudia Maraston\altaffilmark{7}, 
Kirpal Nandra,\altaffilmark{4},
Joel R. Brownstein\altaffilmark{9}, 
Donald P. Schneider\altaffilmark{10},
SDSS-IV team, and SPIDERS team}

\altaffiltext{1}{Leibniz-Institut f\"ur Astrophysik Potsdam (AIP), An der Sternwarte 16, 
                 D-14482 Potsdam, Germany}
\altaffiltext{2}{Universidad Nacional Aut\'onoma de M\'exico (UNAM), Instituto de Astronomía, AP 106,  Ensenada 22860, BC, M\'exico}
\altaffiltext{3}{IAASARS, National Observatory of Athens, GR-15236 Penteli, Greece}
\altaffiltext{4}{Max-Planck-Institut f\"ur extraterrestrische Physik, Gie\ss enbachstra\ss e 1, 
                 D-85748 Garching, Germany}
\altaffiltext{5}{National Astronomical Observatory of Japan, Mitaka, Tokyo 181-8588, Japan}
\altaffiltext{6}{OmagaLambdaTec, Lichtenbergstra\ss e 8, D-85748 Garching, Germany}
\altaffiltext{7}{Institute of Cosmology and Gravitation, Burnaby Road, Portsmouth, PO1 3FX, UK}
\altaffiltext{8}{University of California, San Diego, Center for Astrophysics and
                 Space Sciences, 9500 Gilman Drive, La Jolla, CA 92093-0424, USA}
\altaffiltext{9}{Department of Physics and Astronomy, University of Utah, 115 S. 1400 E., Salt Lake City, UT 84112, USA}
\altaffiltext{10}{Department of Astronomy and Astrophysics, The Pennsylvania
  State University, University Park, PA 16802, USA}
\altaffiltext{11}{Sabbatical leave from UNAM at AIP.}
\email{mkrumpe@aip.de}

\begin{abstract}
    
This is the fifth paper in a series of investigations of the clustering properties
of luminous, broad-emission-line active galactic nuclei (AGN) identified in the \textit{ROSAT} All-Sky Survey (RASS) and Sloan Digital Sky Survey (SDSS). In this work we measure the cross-correlation function (CCF) between RASS/SDSS DR14 AGN with the SDSS CMASS galaxy sample at $0.44<z<0.64$. We apply halo occupation distribution (HOD) modeling to the CCF along with the autocorrelation function of the CMASS galaxies. We find that X-ray and
optically selected AGN at $0.44<z<0.64$ reside in statistically identical
halos with a typical dark matter halo mass of  $M_{\rm DMH}^{\rm typ,AGN} \sim
10^{12.7}\,h^{-1}\,M{_\odot}$. The acceptable HOD parameter space for these
two broad-line AGN samples have only statistically marginal differences caused
by  small deviations of the CCFs in the one-halo-dominated regime on small
scales. In contrast to optically selected AGN, the X-ray AGN sample may contain a larger population of satellites at $M_{\rm DMH} \sim 10^{13}\,h^{-1}\,M{_\odot}$.
We compare our measurements in this work with our earlier studies at lower independent redshift ranges,
spanning a look-back time of 6 Gyr. The comparison over this wider redshift range of $0.07<z<0.64$ reveals: (i) no 
significant difference between the typical DMH masses of X-ray and optically
selected AGN, (ii) weak positive clustering dependencies of $M_{\rm DMH}^{\rm
  typ,AGN}$ with $L_{\rm X}$ and $M_{\rm BH}$, (iii) no significant dependence of $M_{\rm DMH}^{\rm typ,AGN}$ on Eddington ratio, and (iv) the same DMH masses host more-massive accreting black holes at high redshift than at low redshifts.

\end{abstract}



\section{Introduction}
\label{introduction}

According to the consensus cosmological model, primordial density fluctuations
grow and  collapse into gravitationally bound regions called dark matter halos (DMHs). Their gravitational interaction with baryonic matter leads to the formation of galaxies within these DMHs
(e.g., \citealt{white_frenk_1991}). Processes that are still poorly understood  can lead to a flow of matter onto the supermassive black holes (SMBHs) at the centers of the galaxies. Such an object would be observed as an active galactic nucleus (AGN) until the mass flow onto the
SMBH stops. The physical processes leading to the formation and evolution of galaxies (and their AGN phases) are closely connected with their host DMHs. However, these DMHs cannot be observed directly.  
One approach to determine the properties of the DMHs is to measure the
clustering of extragalactic objects hosted by these DMHs. This is commonly
done by measuring the two-point correlation function (2PCF; e.g., \citealt{peebles_1980}). Since AGN can be detected at a variety of wavelength ranges in sufficiently large samples, observational and theoretical studies have investigated the relationship between AGN and their DMHs. AGN clustering measurements (see review by \citealt{krumpe_miyaji_2014}) reveal important physical results on  AGN/galaxy co-evolution, typical DMHs and the full  distribution of DMH masses hosting AGN (e.g., \citealt{porciani_magliocchetti_2004}; \citealt{gilli_daddi_2005,gilli_zamorani_2009}; 
\citealt{yang_mushotzky_2006}; \citealt{coil_georgakakis_2009}; \citealt{ross_shen_2009}; 
\citealt{krumpe_miyaji_2010,krumpe_miyaji_2018}; \citealt{cappelluti_ajello_2010}; 
\citealt{miyaji_krumpe_2011}; \citealt{allevato_2011}; \citealt{mountrichas_georgakakis_2012}; 
\citealt{koutoulidis_plionis_2013}; \citealt{melnyk_elyiv_2018}; \citealt{powell_cappelluti_2018,powell_urry_2020}; \citealt{plionis_koutoulidis_2018}; \citealt{mountrichas_georgakakis_2019}; \citealt{viitanen_allevato_2019}; \citealt{krishnan_almaini_2020}). Simulations and theoretical work 
(e.g., \citealt{springel_2005}; \citealt{booth_schaye_2010};
\citealt{comparat_merloni_2019}; \citealt{georgakakis_comparat_2019}; \citealt{shankar_allevato_2020}) can use these observational constraints, as well as observed clustering dependences with AGN parameters, to improve  models of AGN and galaxy evolution.

As AGN samples are substantially smaller than galaxy samples at low ($z\lesssim0.7$) and moderate redshifts ($0.7\lesssim z \lesssim 1.5$), the AGN autocorrelations function (ACF; measuring the distances between AGN only) has a limited  signal-to-noise ratio (S/N), particularly at small separations. To overcome this observational challenge, the 
AGN cross-correlation function (CCF) with a dense galaxy sample in the same volume can be used; 
\cite{coil_georgakakis_2009} demonstrate the potential of this approach. Due to the substantial increase in the number of AGN-galaxy pairs, the uncertainty in the clustering measurement is reduced compared to the  measurement of the ACF in the same AGN sample. The AGN ACF can then be inferred from the AGN-galaxy CCF. Halo occupation distribution (HOD) modeling can then be used to interpret the observed clustering signal (e.g., \citealt{cooray_sheth_2002}; \citealt{rodriguez_comparat_2017}).

In \citet[hereafter paper I]{krumpe_miyaji_2010}, we use the CCF technique 
to measure the clustering between \textit{ROSAT} All-Sky Survey (RASS) AGN
identified in the Sloan Digital Sky Survey (SDSS) and a large sample of SDSS 
luminous red galaxies (LRGs) at $0.16<z<0.36$. The samples are drawn from a series of SDSS data releases, starting with data release 4 (\citealt{adelman_mccarthy_agueeros_2006}). The high S/N of the measurement allows us to split the sample into low and high X-ray luminosity subsamples. We find a weak X-ray luminosity dependence for luminous, broad-line AGN in which higher-luminosity (median intrinsic $L_{\rm0.1-2.4\,keV} = 3.0\times 10^{44}$ erg s$^{-1}$) AGN cluster more strongly than their lower-luminosity ($9.8\times 10^{43}$ erg s$^{-1}$) counterparts.

In the second paper of this series \citep[hereafter paper II]{miyaji_krumpe_2011}, 
we develop a novel method of applying the HOD modeling 
technique directly to the measured CCF between RASS/SDSS AGN 
and SDSS LRGs. We constrain the distribution of AGN as a function of 
DMH mass instead of quoting only typical DMH masses. The major advantage of this method is that it does not use a phenomenological power-law fit, as is often done to derive typical DMHs. 
We find that models where the AGN fraction among satellite galaxies decreases with 
DMH mass beyond $M_{\rm DMH} \sim 10^{12}$ $h^{-1}$ $M_{\odot}$ are preferred
for luminous, broad-line AGN. This is in contrast to HOD modeling of galaxy samples  
(\citealt{zheng_zehavi_2009}; \citealt{zehavi_zheng_2011}).  

In the third paper \citep[hereafter paper III]{krumpe_miyaji_2012}, we extend the 
cross-correlation measurements to lower and higher redshifts, covering a redshift 
range of $z=0.07-0.50$ and apply the HOD modeling to all CCFs directly. 
We show that the weak X-ray luminosity dependence of 
broad-line AGN clustering is also found if radio-detected AGN are excluded, and that optically and X-ray selected AGN samples in SDSS show no significant difference in their clustering properties. 

In the fourth paper \citep[hereafter paper IV]{krumpe_miyaji_2015}, we explore the physical origin of the weak X-ray luminosity. We find that the clustering strength of luminous broad-line AGN depends on $M_{\rm BH}$ and not on $L/L_{\rm EDD}$. Thus, more-massive SMBHs are hosted by more-massive DMHs than lower-mass SMBHs. 

In this paper we extend the redshift range for our RASS/SDSS AGN clustering measurements further. We compute the CCF between AGN and SDSS CMASS galaxies in the redshift range $0.44<z<0.64$. We investigate the clustering dependencies for optically and X-ray selected AGN in this redshift range as well as for AGN parameters such as $L_{\rm X}$, $M_{\rm i}$, 
$M_{\rm BH}$, $L/L_{\rm EDD}$, and $L_{\rm Bol}$. Combining these new measurements with the previous independent measurements at lower redshift allows us to constrain the evolution of the AGN clustering properties over a look-back time of 6 Gyr. 

This paper is organized as follows. In Section~2 we describe the properties of the 
CMASS galaxy tracer set and the X-ray and optical AGN samples. Section~3 provides details on how we fit the 
H$\alpha$ line profile in the optical SDSS AGN spectra, derive the $M_{\rm BH}$, estimate 
$L/L_{\rm EDD}$, and define our AGN subsamples. In Section~4 we briefly summarize 
the cross-correlation technique, 
how the AGN ACF is inferred from this, and how we derive the clustering 
parameters using HOD modeling. Section~5 provides the results of our clustering measurements. 
Our results are discussed in Section~6, and 
we present our conclusions in Section~7. 

Throughout the paper, all distances are measured in comoving 
coordinates and given in units of $h^{-1}$\,Mpc, where $h= H_{\rm 0}/100$\,km\,s$^{-1}$\,Mpc$^{-1}$, unless 
otherwise stated. We use a cosmology of $\Omega_{\rm m} = 0.3$, $\Omega_{\rm \Lambda} = 0.7$, and 
$\sigma_8(z=0)=0.8$, which is consistent with the {\it Wilkinson Microwave Anisotropy Probe} data release 7 
(Table~3 of \citealt{larson_dunkley_2011}). The same cosmology is used in papers I--IV.
Luminosities and absolute magnitudes are calculated for $h=0.7$.
We use AB magnitudes throughout the paper. The symbol ``$\log$" represents a base-10 logarithm. 
All uncertainties represent 1$\sigma$ 
(68.3\%) confidence intervals unless otherwise stated.


\section{Data}

The data sets used in this study are drawn from the SDSS data releases 12
(\citealt{alam_albareti_2015})
and 14
(\citealt{abolfathi_aguado_2018}).
The data were obtained as part of the SDSS-III
(\citealt{eisenstein_weinberg_2011}) Baryon Oscillation Spectroscopic Survey
(BOSS; \citealt{dawson_schlegel_2013}) using the BOSS spectrograph
(\citealt{smee_gunn_2013}) on the 2.5 m SDSS telescope (\citealt{gunn_siegmund_2006}).
In the following subsections, we explain the 
sample selection of the CMASS (for ``constant mass") SDSS sample (DR12), which serves as a 
tracer set of underlying dark matter density. We also give a description of the 
X-ray selected RASS/SDSS AGN (DR14) and optical SDSS AGN (DR14) samples.  

\subsection{SDSS CMASS Galaxy Sample} 
\label{desc_LRG}

Similar to LRGs, the target selection of CMASS
galaxies is based on colors from the imaging survey of SDSS. However, 
the selection cuts for the CMASS sample are fainter and bluer than those of the LRG
sample. The CMASS sample is approximately stellar mass limited and covers a 
redshift range of $z\sim0.43-0.7$. 
For more details, see, e.g., \cite{dawson_schlegel_2013},
\cite{guo_zehavi_2013}, \cite{maraston_2013}, \cite{ross_beutler_2017}.

The original spectroscopic CMASS sample can be downloaded from: {\tt
 https://data.sdss.org/sas/dr12/boss/lss/}. 
For this work, we focus only on data from the SDSS North Cap ($100\le\rm{R.A.}\le270$), exclude the
South Cap, and use the galaxy catalog ``galaxy\_DR12v5\_CMASSLOWZTOT\_North.fits''.
To minimize possible effects of redshift evolution in the clustering signal, 
we limit the sample in redshift range by using only
spectroscopically identified CMASS galaxies with $0.44<z<0.64$. 
To minimize systematic uncertainties and facilitate interpretation, we require
the galaxy tracer set (CMASS) to exhibit a constant clustering strength over the full  redshift range of interest. Galaxy clustering studies (e.g., \citealt{meneux_guzzo_2009},
\citealt{lawsmith_eisenstein_2017}) show that the clustering strength 
depends on stellar mass in particular for red galaxies. Thus, restricting the 
stellar mass range will lead to a nearly constant clustering strength in a
narrow redshift range. 

\begin{figure}
  \centering
 \resizebox{\hsize}{!}{ 
   \includegraphics[width=8cm,height=5cm]{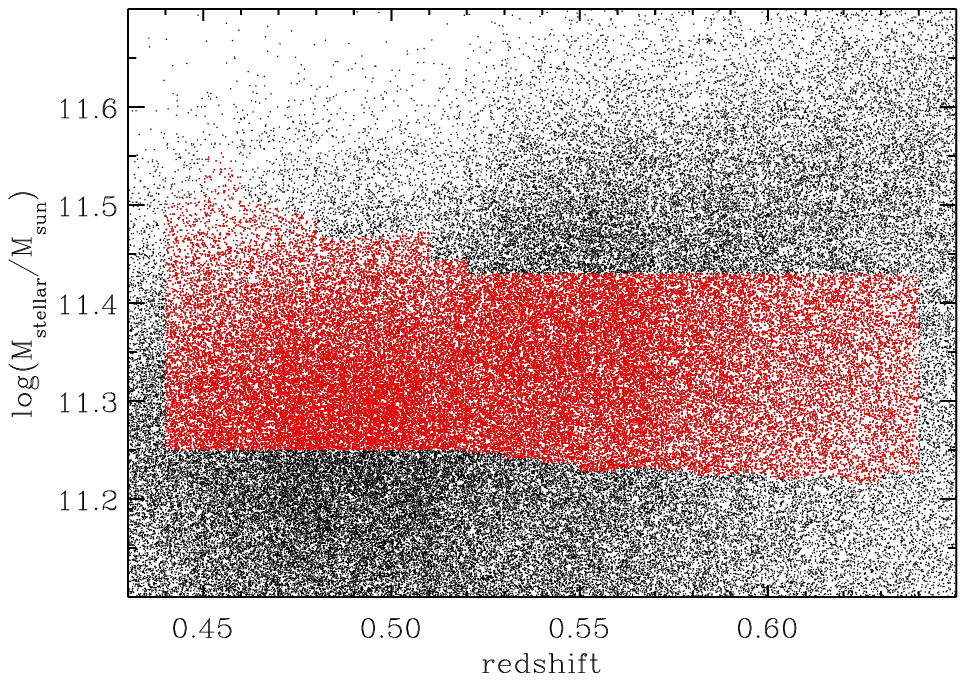}}
      \caption{Stellar mass vs. redshift for the full spectroscopic CMASS sample (black
        data points) and the CMASS subsample (red) used for 
        cross-correlation measurements with AGN.}
      \label{stellarM_z_CMASS}
\end{figure}

\cite{comparat_maraston_2017} performed spectral fitting of stellar population models on all SDSS spectra classified as galaxies in DR14 (including
CMASS and LRGs) and derived galaxy physical properties; stellar masses in particular are relevant here. Fits are performed with the full spectral fitting code Firefly (\citealt{wilkinson_2017}; \texttt{https://www.icg.port.ac.uk/firefly/}) and the stellar population models of \cite{maraston_2011} (hereafter M11), which are provided for several choices of stellar initial mass functions and input stellar library. We use their results for the M11-ELODIE population model, calculated assuming the ELODIE stellar library (\citealt{prugniel_soubiran_2007}) for a Salpeter stellar initial mass function
(\citealt{salpeter_1955}).
The CMASS stellar mass ($M_{\rm stellar}$) as a
function of redshift is shown in Fig.~\ref{stellarM_z_CMASS} (black and red
data points). Initially, we limit the stellar mass to 
$11.25 < \rm{log} (M_{stellar}/M_{\odot})<11.43$, which is a compromise between
a narrow stellar mass range and maximizing the number of  sources. 
The median stellar mass of this sample is $\langle \rm{log} (M_{stellar}/M_{\odot})\rangle=11.33$.

When we split the CMASS sample into low ($0.44<z<0.54$) and high
($0.54\leq z<0.64$) redshift subsamples, the high-z CMASS sample shows on average a slightly
higher clustering strengths on scales $\gtrsim$1 $h^{-1}$ Mpc. 
This is because the median value of the stellar
mass increases slightly from $z=0.44$ to $z=0.64$. This results from an incompleteness
in the CMASS sample stellar mass as a function of redshift as reported in, e.g., \cite{leauthaud_bundy_2016} and \cite{rodriguez_chuang_2016} (see their Fig.~3).
To remove this bias, we compute the median
stellar mass in bins of $\Delta z$=0.01 over the redshift range of
$z=0.44-0.64$ and add objects either below or above the primary chosen stellar
mass range in each individual bin until each redshift bin has an identical
median stellar mass of $\langle \rm{log}
(M_{stellar}/M_{\odot})\rangle=11.33$.
This results in the selection of 148,686
spectroscopic CMASS galaxies, as show in red in Fig.~\ref{stellarM_z_CMASS}.

As described in Appendix~\ref{smallscales} we correct for fiber collisions in the spectroscopic sample when calculating the correlation function. This increases the number of objects in the final CMASS sample to 150,898 galaxies. We consider only the CMASS galaxy sample from the SDSS North Cap
($\sim$6940 deg$^2$). The median redshift is $\langle z \rangle=0.53$ and the 
comoving number density is $(8.00 \pm 0.13) \times
10^{-5}$  $h^{3}$ Mpc$^{-3}$.  Uncertainties on the number densities
are determined using jackknife resampling, the details of which are given in Sect.~\ref{errorAnalysis} below.

\subsubsection{Random Catalogs}
The computation of correlation functions requires a random catalog.  We use the random catalogs provided by 
the SDSS consortium  (e.g., \citealt{reid_ho_2016}). 
The random catalogs for the CMASS sample in the North Cap can be
found at {\tt https://data.sdss.org/sas/dr12/boss/lss/}. As the 
data catalogs for the CMASS sample use weights to account for
all observational biases (see Appendix~\ref{smallscales}), the random catalogs can be used as provided.

For the random catalogs, we use 250 times as many objects as in the observed 
samples. This ensures that we 
minimize uncertainties due to statistical effects of the random catalog, in
particular for pairs at the smallest scales measured. We verify that the
randomly distributed objects fall within the same
SDSS DR12 footprint as the observed CMASS galaxies. 

As in papers I--IV, the corresponding redshifts for the random objects are
drawn from the smoothed redshift distribution of the observed CMASS
sample by applying a least-squares (\citealt{savitzky_golay_1964})
low-pass filter. The same jackknife subarea definition is used as that for the observed sample (see Sect.~\ref{errorAnalysis}). 


\subsection{AGN Samples}
\label{AGN_sample}

The goal of this paper is to compare the clustering properties of X-ray
and optically selected broad-line AGN samples. SDSS is perfectly suited for this purpose, as
it included follow-up spectroscopy of extragalactic X-ray sources 
and an
optical AGN identification program based primarily on color selection. In
addition, we aim to study AGN clustering properties as a function of various AGN parameters. 
The following subsections provide more details on the X-ray and optically selected 
AGN samples and how we create the different broad-line AGN subsamples.

\subsubsection{X-ray RASS/SDSS AGN}
\label{RASS_sample}
The X-ray selected AGN sample is based on SDSS-IV DR14 SPIDERS (SPectroscopic
IDentification of eROSITA Sources; \citealt{blanton_bershady_2017}).
The X-ray telescope eROSITA (\citealt{predehl_andritschke_2021}) on board of the 
SRG observatory (\citealt{sunyaev_arefiev_2021}) was successfully 
launched in 2019 July. However, due to the delay of the launch, SDSS DR14 SPIDERS targets could not be selected 
using eROSITA. Instead the precursor  \textit{ROSAT} mission (\citealt{truemper_1982})
was used as the main input for target selection. 
The \textit{ROSAT} All-Sky Survey (RASS, \citealt{voges_aschenbach_1999}) is 
an all-sky survey in the soft (0.1--2.4 keV) X-ray regime. 
Using this survey, \cite{voges_aschenbach_1999} and
\cite{voges_aschenbach_2000} presented the RASS bright and faint source
catalog, respectively. A goal of the ongoing SPIDERS project is to obtain highly
complete and reliable optical identifications for these sources with the SDSS
telescope. 

We make use of SDSS DR14 and consider only X-ray sources that are listed in
the revised version of the RASS catalog presented by
\cite{boller_freyberg_2016} (2RXS). 
Several major improvements and bug fixes compared to the previous version of
RASS resulted in the 
deepest and cleanest X-ray all-sky survey collected by \textit{ROSAT}. 
In particular, 2RXS aims for
a significant reduction of the number of spurious sources.

We supplement the DR14 SPIDERS 2RXS sample \citep{coffey_salvato_2019, comparat_merloni_2020}
with RASS sources lying outside the eBOSS/DR14 footprint but within the footprint 
of earlier SDSS spectroscopic coverage. We cross-match the AllWISE counterparts 
to 2RXS sources \citep{salvato_buchner_2018} to the SDSS-DR13 photometric catalog, 
choosing the brightest counterpart (in the $r$ band)  within a 3 arcsec radius 
of the AllWISE position. The optical counterparts are then matched to the 
SDSS DR14 optical spectroscopic catalog (\citealt{abolfathi_aguado_2018}) using a search radius of 1 arcsec.

The reported luminosities are 0.1--2.4 keV $k$-corrected rest-frame luminosities, 
assuming a photon index of $\Gamma$=2.4. They are corrected for Galactic absorption.
We restrict the X-ray selected AGN sample to the same redshift range as our
CMASS galaxy sample ($0.44<z<0.64$) and to the same SDSS footprint. 
This yields a total of $\sim$2130 in the North Cap ($100\le\rm{R.A.}\le270$). We apply an additional selection 
to this sample in Sect.~\ref{sec:line} below and obtain a final sample of 1701 broad-line X-ray AGN.

\subsubsection{Optical SDSS AGN}
\label{Paris_sample}

The optical AGN sample is drawn from the SDSS Quasar Catalog DR14 (\citealt{paris_petitjean_2018}), which contains all AGN observed as part of SDSS I--IV.
The newly discovered SDSS-IV/eBOSS AGN arise from the
target selection presented by \cite{myers_palanque-delabrouille_2015}. 
\cite{paris_petitjean_2018} defined an AGN as an object with a luminosity of
$M_{\rm i}[z=2]< -20.5$ that also has at least one emission line with FWHM $>
500$ km s$^{-1}$ or has interesting or complex absorption features. 
Given the large number (526,356) of objects in the catalog, visual
inspection of the optical spectra of these sources is not  feasible. The
automatic classification pipeline determines the redshift based on the maximum
peak of the Mg II emission line, which is observable in eBOSS spectra from
$z=0.3-2.5$.

As with the X-ray selected AGN sample, we limit the optical AGN sample to the same
SDSS footprint and redshift range ($0.44<z<0.64$) as our CMASS galaxy sample. This
results in a total of 11,298 optical AGN in the North Cap. Note that we further down-select  
this sample in Sect.~\ref{sec:line} and obtain 10,994 broad-line optical AGN.

There is a significant overlap between the X-ray and optical AGN samples: 95.7\% of the X-ray selected AGN are also selected by the optical AGN  selection. 
However, 
the two samples do reflect different AGN selection criteria and AGN populations.  
The RASS sample is essentially flux-limited. The optical AGN sample is based on a  heterogeneous AGN selection based on  a combination of nonstellar optical color selection and radio point-source detection. As the host galaxy can contribute significantly to the total optical emission, 
there is a selection bias against detecting low-luminosity AGN in galaxies where the optical host light is substantial.

\begin{figure}
  \centering
\resizebox{\hsize}{!}{  \includegraphics[]{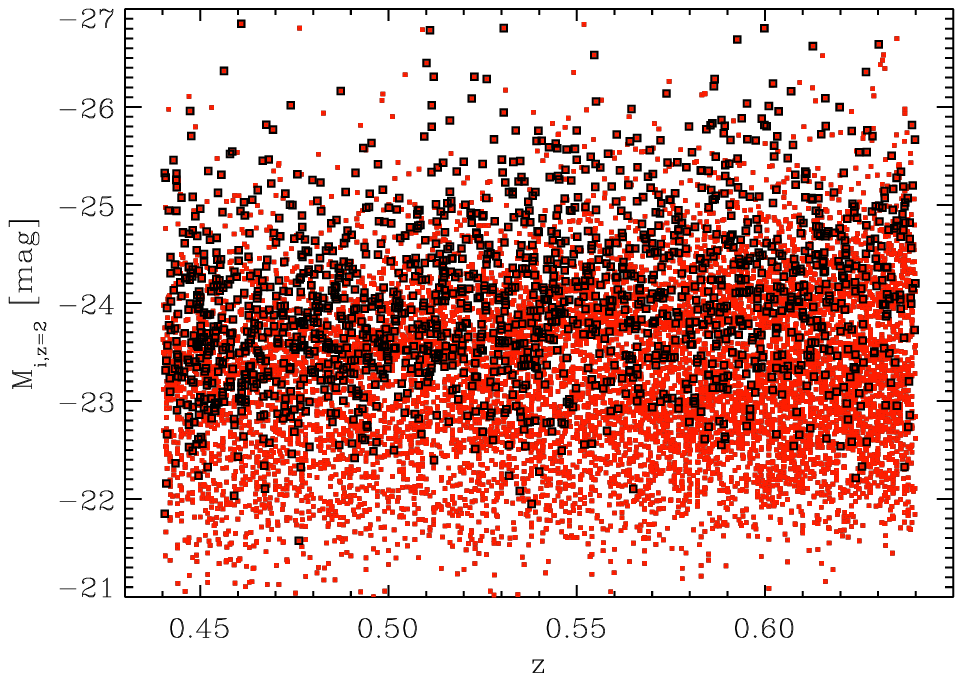}} 
      \caption{Comparison of the X-ray selected (RASS/SDSS) BLAGN sample
        (black squares) and the optically selected (SDSS) BLAGN sample (red points), showing  absolute $i$-band magnitude (k-corrected to $z = 2$) versus redshift. 
      }
\label{Xray_optical_comparison}
\end{figure}

Figure~\ref{Xray_optical_comparison} shows the overlap between and differences in the two selection methods. The X-ray detected AGN have on average higher $M_i(z=2)$ values than the average of the optical broad-line AGN (BLAGN) sample. However, there are also a substantial number of optical BLAGN with very high $M_i(z=2)$ values that are not detected in X-rays: 30\% percent of optical BLAGN brighter than $M_i(z=2) = -24$ mag are not detected in  X-rays. For BLAGN brighter than $M_i(z=2) = -25$ mag, this fraction increases to almost 50\%. Given the substantial differences in these samples, we aim to test whether the different BLAGN selections lead to differences in the clustering properties of the samples.

\section{Creating Broad-line AGN Samples}

For the X-ray selected and optical AGN samples, we fit the individual
SDSS optical spectra. In the redshift range of
$0.44<z<0.64$, the H$\beta$ line is always in the available SDSS spectral
wavelength range, except for objects in which this wavelength range is masked
out based as a result of quality checks. We use the H$\beta$ line properties to
include or exclude an object in our broad-line AGN samples. 
In \cite{krumpe_miyaji_2015}  we used the H$\alpha$ line to make this 
distinction as the AGN sample originated from a lower redshift range ($0.16<z<0.36$).
Using the H$\beta$ line properties not only allows for the generation of
a broad-line AGN sample but is also used to estimate the $M_{\rm BH}$ and 
$L/L_{\rm EDD}$. The details of the procedure are given in the following
subsections.

\subsection{H$\beta$ Spectral Line Measurements} \label{sec:line}
We model the spectral region around the H$\beta$ line for all SDSS optical
spectra.  
For the X-ray sample, we visually inspect each fit to verify its robustness 
and improve the fit if necessary. For the larger optical AGN sample, we visually inspect 
a subset, focusing on potential  problematic cases. 

In particular, the X-ray sample includes not only broad-line AGN but narrow-line AGN, inactive  galaxies, and stars as well. We use 
the spectral class given by the SDSS pipeline \citep{bolton_schlegel_2012} as an 
initial estimate of the nature of the X-ray source. For objects classified 
as {\tt QSO} or {\tt BLAGN}, we fit a broad-line model to the SDSS spectrum. 
In other cases, we first test if a potential line with a peak S/N$>3$ per 
pixel is present at the location of H$\beta$. If so, we fit the spectrum 
with a narrow-line-only and a narrow+broad-line model. 
We use the latter model only if it leads to an improvement in the reduced $\chi^2$ by at least 
25\%. This threshold is empirical and based on our visual quality checks of
the fits. Otherwise we classify the object as a narrow-line object.

For the optical AGN sample, we initially fit every object with a broad-line model.
For cases with a low amplitude of broad H$\beta$ or cases identified during 
the visual inspection, we refit the spectrum with both a narrow-line and 
narrow+broad-line model and evaluate if the addition of a broad component is 
justified as described above.
Based on this procedure, we identify 1701 and 
10,994 objects that require a broad-line component in the fit in the X-ray 
and the optical AGN DR14 sample, respectively.

\begin{figure}
  \centering
\resizebox{\hsize}{!}{  \includegraphics[]{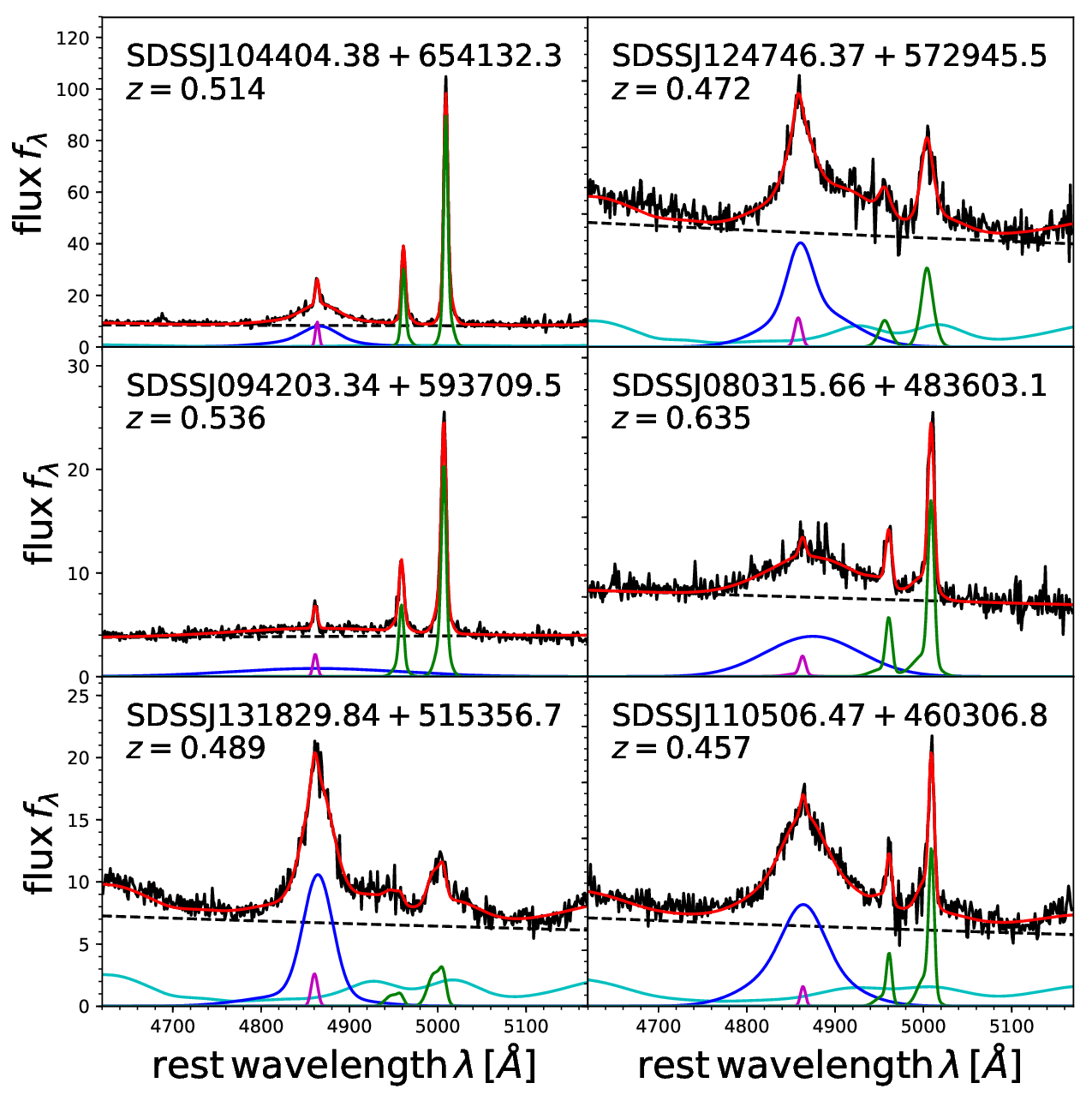}} 
      \caption{Example SDSS spectra (black solid line) of typical S/N sources and 
the best-fit models (red solid line) of the H$\beta$ region. We show the 
individual model components, including a power law (black dashed line), the 
iron template (cyan line), the broad (blue) and narrow (magenta) H$\beta$ 
lines, and the [\ion{O}{3}] lines (green).}
     \label{fig:spec_fit}
\end{figure}

Our procedure for continuum and emission line fitting uses code originally presented in \citet{schulze_schramm_2017,schulze_silverman_2018}. 
It builds on a Levenberg-Marquardt least-squares minimization as implemented 
in MPFIT \citep{markwardt_2009}. We first correct the spectra for galactic
extinction using the extinction map from \citet{schlegel_finkbeiner_1998} and the
reddening curve from \citet{cardelli_clayton_1989} and shift the spectra to their rest frame.
We then fit and subtract a local pseudo-continuum over the wavelength windows $4435-4680$ \AA{} and $5100-5535$ \AA{}. The model consists
of a power-law continuum and an optical iron template \citep{boroson_green_1992}, 
broadened by a Gaussian. The pseudo-continuum subtracted spectrum is then fit with
an emission line model over the range $4700-5100$ \AA\,. 
For the narrow-line model, we fit the narrow H$\beta$ and the narrow  
[\ion{O}{3}] $\lambda\lambda4959,5007$ lines, 
each fit by a single Gaussian. The [\ion{O}{3}]
$\lambda\lambda4959,5007$ doublet is fit with two Gaussians for each line, one for the
line core and one for the blue wing often present in this line 
(e.g., \citealt{mullaney_alexander_2013}). The two [\ion{O}{3}] lines are coupled  
in shape and their line ratio is fixed to 3.0. The line width and velocity centroids of the
core components are tied together for H$\beta$ and the 
[\ion{O}{3}]  lines. In the broad-line model we add a broad H$\beta$ line, 
which we model with up to three Gaussians with FWHM$>1000$~km s$^{-1}$ each. We 
stress that we do use these to capture the often non-Gaussian profile shape of 
the broad H$\beta$ line and do not assign any physical nature to the individual 
components.

We measure the FWHM of the broad H$\beta$ line (corrected for the instrumental
resolution) and the continuum flux at 5100\AA{} from the best-fit model. In
Figure~\ref{fig:spec_fit} we show spectra and the best fit models for
six representative objects with typical S/N. For spectra with low S/N the uncertainties and
systematic errors on the line width measurements can become substantial, compared to the
systematic uncertainty of $\sim0.3$~dex of the virial method to estimate black
hole masses. Therefore, we use a lower S/N threshold for the sample, similar to paper~IV. We define the S/N over the H$\beta$ range as the
median S/N per pixel over the range $4750-4950$ \AA{}.
In Appendix~\ref{reliability} we show that H$\beta$ FWHM measurements with
S/N$>$5 should be used to determine robust results and  allow for a meaningful estimate of $M_{\rm BH}$. In Appendix~\ref{comparing} we compare our H$\beta$ measurements with those from other studies and find good agreement.
The broad-line X-ray and optically selected AGN samples used here (H$\beta$-S/N$>$5) have median S/N values of 14.1 and 10.8, respectively.

\subsection{Estimating Black Hole Masses and Eddington Ratios} \label{sec:mbh}
For broad-line AGN, black hole masses can be estimated using the
viral method \citep[e.g.][]{mclure_jarvis_2002, vestergaard_peterson_2006}. This technique 
builds on the assumption of virialized motions of the broad-line region (BLR)
gas \citep{peterson_wandel_2000}, where the broad-line width serves as an estimate of
the gas velocity, and uses an empirical scaling relation between BLR size and
continuum luminosity \citep{kaspi_smith_2000,bentz_peterson_2009}. The typical uncertainty on \mbh using 
the virial method for individual objects is $\sim$0.3--0.4~dex. The method is
particularly powerful when studying the statistical properties of large samples of broad-line AGN \citep[e.g.][]{mclure_dunlop_2004,kelly_shen_2013,schulze_bongiorno_2015}. 

The virial method is most directly calibrated for the H$\beta$ line; thus, this
line typically provides the most reliable black hole mass estimate. The
\ion{Mg}{2} line, which is covered by the SDSS spectra in our sample, is also known
to be a reliable black hole mass estimator
\citep{trakhtenbrot_netzer_2012,mejia_trakhtenbrot_2016}. However, the \ion{Mg}{2} mass estimator
is calibrated to H$\beta$, so when both lines are available (as in our
sample), H$\beta$ is the preferred choice.  We therefore do not fit the
\ion{Mg}{2} line for this study. 
We estimate black hole masses from H$\beta$ using the formula given by
\citet{vestergaard_peterson_2006}:
\begin{equation}
\mbh (\rm{H}\beta)= 10^{6.91} \left( \frac{L_{5100}}{10^{44}\,\mathrm{erg\,s}^{-1}}\right)^{0.5} \left( \frac{\mathrm{FWHM}}{1000\,\mathrm{km\,s}^{-1} }\right)^2 M_\odot    \label{eq:mbhHb}
\end{equation} 

For the broad luminosity range of our AGN sample, the host galaxy
contamination to the continuum luminosity $L_{5100}$ is not
negligible. \citet{shen_richards_2011} showed that host galaxy contamination becomes
significant at $L_{5100}<10^{45}$ erg s$^{-1}$. We account for the host
contribution in an average sense by applying the formula for the average host
contamination given by \citet{shen_richards_2011}. 
 
We estimate the bolometric luminosity for our sample from the (host-corrected)
continuum luminosity $L_{5100}$. Specifically, we use a constant bolometric
correction factor $\rm{BC_{5100}}=7.0$ \citep{netzer_trakhtenbrot_2007}, which is consistent on
average with the luminosity-dependent bolometric correction of
\citet{marconi_risaliti_2004} and excludes reprocessed emission from the
mid-IR. \citet{schulze_silverman_2018} demonstrated the consistency of this
bolometric correction factor with $L_{\rm bol}$ obtained by direct integration
of the spectral energy distribution (SED) and other common bolometric
luminosity indicators. Combining our estimates of \mbh\, and $L_{\rm bol}$
provides the Eddington ratio $L/L_{\rm EDD}$ for AGN in our sample, where
$L_{\rm{EDD}}\cong1.3\times 10^{38} (\mbh / M_\odot)$~erg s$^{-1}$ is the
Eddington luminosity. 

We also compute the absolute magnitude as the $z=2$ $k$-corrected $i$-band 
magnitude for both the X-ray and the optical AGN samples following 
\cite{richards_strauss_2006} using the $k$-correction from their Table~4.
This allows us to better compare  the optical and X-ray AGN samples, as
for the latter, no $M_i(z=2)$ is provided. For the optical AGN we compare our 
$M_i(z=2)$ values with those in  \cite{paris_petitjean_2018} and find excellent agreement. In 68\% of the sample, the magnitudes differ by less then 0.056
mag, and in 95\% of the sample, they differ by less then 0.085 mag.

\subsection{Defining Broad-line X-ray AGN Subsamples}

\label{matchingSamples}
\begin{figure}
  \centering
 \resizebox{\hsize}{!}{ 
  \includegraphics[width=8cm,height=5cm]{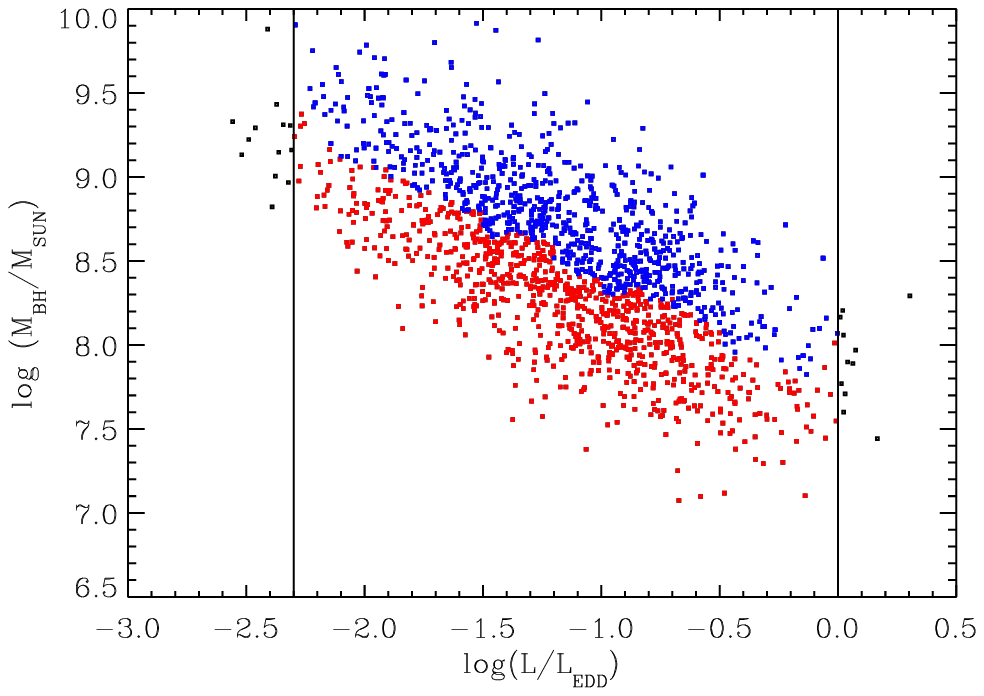}} 
      \caption{$M_{\rm BH}$ vs. $L/L_{\rm EDD}$ for broad-line 
               X-ray RASS/SDSS AGN with $S/N$ at H$\beta$ $\geq 5$.  Blue symbols show the high $M_{\rm BH}$
               sample, while red symbols show the low $M_{\rm BH}$ sample
               (the two samples have identical distributions in $L/L_{\rm EDD}$). 
               Vertical lines mark the restriction of
               the parameter space to remove extreme objects.}
         \label{RASS_AGN_MBH_LLEDD}
\end{figure}

Due to the relatively low number of sources,
we decide to split the X-ray selected broad-line AGN sample into only two
subsamples as a function of various AGN parameters. 
First we divide the sample into low and high X-ray luminosity subsamples at the
median log ($L_{\rm X}/[\rm{erg\,s}^{-1}])= 44.82$ of the full sample. The properties of all X-ray AGN
samples as well as subsamples are given in Table~\ref{tab:overview}. 

The observed X-ray luminosity is
a combination of the physical parameter $M_{\rm BH}$ and $L/L_{\rm EDD}$.
However, as in other redshift ranges (see paper IV), the RASS/SDSS AGN do
not uniformly populate the $L/L_{\rm EDD}$ -- $M_{\rm BH}$ plane 
(Fig.~\ref{RASS_AGN_MBH_LLEDD}). Higher $L/L_{\rm EDD}$ are usually found in AGN with lower $M_{\rm BH}$. 
The absence of objects in the lower-left corner of the plane is an
observational bias. 
To remove the correlation between $M_{\rm BH}$ and $L/L_{\rm EDD}$ and test
the clustering properties of each AGN parameter independently, we create
subsamples that depend on one parameter only, while the distribution of the second
parameter in both subsamples is identical. This ``matching'' of the subsamples is a commonly used method in clustering 
measurements (e.g., \citealt{coil_georgakakis_2009}) and was used in
Paper IV. Thus we split the AGN sample into low and high $M_{\rm BH}$ samples
with identical distributions in $L/L_{\rm EDD}$ and vice versa. 
To do so, we first remove extreme objects by  considering only 
objects with $-2.3 < {\rm log} (L/L_{\rm EDD}) < 0.0$ for the $M_{\rm BH}$
split and $7.4 < {\rm log} (M_{\rm BH}/M_{\odot}) < 9.6$ for the 
$L/L_{\rm EDD}$ split. We then determine the number of
objects in each bin in the parameter we are matching on, using a bin width of 0.1 for both parameters. 
Then in each bin, we split the sample at the median value of this bin and 
allow multiple draws of objects, to result in an identical number of objects
in the low and high subsamples in each bin. As an example, we show the matched 
$M_{\rm BH}$ division in Fig.~\ref{RASS_AGN_MBH_LLEDD}. The $L/L_{\rm EDD}$
distribution of the low and high $M_{\rm BH}$ samples are identical.

We also split the sample at the median log $(L_{\rm Bol}/[\rm{erg\,s}^{-1}])= 45.43$ and create low and high $L_{\rm Bol}$ subsamples. 
Finally, we split the X-ray broad-line AGN sample into
faint and bright $M_i(z=2)$ subsamples using the median $M_i(z=2)=-24.07$ mag of
the full AGN sample with reliable $M_{\rm BH}$ estimates. 

\subsection{Defining Broad-line Optical AGN Subsamples}

\begin{figure}
  \centering
 \resizebox{\hsize}{!}{ 
  \includegraphics[width=8.0cm,height=6.0cm]{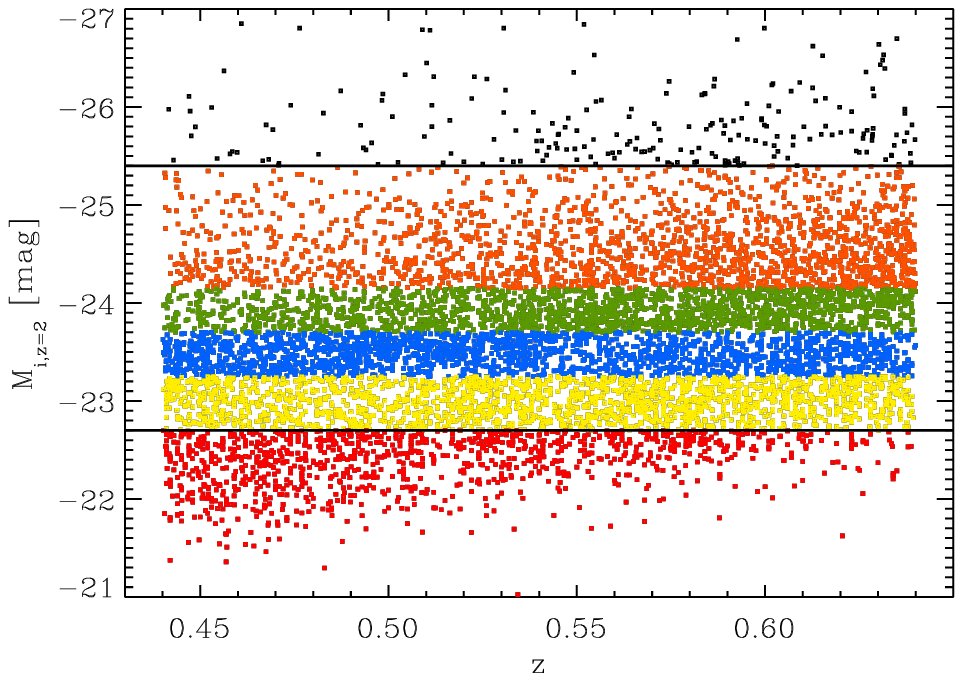}} 
      \caption{Absolute $i$-band magnitude ($k$-corrected to $z=2$) vs. redshift
        for the broad-line optical AGN with $S/N$ at H$\beta$ $\geq 5$. The
        different colors represent the different $M_i(z=2)$ subsamples. 
       The few brightest objects with
        $M_i(z=2)<-25.4$ mag are not considered for the $M_i(z=2)$
        subsamples.} 
         \label{OPT_AGN_MI_z}
\end{figure}

For the optically selected SDSS AGN, we follow a very similar approach as for
the X-ray selected AGN. The major difference is that this sample has more than
five times as many objects than the X-ray AGN sample. Thus we split the
optical AGN sample in four, instead of two, subsamples in each parameter of interest. 

We again create AGN subsamples with respect to $M_{\rm BH}$, $L/L_{\rm EDD}$ (with matched distributions 
in the other parameter of interest).
We create these matched samples using objects 
with $-2.4 < {\rm log} (L/L_{\rm EDD}) < 0.0$ for the $M_{\rm BH}$
split and $7.0 < {\rm log} (M_{\rm BH}/M_{\odot}) < 9.6$ for the 
$L/L_{\rm EDD}$ split. 
We create four $L_{\rm Bol}$ subsamples by determining the 
values of $L_{\rm Bol}$ in the optical broad-line AGN sample 
(with reliable $M_{\rm BH}$ measurements) that correspond to 25\%, 50\%, and 75\% objects of the full sample. 

We also split the optical AGN sample in subsamples as a function of $M_i(z=2)$.
In order to have well-selected subsamples with similar redshift distributions,
we limit the sample first at $M_i(z=2)=-22.7$ mag and
$M_i(z=2)=-25.4$ mag. Figure~\ref{OPT_AGN_MI_z} shows the four resulting
subsamples, each of which have a  similar number of objects. We keep the objects 
with $M_i(z=2)>-22.7$ mag as an additional fifth $M_i$ subsample. 
The properties of all optical AGN subsamples are listed in
Table~\ref{tab:overview}.

\begin{figure*}
 \begin{minipage}[b]{0.48\textwidth}
\centering
 \resizebox{\hsize}{!}{
  \includegraphics[width=8cm,height=5cm]{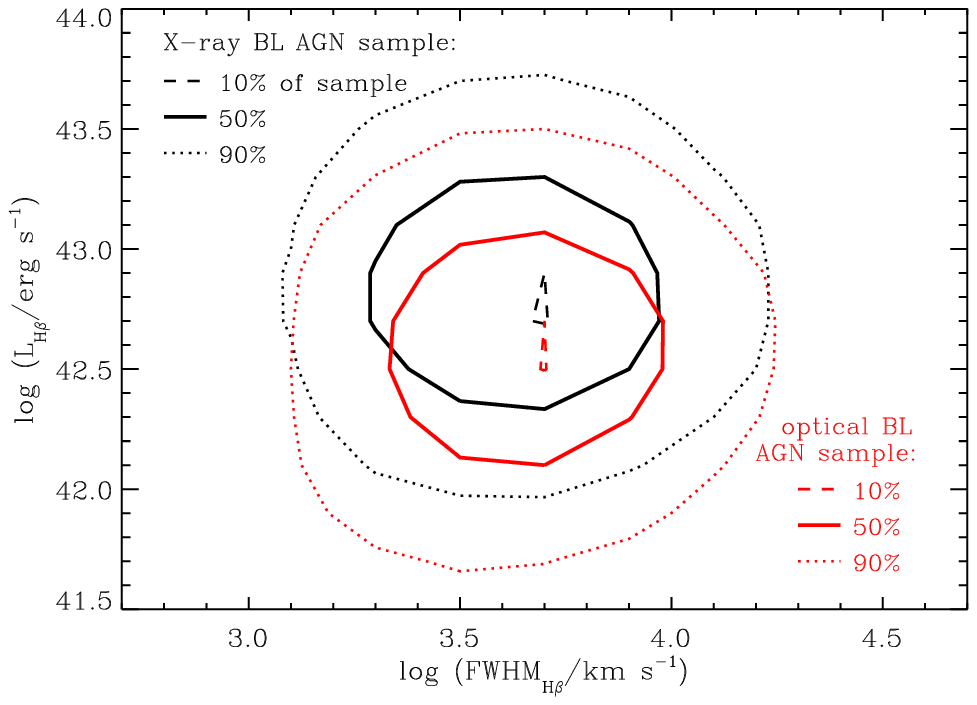}}
\end{minipage}
\hfill
\begin{minipage}{0.48\textwidth}
\vspace*{-4.9cm}
\centering
\resizebox{\hsize}{!}{
  \includegraphics[width=8.0cm,height=5.0cm]{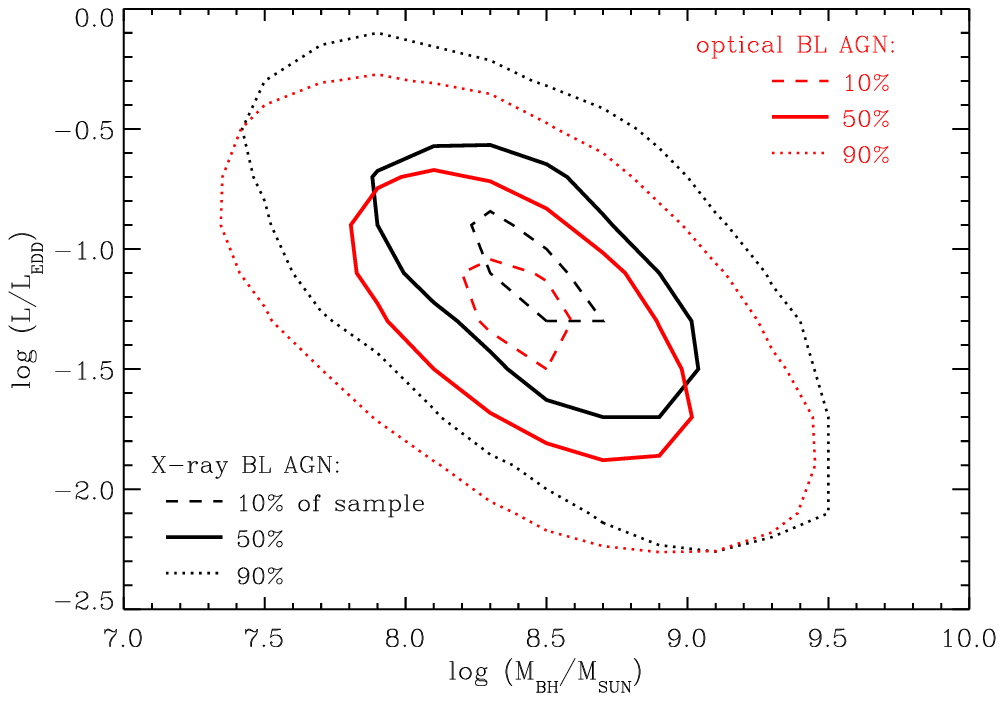}}
\vspace*{-0.0cm}
\end{minipage}
\caption{Comparison between broad-line X-ray selected RASS/SDSS AGN (black lines) and  
broad-line optically selected SDSS AGN (red lines) with S/N at H$\beta$ $>5$. The contours show the location of 10\% (dashed lines), 
50\% (solid lines), and 90\% (dotted line) of 
the full sample, respectively. The left figure shows the observed parameters of the  
luminosity and FWHM of the H$\beta$ line, while the right figure 
shows the derived parameters $L/L_{\rm EDD}$ and $M_{\rm BH}$. 
In both figures a boxcar smoothing with a width of two has been applied.}
\label{contour_compare}
\end{figure*}

When comparing the X-ray and optical broad-line AGN samples (see
Fig.~\ref{contour_compare}) in the space of the observed parameters $L_{\rm H\beta}$ vs. FWHM (left panel) and the 
derived parameters $L/L_{\rm EDD}$ vs. $M_{\rm BH}$ (right panel),
the two samples cover a similar though not identical parameter space. 
Unlike our previous lower-redshift samples ($0.16 \le z \le 0.36$; paper IV), here the X-ray selected AGN sample 
does not extend to lower $L_{\rm H\beta}$. Instead the optical selection is
able to detect more objects with lower H$\beta$ luminosities. {\it ROSAT}'s
flux sensitivity is limited, and so with increasing redshift, even luminous AGN will drop below the RASS flux limit. This is reflected in the total number of selected objects, as well as  when directly comparing the number of objects with log ($L_{\rm H\beta}/[\rm {erg ^{-1}}]) \geq 43.5$, where we have 86 X-ray AGN and 141 optical AGN.  

Differences in the observed parameter space naturally translate into differences 
in the derived $M_{\rm BH}$--$L/L_{\rm EDD}$ space 
(Fig.~\ref{contour_compare}, right panel). The X-ray AGN sample tends
to select objects with somewhat higher $M_{\rm BH}$ and higher $L/L_{\rm EDD}$. This is
not surprising as the X-ray luminosity is increasing from the lower-left to
the upper-right corner in this parameter space. The optical  
sample extends to lower $M_{\rm BH}$ and lower 
$L/L_{\rm EDD}$.

%

\section{Method}

\subsection{Clustering Measurements}
\label{pimaxvalue}
We measure the two-point correlation function $\xi(r)$ (\citealt{peebles_1980}), which measures 
the excess probability $dP$ above a Poisson distribution. The auto-correlation 
function (ACF) measures the spatial clustering of objects in the same sample, 
while the cross-correlation function (CCF) measures the clustering of objects in 
two different samples. We use the same approach as 
described in detail in papers I--IV . 
Here we only repeat the most essential elements of our method.

We use the correlation estimator of \cite{davis_peebles_1983} in the form
\begin{equation}
\label{DD_DR}
 \xi(r)= \frac{DD(r)}{DR(r)} -1\ ,
\end{equation}
where $DD(r)$ is the number of data-data pairs with a separation $r$, and $DR(r)$ 
is the number data-random pairs. Both pair counts have been normalized by the number density of
data and random points. For our purposes, the use of this simple estimator has a number of major advantages and results in only 
a marginal loss in the S/N when compared to more advanced estimators 
(e.g., \citealt{landy_szalay_1993}). The estimator in Equation~(\ref{DD_DR}) 
requires the generation of a random catalog for the tracer set only.

To separate the effects of redshift distortions, the correlation 
function is measured as a function of two components of the separation vector 
between two objects, one perpendicular to ($r_p$) and the other along 
($\pi$) the line of sight. $\xi(r_p,\pi)$ is thus extracted by counting 
pairs on a two-dimensional grid of separations $r_p$ and $\pi$. 
We obtain the  projected correlation function $w_p(r_p)$ by integrating $\xi(r_p,\pi )$ along
the $\pi$-direction.

We measure $r_p$ in the range of 0.05--40 $h^{-1}$ Mpc in 15 logarithmic bins, 
identical to those used in papers III and IV.
We compute $\pi$ in bins of 5 $h^{-1}$ Mpc in the range $\pi=0-200$ $h^{-1}$ Mpc.
To derive $\pi_{\rm max}$, we compute $w_p(r_p)$ for a set of $\pi_{\rm max}$ ranging 
from 10--160 $h^{-1}$ Mpc in steps of 10 $h^{-1}$ Mpc. We then fit $w_p(r_p)$ over 
a $r_p$ range of 0.35--30 $h^{-1}$ Mpc with a fixed $\gamma = 1.9$ 
(based on \citealt{krumpe_miyaji_2010})
and determine the 
correlation length $r_{\rm 0}$ for the individual $\pi_{\rm max}$
measurements. 
With increasing $\pi_{\rm max}$ more clustering signal is included ,though at some $\pi_{\rm max}$ value, the estimated $w_p(r_p)$ stops increasing and the uncertainty of $w_p(r_p)$ increases afterwards.
For the CMASS galaxy ACF and all AGN--CMASS CCFs, the correlation signal
saturates at $\pi_{\rm max,ACF}=80$ $h^{-1}$ Mpc and $\pi_{\rm max,CCF}=40$ $h^{-1}$
Mpc, respectively. 
Previous CMASS ACF studies also use $\pi_{\rm max,ACF}=80-100$ $h^{-1}$ Mpc 
(e.g., \citealt{white_blanton_2011, guo_zehavi_2013, nuza_sanchez_2013,guo_zheng_2015}). 

Spectroscopic broad-line AGN redshifts have on average larger uncertainties than spectroscopic redshifts of galaxies. Typically the redshift  uncertainties are $\delta z \sim 0.01$. Since we are 
integrating over $\pi_{\rm max,CCF}=40$ $h^{-1}$ Mpc, these small redshift uncertainties have 
a negligible impact on the clustering signal.

\subsection{Error Analysis}
\label{errorAnalysis}

The error analysis is identical to that used in papers I--IV.
We use the jackknife resampling technique to estimate the measurement errors
based on the covariance matrix $M_{ij}$, which reflects the degree to which bin $i$ is 
correlated with bin $j$. 

We divide the survey area into $N_{\rm T}=100$ subsections, which have roughly
an equal area of $\sim$70 square degrees. At the median redshift of
the CMASS sample ($\langle z \rangle=0.55$), each subarea spans a
physical scale at least four times larger than our largest scales studied in this paper. 

The $N_{\rm T}$ jackknife-resampled correlation functions define the 
covariance matrix: 
\begin{eqnarray}
\label{jackknife}
 M_{ij} = \frac{N_{\rm T} -1}{N_{\rm T}} \left[\sum_{k=1}^{N_{\rm T}} \bigg(w_k(r_{p,i})-<w(r_{p,i})>\bigg)\right.\nonumber\\
          \times \bigg(w_k(r_{p,j})-<w(r_{p,j})>\bigg)\bigg] \,  
\end{eqnarray}
We calculate $w_p(r_p)$ $N_{\rm T}$ times, where each jackknife sample excludes one 
section; $w_k(r_{p,i})$ and $w_k(r_{p,j})$ are from the $k$th jackknife
samples of the CMASS ACF and AGN CCFs, respectively, and $<w(r_{p,i})>$ and $<w(r_{p,j})>$ are the averages
over all of the jackknife samples. The uncertainties represent 1$\sigma$
(68.3\%) confidence intervals.


\subsection{HOD Modeling}

HOD modeling is a popular method for interpreting correlation function results. In this model the sample objects are assumed to be in DMHs where the average number of objects per halo follows a parameterized function of halo mass separately for the central halo and satellite halos. The correlation function can be  modeled as a sum of the contribution of pairs from the same DMH (one-halo term) and different DMHs (two-halo term).
We interpret our results using the HOD modeling following
an approach similar to that presented in papers II and IV as well as in
\citet{krumpe_miyaji_2018}.
We use the HOD approach to obtain linear bias parameters, as well as to 
investigate differences in the measured CCFs among various AGN subsamples 
with CMASS galaxies, probing  differences beyond using the bias parameters alone. 

As we do 
not intend to present full constraints of the AGN HODs in this paper, the HOD model
of the AGN is deliberately designed to be simple. In order to enable historical comparisons 
with the results from our previous work, we use the same expressions for the $b$--$M_{\rm DMH}$ relation \citep{tinker_weinberg_2005} and the halo mass function \citep{sheth_mo_2001}, as in our previous work.
In \citet{krumpe_miyaji_2018}, we use an improved version of our code that takes into account the effects of halo-halo collisions and scale-dependent bias \citep{tinker_weinberg_2005}. Practically, this removes the need for excluding the transition range between one-halo and two-term-dominated regimes in the HOD modelings (see papers II--IV).

The outline of our approach is as follows:
\begin{enumerate}
\vspace*{-0.2cm}\item We first apply the HOD modeling technique to the ACF of the CMASS galaxies and
  determine central and satellite HODs using a correlated $\chi^2$-fit with a model with four free parameters.
  The number density constraint is also included.
  The best-fitting parameter search and the associated confidence regions are generated using a Markov Chain Monte Carlo (MCMC) method.
\item In order to model the CCF between the CMASS galaxies and AGN samples, two HODs are required: the HOD of the CMASS galaxies derived in the previous step and that of the AGN. As the CMASS galaxy sample is much larger than the AGN samples  
(and therefore the statistical significance of the CMASS galaxy ACF is much higher than that of the CMASS galaxy--AGN CCF), we use the best-fit CMASS galaxy HOD derived above for calculating the uncertainties of the AGN HOD. Due to the lower S/N of the CCF measurements between AGN and CMASS galaxies, some parameters remain unconstrained when applying our HOD approach. We therefore fix these parameters to reasonable values. We use an MCMC parameter search when there are three free parameters to fit. If we fit only two free parameters, we apply a simple grid search. Details are given below. 
\end{enumerate}

\begin{figure*}

\begin{center}
\hbox{\hspace*{-0.4cm}
 \includegraphics[width=8cm,height=8cm]{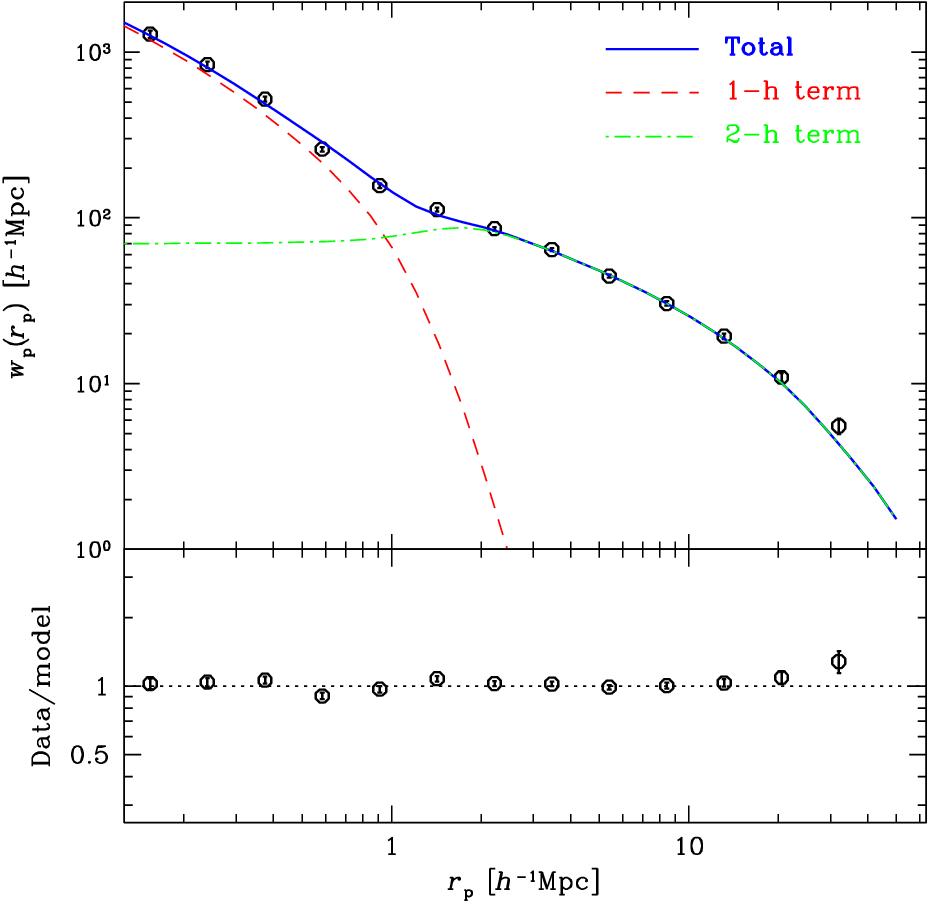}
\hspace{0.2cm}
 \includegraphics[width=9.9cm,height=8.0cm]{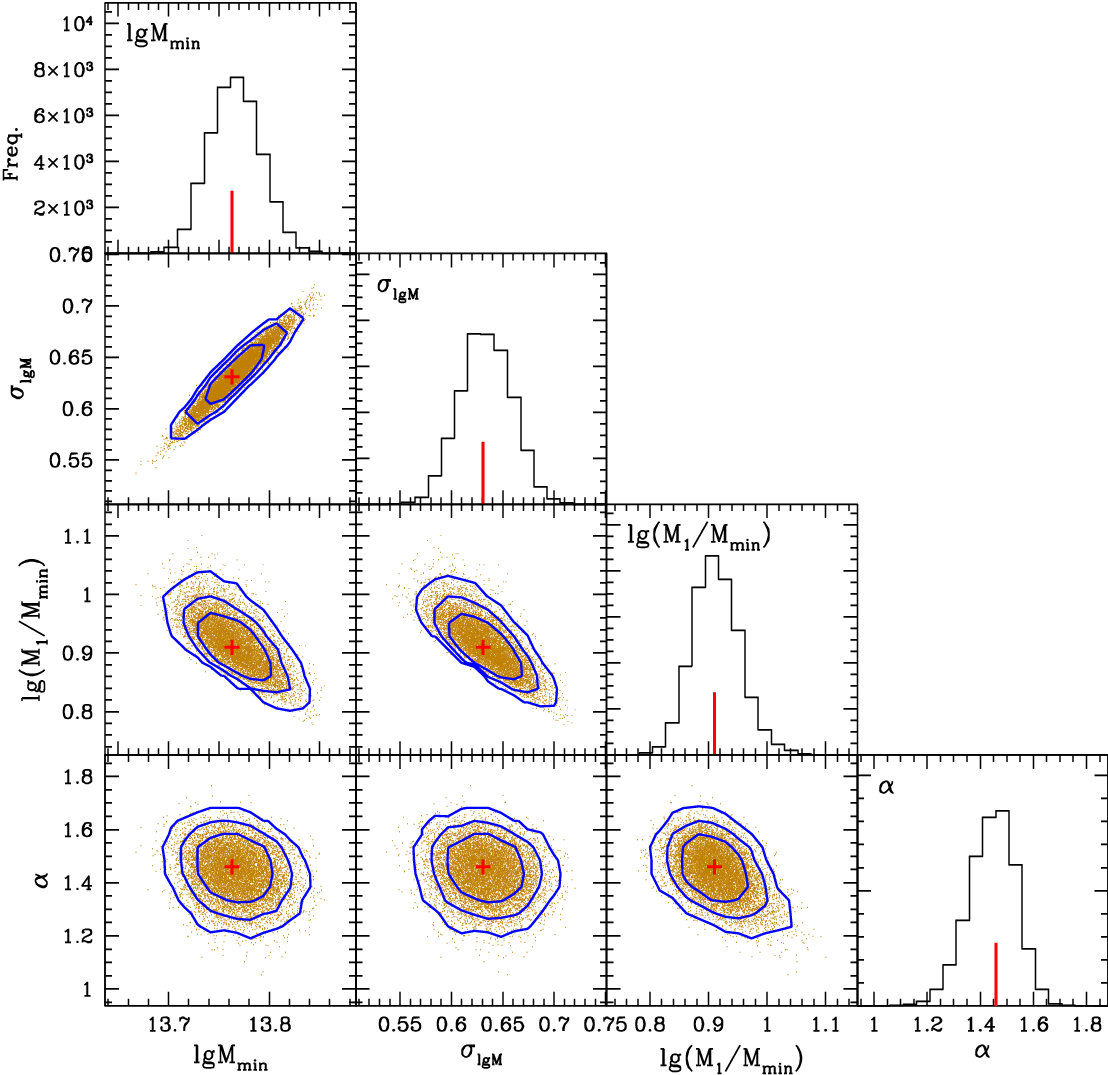}
}
\end{center}
\caption{{\it Left:}
The CMASS galaxy ACF (black circles) with 1$\sigma$ error bars is shown, along with
the best fit HOD model (blue solid line). The one-halo and two-halo terms
of the HOD model are shown with red short dashed and green long dashed
lines, respectively. 
The lower panel shows the fit residuals in terms of the
data/model ratio.
{\it Right:} The MCMC points and confidence contour matrix of the CMASS HOD model parameters are shown, along with the marginal probability distribution of each parameter in the diagonal panels.
The best fit value in a two parameter space is indicated with a red cross, while the best fit value for one parameter is marked with a red vertical line.
Individual dots indicate the MCMC chain points in the corresponding two parameter space. The three contours in each panel correspond to
$\Delta \chi^2=\chi^2-\chi^2_{\rm min}$ of 4.8, 7.9, and 12.1, below which 
68, 90, and 98\% of the chain points fall respectively. 
In these figures parameters name have been shortened for labeling purposes by omitting spaces and expressing 
$\log$ as ${\rm lg}$. }
\label{fig:CMASSacfhod}
\end{figure*}

\subsubsection{HOD of CMASS Galaxies}
\label{sec:CMASShod}

For the HOD modeling of the CMASS galaxy sample, we use the five-parameter model of \citet{zheng_coil_2007}:
\begin{eqnarray}
\langle N_{\rm G,c}\rangle (M_{\rm DMH}) &=& 
\frac{1}{2}\left[1+\mathrm{erf}\left(\frac{\log M_{\rm DMH}-\log M_{\rm
min}}{\sigma_{\log M}}\right)\right]\nonumber\\
\langle N_{\rm G,s}\rangle (M_{\rm DMH}) &=& \langle N_{\rm G,c}\rangle (M_{\rm DMH})\;
\left(\frac{M_{\rm DMH}-M_0}{M_1^\prime}\right)^{\alpha_{\rm s}},
\label{eq:zheng07hod}
\end{eqnarray}
where the model fit parameters are $\log M_{\rm min}$, $\sigma_{\log M}$, $\log (M_1^\prime/M_{\rm min})$, and $\alpha_s$ (satellite slope). Since $M_0$ is poorly constrained for our sample, we fix  this to $M_0=0$, reducing the number of free parameters to four. For $M_0=0$ the convention is to use the symbol $M_1$ (which is the halo mass at which $\langle N^s(M_{\rm DMH})\rangle = 1$) instead of $M_1^\prime$. We therefore use the variable $M_1$ hereafter.  

Best-fit parameters are searched in the range of $0.1<r_{\rm p}{\rm [h^{-1}Mpc]}<40$ and by minimizing the correlated $\chi^2$:
\begin{eqnarray}
\chi^2 &=& \sum_{ij}\{[w_{\rm p}(r_{{\rm p},i})-w_{\rm p}^{\rm mdl}(r_{{\rm p},i})]M^{-1}_{ij}\times\nonumber\\
       & & [w_{\rm p}(r_{{\rm p},j})-w_{\rm p}^{\rm mdl}(r_{{\rm p},j})]\} +\nonumber\\
       & &  (n_{\rm G}-n_{\rm G}^{\rm mdl})^2/\sigma_{n_{\rm G}}^2,
\label{eq:acf_chi2}
\end{eqnarray}
where the quantities from the model are indicated by a superscript ``mdl'', $M_{ij}$ is the covariance 
matrix (Eq.~\ref{jackknife}), $n_{\rm G}$ is the number density of CMASS galaxies, and $\sigma_{n_{\rm G}}$ is the 1$\sigma$ error. 
We find 
$n_{\rm G}=(8.00 \pm 0.18) \times 10^{-5}$  $h^{3}$ Mpc$^{-3}$ 
over $0.44<z<0.64$, where the 1$\sigma$ error is estimated using jackknife resampling.

The best-fit parameter search and  determination of the confidence contours are made using an MCMC method with the MCMC-F90 library by Marko Laine\footnote{\url{http://helios.fmi.fi/~lainema/mcmcf90/}}, which we have modified and linked to our HOD model calculation software.
Table~\ref{tab:cmassacfhod} provides the best-fit parameters and their 90\% confidence errors (or 95\% upper limits).

\begin{table}
\caption{Best-fit HOD Parameters of the CMASS ACF}
\label{tab:cmassacfhod}
\begin{center}
\begin{tabular}{ccc} 
\hline
              &             & 90\% Confidence Range \\
Fit Parameter & Best Fit    & (5th--95th Percentiles)\\\hline
$\log M_{\rm min}$     &     13.76 &  13.73  --  13.81 \\
$\sigma_{\log M}$      &     0.63  &  0.59   --  0.68  \\
$\log (M_1/M_{\rm min})$ &  0.91   &  0.85  -- 0.98\\
$\alpha_{\rm s}$                      & 1.46   &  1.29  --   1.58\\
$b \rm{(linear)}    $         & 2.24   &  2.21 -- 2.28  \\\hline
\end{tabular}
\end{center}
\end{table}

\subsubsection{HOD of AGN Samples}

We fit HOD models to the CCFs between the CMASS galaxy and the AGN (sub)samples. 
To calculate
the expected CCF $w_{\rm p}(r_{\rm p})$, we use the best-fit HOD of the CMASS galaxies derived above
and a model of the AGN HOD. 
For the AGN samples, we use the same
HOD form as for the galaxy HOD, with an additional global normalization factor:
\begin{eqnarray}
\langle N_{\rm A,c}\rangle (M_{\rm DMH}) &=& 
f_{\rm A}\frac{1}{2}\left[1+\mathrm{erf}\left(\frac{\log M_{\rm DMH}-\log M_{\rm min}}{\sigma_{\log M}}\right)\right]\nonumber\\
\langle N_{\rm A,s}\rangle (M_{\rm DMH}) &=& \langle N_{\rm A,c}\rangle (M_{\rm DMH})\;
\left(\frac{M_{\rm DMH}}{M_1}\right)^{\alpha_s},
\label{eq:agnhod}
\end{eqnarray}
where $f_{\rm A}$ is the AGN fraction (duty cycle) among central galaxies at $M_{\rm DMH}\gg M_{\rm min}$; this global normalization factor scales both the central and satellite galaxies. 
For the AGN samples $\sigma_{\log M}$ can be poorly constrained,\footnote{When we fit $\sigma_{\log M}$ as a free parameter, the constraints are poor, with the smallest $\chi^2$-value at $\sigma_{\log M} \sim 0$.} and therefore we fix this parameter to 0. In this case, the factor
$\frac{1}{2}\left[1+\mathrm{erf}\left(\frac{\log M_{\rm DMH}-\log M_{\rm min}}{\sigma_{\log M}}\right)\right]$ becomes $\Theta (\log M_{\rm DMH}-\log M_{\rm min})$
where $\Theta(x) $ is a step function that has
the value of 0 at $x<0$ and 1 at $x\geq 0$, respectively.
 Thus there are three free parameters for the AGN HODs: $\log M_{\rm min}$, $\log (M_{\rm 1}/M_{\rm min})$ and $\alpha$.
 We note that the central HOD does not have to saturate at a constant value for AGN samples, unlike in the case of HODs for mass- or
 luminosity-thresholding galaxy samples, where it is usually assumed that the centers of the most-massive DMHs are occupied by a massive or luminous galaxy. Thus, in general, Eq.~\ref{eq:agnhod} may be too restrictive for an AGN HOD, and it would be ideal to introduce separate high-mass slopes for central and satellite AGN. A cosmological simulation-based forward model by \citealt{aird_coil_2021} predicts that the central AGN HOD decreases at large halo masses in many AGN samples. Such cases cannot be fully represented by Eq.~\ref{eq:agnhod}. 
 If the central HOD does decrease with increasing halo mass, forcing to fit using Eq.~\ref{eq:agnhod} could cause a misestimation of $\alpha_{\rm s}$. However, using the simple form here provides a guide to the satellite HOD behavior and can highlight  differences between different samples.

We do not use AGN density constraints for the $\chi^2$-fits and parameter searches, as the various observational biases in the AGN selection make the number density estimation very uncertain. 
The value of $f_{\rm A}$, which sets the global normalization of the AGN HOD, does not affect the resulting CCF.
$f_{\rm A}$ can be determined following the CCF fit using the AGN density constraint. 

We apply the MCMC method for the AGN HOD parameter search to the CCFs between CMASS galaxies and the total broad-line X-ray and optical AGN samples. We repeat the analysis for  objects with S/N at H$\beta \geq 5$. We fit the range of $0.2<r_{\rm p}{\rm [h^{-1}\,Mpc]}<25$ and $0.1<r_{\rm p}{\rm [h^{-1}\,Mpc]}<25$ for the X-ray and optical AGN samples, respectively.
The lower bounds are determined by the requirement that both samples have 15 pairs or more in the smallest $r_{\rm p}$ bin. At $r_{\rm p}{\rm [h^{-1}\,Mpc]}>25$ redshift space distortions (RSD)  substantially affect the measured $w_{\rm p}(r_p)$ values. 
Measurements at $r_{\rm p} {\rm [h^{-1}\,Mpc]} >25$ have minimal contribution to the power-law fits  used to determine $\pi_{\rm max}$ as discussed above; therefore, we do not include them in the HOD model fits. 
The latest version of our HOD code accounts for RSD to the two-halo term based on the linear theory \citep{kaiser87} following the recipe by \citet{vandenbosch13}. Using it is computationally prohibitive for our MCMC chain searches. Instead we calculate the RSD corrected $w_{\rm p}(r_{\rm p})$ model for the best-fit parameters to determine the potential level of impact (Sect. \ref{sec:results}.)

\begin{figure*}
\begin{center}
\hbox{
 \includegraphics[height=9cm,clip=true]{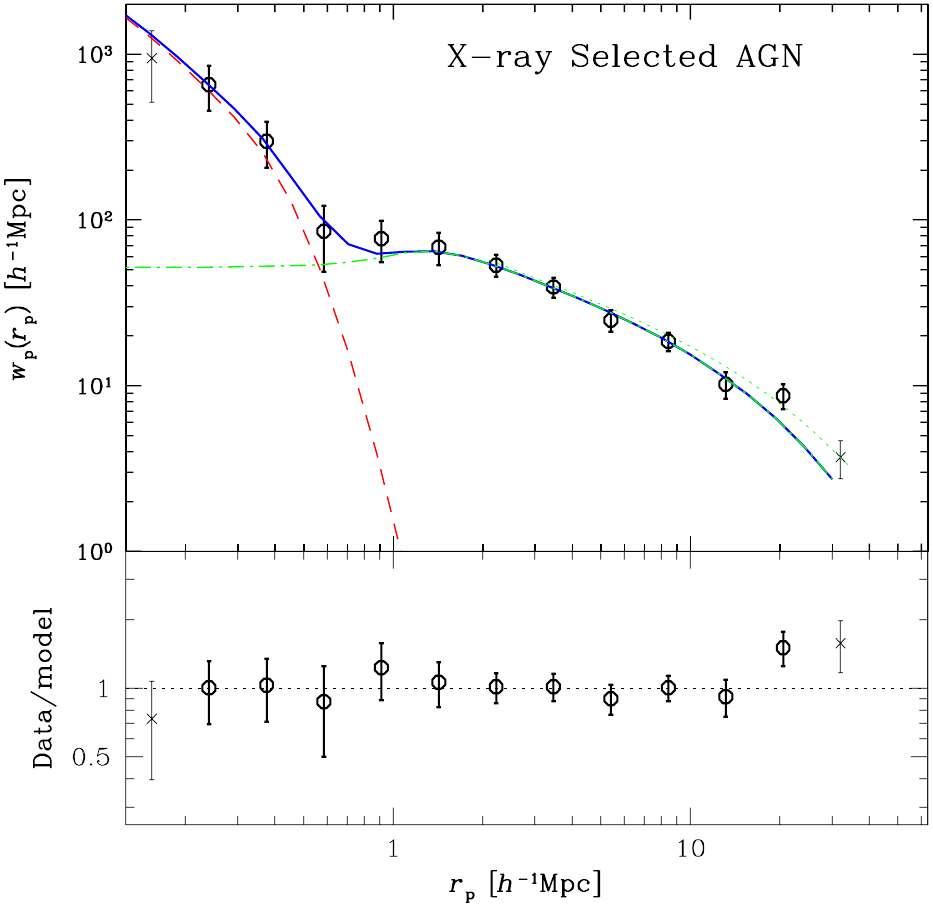}
 \includegraphics[height=9cm,clip=true]{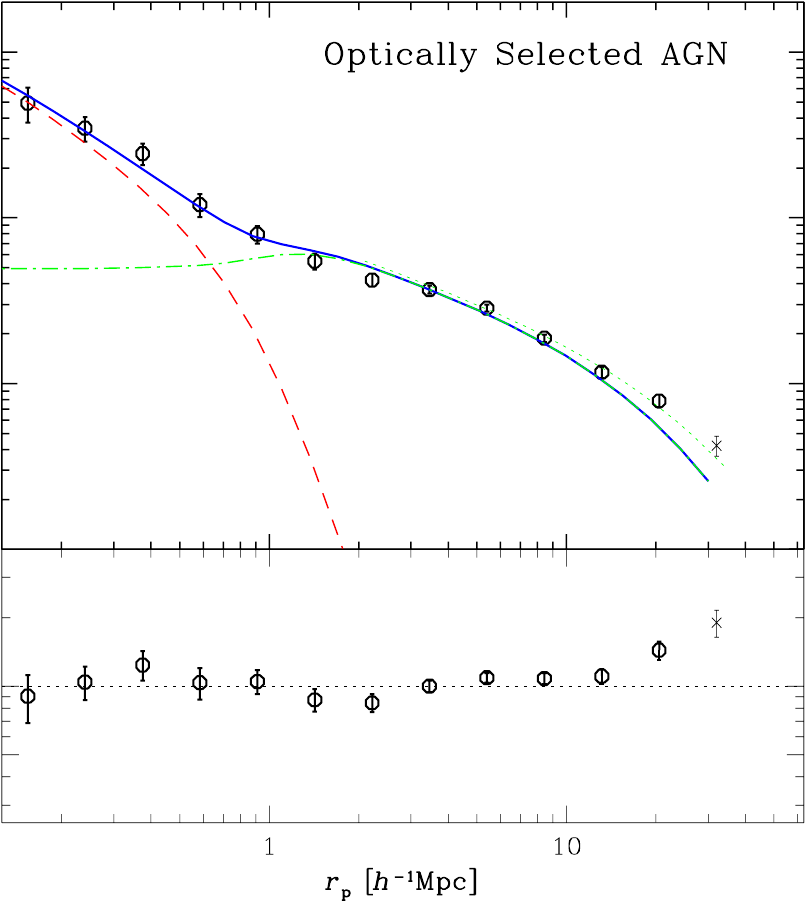}
}
\end{center}
\caption{
The measured cross-correlation function between the X-ray selected (left) and the optically-selected AGN sample (right) with the CMASS galaxy sample. In both cases we show the AGN samples corresponding to "BL, H$\beta$-S/N$ > $5" in Table~\ref{tab:overview}. The data points indicated by open circles and x-points represent the measurements included and excluded from the HOD model fitting process, respectively. The best fit HOD model is shown with the one-halo (red dashed) term, two-halo (green dashed) term, and the sum of these terms (blue solid lines). The best fit two-halo term model with the RSD correction is shown with a green short-dashed line.}
\label{fig:CMASSccf}
\end{figure*}

\begin{figure*}
\begin{center}
\hbox{\hspace*{-0.4cm}
\includegraphics[width=9.0cm,height=8cm]{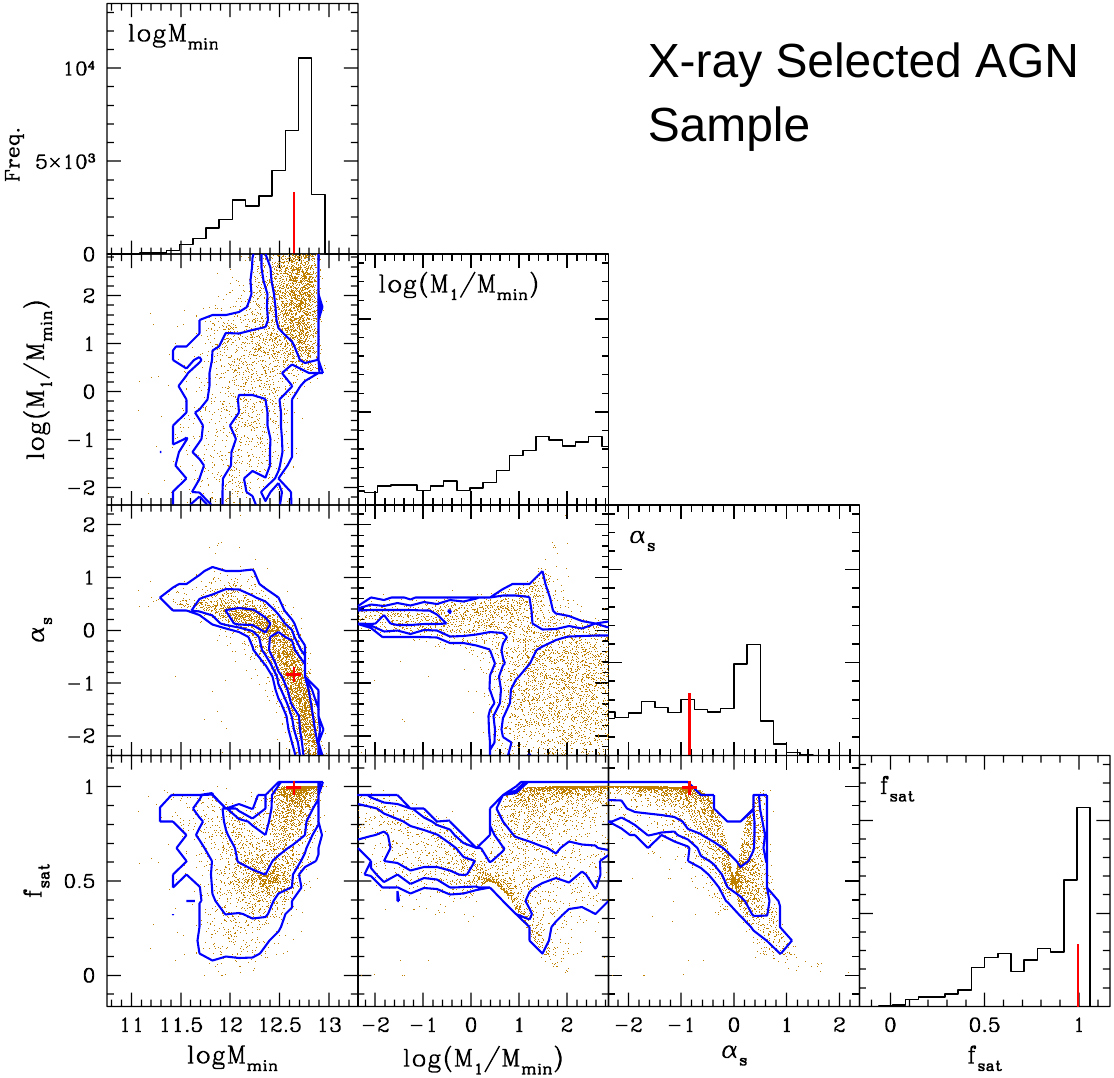}
\hspace{0.2cm}
 \includegraphics[width=9.0cm,height=8.0cm]{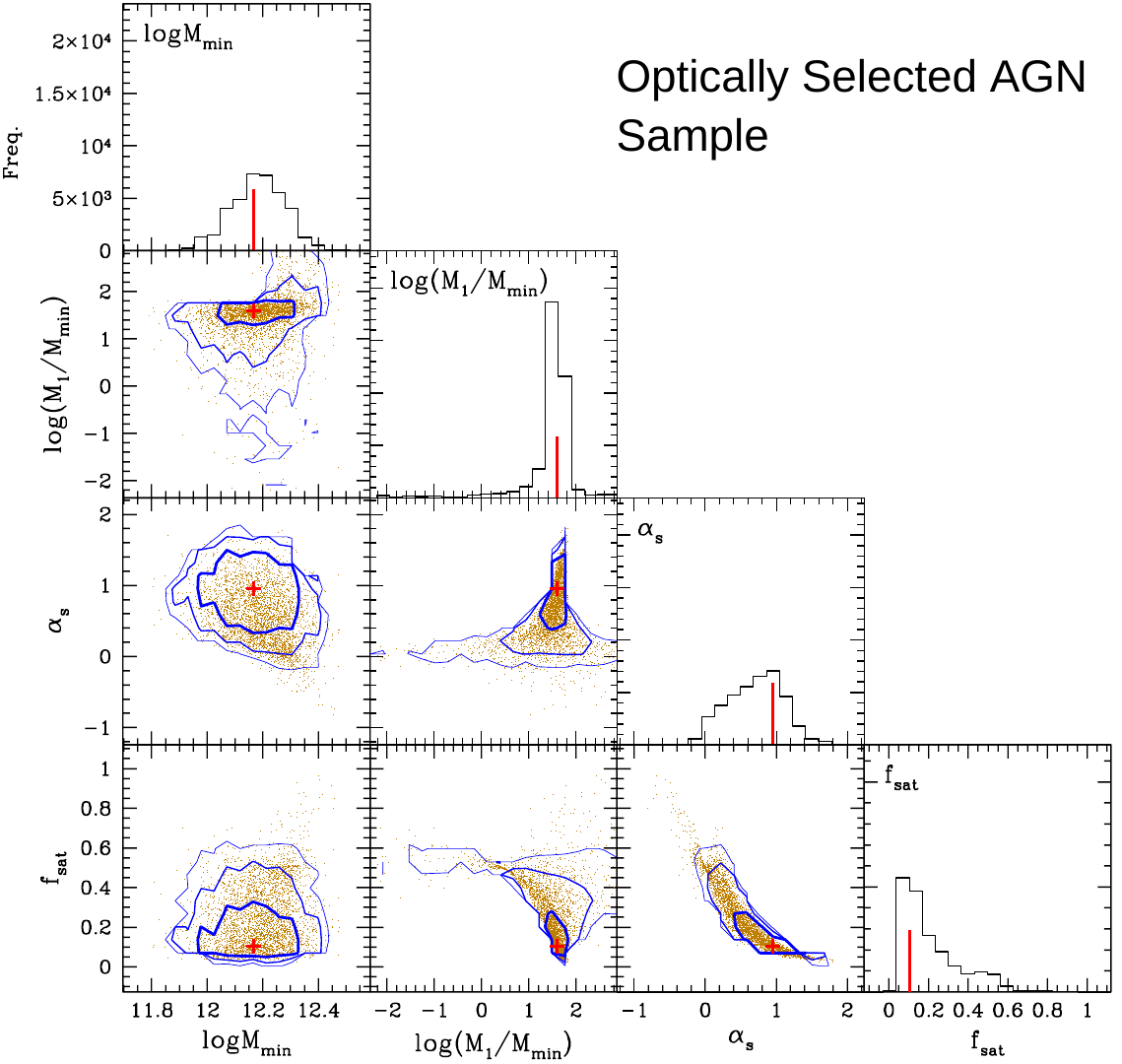}
}
\end{center}
\caption{
MCMC results with confidence contours for the parameters in Eq.~\ref{eq:agnhod} from the CCF HOD analysis of the broad-line X-ray ({\it left}) and optically selected AGN  ({\it right}) samples. Constraints on the satellite fraction $f_{\rm sat}$ are also derived from the model (see Eq.~\ref{eq:fsat}).
In both cases we use the sample with a S/N ratio at H$\beta$ of $>$5. The nomenclatures are the same as those in Fig.~\ref{fig:CMASSacfhod}, except that $\Delta \chi^2$-levels are 3.6, 6.2, and 10.4 for the left panel and 4.5, 7.7, and 10.9 for the right panel, respectively. These correspond to the 68, 90, and 98\% of the chain points fall because of the non-Gaussian probability distributions.}
\label{fig:ccfhodconts}
\label{fig:CMASSccfhod}
\end{figure*}

\begin{figure}
  \centering
 \resizebox{\hsize}{!}{ 
   \includegraphics[]{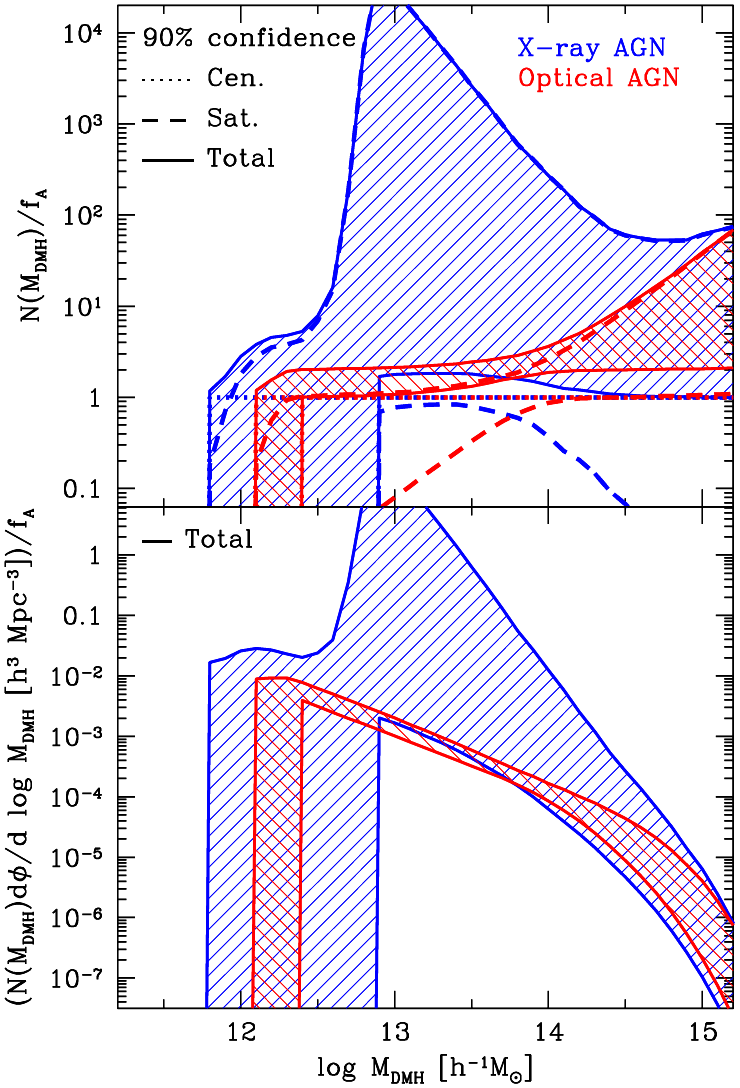}}
      \caption{Comparison of the HODs determined for the X-ray (blue) and optical (red) AGN samples. Dotted, dashed, and solid lines show the 90\% confidence ranges  of the models in the MCMC chain evaluated at each $\log M_{\rm DMH}$ for $\langle N_{\rm A,c}\rangle (M_{\rm DMH})$,  $\langle N_{\rm A,s}\rangle (M_{\rm DMH})$, and their sum, respectively. The total HOD constraint ranges are shown with hatches. Due to the functional form assumed for $\langle N_{\rm A,c}\rangle (M_{\rm DMH})$, the 5\% and 95\% percentile lines overlap with each other at high $M_{\rm DMH}$. Furthermore, the central, satellite, and total 90\% limit lines also overlap at $M_{\rm DMH}=M_{\rm min}$ (vertical part). 
      The y-axis in the upper panel shows $N(M_{\rm DMH})/f_{\rm A}$, such that large values may reflect a small $f_{\rm A}$. In the bottom panel the product of the HOD and the DMH mass function $d\phi/d\,\log M_{\rm DMH}$ for the X-ray (blue) and optical (red) AGN samples (central+satellite) is shown. 
      The large difference in the satellite term between these samples at $M_{\rm DMH}\sim 10^{13}\,h^{-1}\,\rm{M}_\odot$ is due to very different $\alpha_{\rm s}$ constraints and is discussed  in Sect.~\ref{implications}. }
      \label{fig:nm_nmphi}
\end{figure}

\subsubsection{Luminosity, $M_{\rm BH}$, and $L/L_{\rm EDD}$ AGN Subsamples}

One of the main goals of this work is to investigate the differences in AGN HOD properties as a function of AGN luminosity, $M_{\rm BH}$, $L/L_{\rm EDD}$, and $M_i$, for which we create
AGN subsamples. 
The constraints on ${M_1^\prime}/M_{\rm min}$ are poor when fitting the full X-ray selected AGN sample. For the subsamples we therefore fix ${M_1^\prime}/M_{\rm min}$ to be equal to 5, and we  check the sensitivity of the results to different values of this parameter.
This reduces the number of fit parameters to the most fundamental ones of interest, namely $\log M_{\rm min}$ and $\alpha$. 
Fixing ${M_1^\prime}/M_{\rm min}$ to 10 or 30 only modifies the large-scale bias in most cases by less than $\pm$0.02. In the most extreme case, the change in bias is $\pm$0.05.   
Even in this extreme scenario, this systematic error is well within the 1$\sigma$ uncertainty range of the bias. 

The purpose of reducing the number of fit parameters is two-fold. 
First, for these two parameters we do not need to use a computationally intensive MCMC for each subsample; we can use a two-dimensional grid of models in a tabular form to explore the confidence range.
Second, with fewer free parameters we can highlight the differences in the HODs of the subsamples, as fixing the extra parameter may reduce the confidence regions in the two-parameter confidence contour plots by removing projection effects. 
As our purpose is to investigate the difference between samples, rather than obtaining full HOD constraints, comparing the confidence ranges in the plane of the two essential parameters, while holding other parameters fixed,  is more demonstrative than comparing full confidence contour plot matrices in the full parameter space.

\begin{table*}
\centering
\caption{Properties of the AGN Samples and Derived HOD Model Quantities Using Grid Fitting \label{tab:overview}} 
\tablewidth{0pt}
\begin{tabular}{lcccccccccc}

\hline
{Sample} & {Number} & {Median} & {Median} & {Median}& {Median}&{Median}&{Median} & {$b(z)$}   & {$\log\,M_{\rm DMH}^{\rm typ}$} & {$\log\,\langle M_{\rm DMH}\rangle$} \\
{Name} & {of Objects} & {$z$} & {$M_i$(z=2)}& {$log L_{\rm X}$}& {$log M_{\rm BH}$} & {$log L/L_{\rm EDD}$}& {$log L_{\rm Bol}$} & {(HOD)} & {($h^{-1}$ $M_{\odot}$)} &{($h^{-1}$ $M_{\odot}$)} \\

\hline

\multicolumn{11}{c}{X-ray Broad-line AGN -- RASS/SDSS, North Cap only}\\
total BL sample          & 1701 & 0.53 & -24.03 & 44.82 & 8.48 & -1.14 & 45.43 &  1.37$^{+0.12}_{-0.13}$& 12.80$^{+0.16}_{-0.20}$&12.96$^{+0.12}_{-0.19}$\\
BL, H$\beta$-S/N$ > $5   & 1632 & 0.53 & -24.07 & 44.82 & 8.49 & -1.14 & 45.45 &  1.35$^{+0.14}_{-0.11}$& 12.77$^{+0.19}_{-0.17}$&12.92$^{+0.15}_{-0.19}$\\
{\it low $L_{\rm X}$}     &  816 & 0.50 & -23.76 & 44.66 & 8.39 & -1.19 & 45.28 &  1.20$^{+0.16}_{-0.14}$& 12.52$^{+0.26}_{-0.30}$  &12.65$^{+0.25}_{-0.29}$ \\
{\it high $L_{\rm X}$}    &  816 & 0.55 & -24.43 & 45.00 & 8.60 & -1.06 & 45.62 &  1.55$^{+0.14}_{-0.21}$& 13.03$^{+0.15}_{-0.27}$  &13.27$^{+0.09}_{-0.14}$\\
 & & & & & & & & & &    \\ 
{\it low $M_{\rm BM}$}    &  833 & 0.51 & -23.64 & 44.72 & 8.23 & -1.14 & 45.21 & 1.24$^{+0.16}_{-0.15}$& 12.60$^{+0.24}_{-0.32}$  &12.71$^{+0.23}_{-0.28}$ \\
{\it high $M_{\rm BM}$}   &  833  & 0.55 & -24.63& 44.93 & 8.75 & -1.14 & 45.71 & 1.57$^{+0.18}_{-0.16}$& 13.05$^{+0.19}_{-0.19}$  &13.27$^{+0.12}_{-0.13}$\\
 & & & & & & & & & &    \\ 
 {\it low $L/L_{\rm EDD}$} &  829  & 0.51 & -23.62 & 44.73 & 8.49 & -1.42 & 45.21 & 1.42$^{+0.18}_{-0.13}$& 12.87$^{+0.22}_{-0.19}$ &12.97$^{+0.20}_{-0.17}$ \\
{\it high $L/L_{\rm EDD}$}&  827  & 0.54 & -24.63 & 44.92 & 8.49 & -0.83 & 45.71 & 1.28$^{+0.15}_{-0.18}$ & 12.66$^{+0.22}_{-0.34}$ &12.93$^{+0.19}_{-0.31}$\\
 & & & & & & & & & &    \\ 
{\it low $L_{\rm Bol}$} &  817  & 0.51 & -23.60 & 44.71 & 8.35 & -1.31 & 45.19 & 1.42$^{+0.16}_{-0.18}$& 12.87$^{+0.19}_{-0.27}$ &12.97$^{+0.17}_{-0.24}$\\
{\it high $L_{\rm Bol}$}&  815  & 0.55 & -24.64 & 44.94 & 8.62 & -0.93 & 45.71 & 1.35$^{+0.16}_{-0.17}$& 12.77$^{+0.21}_{-0.29}$ &13.10$^{+0.13}_{-0.21}$\\
 & & & & & & & & & &    \\ 
{\it faint $M_i$}        &  815  & 0.51 & -23.58 & 44.71 & 8.36 & -1.30 & 45.19 & 1.33$^{+0.16}_{-0.17}$& 12.74$^{+0.22}_{-0.30}$ &12.84$^{+0.22}_{-0.25}$\\
{\it luminous $M_i$}     &  817  & 0.55 & -24.64 & 44.94 & 8.62 & -0.94 & 45.71 & 1.44$^{+0.14}_{-0.17}$ &12.90$^{+0.16}_{-0.26}$ &13.08$^{+0.13}_{-0.25}$\\\hline
\multicolumn{11}{c}{Optical Broad-line AGN (Paris et al. sample) -- SDSS, North Cap only }\\
total BL sample          & 10994 & 0.56 & -23.37 & -- & 8.35 & -1.31 & 45.14 & 1.28$^{+0.04}_{-0.04}$& 12.66$^{+0.06}_{-0.06}$ & 12.89$^{+0.08}_{-0.05}$ \\  
BL, H$\beta$-S/N$ > $5   & 8889 & 0.55 & -23.58 & -- & 8.40 & -1.26 & 45.24 & 1.37$^{+0.07}_{-0.06}$& 12.80$^{+0.10}_{-0.09}$ & 12.99$^{+0.04}_{-0.08}$ \\  
 & & & & & & & & & &    \\ 
0--25\% $M_{\rm BH}$ & 2243 & 0.52 & -22.78 & -- & 7.93 & -1.26 & 44.81 & 1.29$^{+0.08}_{-0.09}$& 12.68$^{+0.12}_{-0.16}$ & 12.92$^{+0.11}_{-0.20}$\\  
25--50\% $M_{\rm BH}$ & 2243 & 0.54 & -23.40 & -- & 8.29 & -1.26 & 45.13 & 1.39$^{+0.10}_{-0.10}$& 12.83$^{+0.13}_{-0.15}$ & 13.08$^{+0.09}_{-0.13}$\\  
50--75\% $M_{\rm BH}$ & 2243 & 0.55 & -23.82 & -- & 8.51 & -1.26 & 45.36 & 1.42$^{+0.09}_{-0.11}$& 12.87$^{+0.11}_{-0.16}$ & 12.96$^{+0.13}_{-0.13}$\\  
75--100\% $M_{\rm BH}$ & 2243 & 0.57 & -24.43 & -- & 8.82 & -1.26 & 45.64 & 1.39$^{+0.13}_{-0.12}$& 12.83$^{+0.17}_{-0.19}$ & 13.15$^{+0.09}_{-0.12}$\\  
 & & & & & & & & & &    \\ 
0--25\% $L/L_{\rm EDD}$ & 2242 & 0.53 & -22.82 & -- & 8.40 & -1.68 & 44.84 & 1.29$^{+0.11}_{-0.11}$& 12.68$^{+0.16}_{-0.20}$ & 13.05$^{+0.11}_{-0.10}$\\  
25--50\% $L/L_{\rm EDD}$ & 2242 & 0.54 & -23.37 & -- & 8.40 & -1.37 & 45.15 & 1.50$^{+0.10}_{-0.12}$& 12.97$^{+0.12}_{-0.15}$ & 13.05$^{+0.12}_{-0.12}$\\  
50--75\%  $L/L_{\rm EDD}$ & 2242 & 0.55 & -23.82 & -- & 8.40 & -1.14 & 45.37 & 1.50$^{+0.09}_{-0.12}$& 12.97$^{+0.11}_{-0.15}$ & 13.06$^{+0.11}_{-0.14}$\\  
75--100\%  $L/L_{\rm EDD}$ & 2242 & 0.56 & -24.47 & -- & 8.40 & -0.82 & 45.66 & 1.33$^{+0.11}_{-0.10}$& 12.74$^{+0.16}_{-0.16}$ & 12.93$^{+0.14}_{-0.17}$\\  
 & & & & & & & & & &    \\ 
0--25\% $L_{\rm Bol}$ & 2223 & 0.52 & -22.73 & -- & 8.11 & -1.47 & 44.80 & 1.30$^{+0.09}_{-0.12}$& 12.70$^{+0.13}_{-0.22}$ & 12.98$^{+0.12}_{-0.14}$\\  
25--50\% $L_{\rm Bol}$ & 2223 & 0.55 & -23.35 & -- & 8.35 & -1.35 & 45.12 & 1.36$^{+0.13}_{-0.08}$& 12.78$^{+0.18}_{-0.12}$ & 13.03$^{+0.10}_{-0.11}$\\  
50--75\%  $L_{\rm Bol}$ & 2222 & 0.55 & -23.83 & -- & 8.47 & -1.22 & 45.37 & 1.53$^{+0.13}_{-0.04}$& 13.01$^{+0.14}_{-0.05}$ & 13.08$^{+0.14}_{-0.04}$\\  
75--100\%  $L_{\rm Bol}$ & 2221 & 0.57 & -24.52 & -- & 8.64 & -1.01 & 45.67 & 1.31$^{+0.09}_{-0.13}$& 12.71$^{+0.13}_{-0.23}$ & 13.09$^{+0.09}_{-0.15}$\\  
 & & & & & & & & & &    \\ 
$M_{\rm i} > -22.7$       & 1097 & 0.50 & -22.44 & -- & 8.05 & -1.48 & 44.69 & 1.33$^{+0.10}_{-0.12}$& 12.74$^{+0.14}_{-0.20}$ & 13.10$^{+0.11}_{-0.14}$\\ 
$ -22.7<M_{\rm i}<-23.25 $& 1853 & 0.55 & -23.01 & -- & 8.27 & -1.41 & 44.96 & 1.30$^{+0.13}_{-0.12}$& 12.7$^{+0.18}_{-0.22}$ & 12.85$^{+0.16}_{-0.13}$\\ 
$-23.25<M_{\rm i}<-23.70$ & 1997 & 0.53 & -23.48 & -- & 8.40 & -1.32 & 45.20 & 1.55$^{+0.08}_{-0.13}$& 13.03$^{+0.09}_{-0.16}$ & 13.15$^{+0.07}_{-0.14}$\\ 
$-23.70<M_{\rm i}<-24.15$ & 1871 & 0.56 & -23.91 & -- & 8.47 & -1.21 & 45.40 & 1.39$^{+0.10}_{-0.13}$& 12.83$^{+0.13}_{-0.20}$ & 13.03$^{+0.11}_{-0.17}$\\ 
$-24.15<M_{\rm i}<-25.4$  & 1853 & 0.55 & -24.51 & -- & 8.58 & -1.03 & 45.66 & 1.33$^{+0.15}_{-0.08}$& 12.74$^{+0.21}_{-0.12}$ & 12.94$^{+0.14}_{-0.18}$\\ 
\hline
\end{tabular}
\tablecomments{\small All samples span a redshift range of $0.44<z<0.64$. For the HOD analysis of all (sub)samples presented in this table, we use a common, conservative fitting range of $r_p \geq 0.9$ $h^{-1}$ Mpc. All bias values are computed at $z=0.53$.}
\end{table*}

\begin{table*}
\centering
\caption{Best-fit HOD Parameters of the Full AGN Samples Using MCMC}
\label{CCF_MCMC_results}
\begin{tabular}{lcccccc} 
\hline
Sample & $b(z)$  & $\log\,M_{\rm DMH}^{\rm typ}$ & $\log\,\langle M_{\rm DMH}\rangle$ & $\log\,M_{\rm min}$ &$\alpha_{\rm s}$ & $\log (M_{\rm 1}/M_{\rm min})$\\
Name & (HOD, MCMC) & ($h^{-1}$ $M_{\odot}$) &($h^{-1}$ $M_{\odot}$) & ($h^{-1}$ $M_{\odot}$) & & \\\hline
\multicolumn{7}{c}{X-ray Broad-line AGN -- RASS/SDSS, North Cap only}\\
Full BL sample         &  1.43$^{+0.08}_{-0.16}$& 12.88$^{+0.10}_{-0.20}$ &  12.53$^{+0.20}_{-0.33}$ & 12.98$^{+0.09}_{-0.10}$ & -0.21$^{+0.50}_{-1.15}$ & 1.07$^{+1.19*}_{-1.05}$ \\
H$\beta$-S/N$ > $5 &  1.42$^{+0.09}_{-0.13}$& 12.89$^{+0.11}_{-0.30}$ & 12.97$^{+0.10}_{-0.11}$ & 12.54$^{+0.19}_{-0.42}$ & -0.35$^{+0.52}_{-1.1}$ & 0.97$^{+1.26*}_{-1.06}$ \\\hline
\multicolumn{7}{c}{Optical Broad-line AGN (Paris et al. sample) -- SDSS, North Cap only}\\
Full BL sample          & 1.24$^{+0.04}_{-0.04}$& 12.72$^{+0.07}_{-0.08}$ & 12.90$^{+0.06}_{-0.06}$ & 11.99$^{+0.10}_{-0.11}$ & 0.99$^{+0.28}_{-0.31}$ & 1.60$^{+0.12}_{-0.18}$ \\
H$\beta$-S/N$ > $5   & 1.32$^{+0.05}_{-0.05}$& 12.72$^{+0.06}_{-0.06}$ & 12.93$^{+0.06}_{-0.06}$ & 12.20$^{+0.09}_{-0.11}$ & 0.54$^{+0.43}_{-0.38}$ & 1.55$^{+0.18}_{-0.38}$\\\hline
\end{tabular}
\tablecomments{The quoted best-fit value is the 50th percentile (median) of the MCMC parameter distribution, and the uncertainties are 68\% confidence values for one parameter, derived from the distribution of the MCMC chain. The fitting ranges for the full samples are $0.2 < r_p <25$ and $0.1 < r_p <25$ [$h^{-1}$ Mpc] for the X-ray and optical AGN samples, respectively. An asterisk (*) for the error value indicates that it is pegged at a parameter search limit.}
\end{table*}

\section{Results}
\label{sec:results} \label{inferringACF}

We compute the CMASS galaxy ACF within $0.44\leq z<0.64$ and at a median stellar mass of $\langle \rm{log} (M_{stellar}/M_{\odot})\rangle=11.33$ (which is constant across the full redshift range). The best-fit HOD model parameters, including the large-scale bias, are given in Table~\ref{tab:cmassacfhod}. The measured ACF and the HOD results are shown in Fig.~\ref{fig:CMASSacfhod}.

We measure the CCFs between different X-ray and optical AGN samples and  
CMASS galaxies. As described above, we run the computationally intensive MCMC fitting  for the full X-ray and optical AGN samples. For the AGN subsamples we determine the best-fit based on a search in a two-dimensional grid. 
Figure~\ref{fig:CMASSccf} shows the resulting CCFs ($w_{\rm p}(r_p)$) and best fit HOD models, as well as residuals from the fits, for the $SN_{\rm H_\beta}>5$ samples. In each panel, the best-fit model corrected for the RSD effect is also shown, which  demonstrates that the deviation of $w_{\rm p}(r_{\rm p})$ at the largest $r_{\rm p}$ can be accounted for by the RSD effect, though this effect is not included in the MCMC procedure due to the computational requirements.  Confidence contour matrices for these results are shown in Fig.~\ref{fig:CMASSccfhod}, with dots representing the corresponding MCMC chain.  We present the best-fit values in Table~\ref{CCF_MCMC_results}.

The confidence contours can differ from the concentration of the MCMC chain points, as the contours are determined from the minimum $\chi^2$-values along the projections and not from the probabilities based on the marginal distributions of the MCMC points in the projected space. Results are also shown for the satellite fraction
\begin{equation}
    f_{\rm sat}=
    \frac{\int \langle N_{\rm A,s}\rangle (d\phi/d\log M_{\rm DMH})d\log M_{\rm DMH}}{\int \left(\langle N_{\rm A,c}\rangle+\langle N_{\rm A,s}\rangle\right)(d\phi/d\log M_{\rm DMH})d\log M_{\rm DMH}},
    \label{eq:fsat}
\end{equation}
where $d\phi/d\log M_{\rm DMH}$ is the halo mass function. The larger uncertainties derived for the X-ray selected sample compared to the optical AGN sample are due to the smaller size of the X-ray sample.

Figure \ref{fig:nm_nmphi} (upper panel) shows the central and satellite HODs with 90\% confidence ranges for the same samples as those in  Fig.~\ref{fig:CMASSccf}. The sum of central and satellite AGN HODs are shown with their 90\% confidence ranges in hatches.  The lower panel shows the sum of the central and satellite HODs multiplied by the halo mass function, again with 90\% confidence ranges shown in hatches.

\subsection{Implications of the HOD Analysis}
\label{implications}

Before discussing our HOD analysis results in detail, we provide guidelines for interpreting the HOD parameterized model in Eq.~\ref{eq:agnhod}:

\begin{itemize}
\item As the DMH mass distribution for AGN is very wide, the best-fit typical DMH should be interpreted carefully. It is useful to compare the typical DMH masses of different AGN samples to look for clustering trends; however, it should not be interpreted to imply that most AGN reside in halos of this single mass.
    \item 
    A positive value of $\alpha_{\rm s}>0$ implies that the number of AGN in satellite galaxies increases with increasing $M_{\rm DMH}$, as might be expected, as this is the behavior of the galaxy HOD. On the other hand, $\alpha_{\rm s}<0$ implies that AGN in satellite galaxies are preferentially found in the lower-mass halos (i.e., modest richness groups) and are rare in high-mass halos (i.e., in rich groups and clusters).
    \item By construction in the model $\log M_{\rm 1}/M_{\rm min}$ becomes unconstrained for $\alpha_{\rm s}=0$.
    \item The two-halo term of the galaxy--AGN CCF is $\propto b_{\rm Gal}
      b_{\rm AGN}$, where the bias value is an indicator of some weighted mean host DMH mass. 
    \item There are two major constraints from the one-halo term: the amplitude of the clustering signal and the spatial extent of the one-halo-term-dominated region. The amplitude reflects the satellite-central and satellite-satellite pair counts within the same halos. The spatial extent reflects the maximum halo mass (i.e., the virial radius) that hosts a substantial satellite population.
    \item The value of $f_{\rm sat}$ should not be over-interpreted, as it is highly model dependent, particularly the behavior of the model near $M_{\rm DMH} \sim M_{\rm min}$ where
    the halo mass function is large.
\end{itemize}

\subsection{Findings from the AGN Grid Fitting Procedure}
In addition to the MCMC method, we  use the grid fitting procedure for all AGN (sub)samples. 
The lower $r_{\rm p}$ value in each sample is determined such that there are at least 15 pairs in a bin. While for the full X-ray and optical AGN samples and most optical AGN subsamples, this is the case for $r_p \geq 0.1-0.3$ $h^{-1}$ Mpc, for the X-ray AGN subsamples, this criterion is met for $r_p \geq 0.9$ $h^{-1}$ Mpc. Thus we choose a conservative lower limit of $r_p = 0.9$ $h^{-1}$ Mpc for all AGN samples where we use the grid fitting method and provide the results in Table~\ref{tab:overview}. 
In this table we list the number of objects per AGN (sub)sample, the
median redshift of the AGN sample, median absolute $i$-band magnitude ($k$-corrected for $z=2$), median X-ray luminosity (rest-frame 0.1--2.4 keV, corrected for Galactic absorption), median black hole mass, median accretion
ratio relative to Eddington, median bolometric luminosity, linear bias,
typical DMH mass of the sample, and mean DMH mass. The constraints on the last
three quantities are derived from the HOD modeling approach and use the above
mentioned fitting ranges. All typical DMH masses of the sample and mean DMH
masses are computed for $z=0.53$. The constraints on the bias, the typical DMH
mass (derived from the bias), and mean DMH mass (which mainly depends on the bias) all effectively result from the two-halo term. Thus, we can obtain constraints without using data in the one-halo-dominated regime ($r_{\rm p}<0.9\,h^{-1} {\rm Mpc}$).

The results from the MCMC method show that the full broad-line X-ray and optical AGN samples have similar large-scale biases and HOD results. The largest difference is found for $\log (M_{\rm 1}/M_{\rm min})$, though it is only 1.1$\sigma$ when considering the combined uncertainties. In particular the exclusion of the AGN for which the $M_{\rm BH}$ estimates are not as  reliable does not significantly change the clustering results.  

When comparing results from the MCMC and grid fitting approaches for the full
X-ray and optical AGN samples, we obtain similar results (within the 1$\sigma$
uncertainties). Differences arise due to having three free parameters in the
MCMC method, while in the grid fitting method, two parameters are fit while
the third is fixed at the same value for all samples. The grid fitting then results in tighter constraints on $\log\,M_{\rm min}$ and $\alpha_{\rm s}$ than the MCMC approach. 

Using the grid fitting procedure, we find differences in the large-scale clustering properties of some X-ray  AGN subsamples in excess of the combined 2$\sigma$ uncertainties.  There is a weak $L_{\rm X}$ clustering dependence with the mean DMH masses, at a significance of 2.1$\sigma$, in that more X-ray luminous AGN cluster more strongly than their lower-luminosity counterparts. We also find a 2.1$\sigma$ positive $M_{\rm BH}$ clustering dependence in the X-ray AGN samples when considering the mean DMH masses. Among the optical AGN subsamples, we find a $>$2$\sigma$ difference in the clustering properties with respect to $L_{\rm Bol}$ (for details, see Sect.~\ref{Lbol_discussion}). For all other parameters (e.g, $L/L_{\rm EDD}$, $M_{\rm i}$), the clustering properties in the optical and X-ray subsamples agree within $<$2$\sigma$. In general, the mean DMH mass is a similar quantity as the typical DMH, but the mean has contributions from the one- and two-halo terms while the typical DHM reflects the large-scale bias value, which has contributions from the two-halo term only.


\section{Discussion}

In discussing our findings we focus both on comparing our results with  
other studies as well as our previous work (in particular papers III and IV) 
to draw conclusions on the evolution of broad-line AGN clustering properties across a redshift range of $z=0.07-0.64$. The upper value of this redshift range corresponds to a look-back time of 6 Gyr, almost half of the lifetime of the universe. Our studies provide four independent measurements across this redshift range. Since the X-ray and optical AGN samples are generated by the same instruments (ROSAT and SDSS) and the  methodology used (cross-correlation approach \& HOD modeling) is identical, systematic effects are expected to be minimized compared to considering other studies that use different samples and methods. However, there are some differences across our studies, including (i) our HOD analysis  improved over the years, (ii) different galaxy tracer sets are used, and (iii) the AGN samples from ROSAT/SDSS differ in some properties across this wide redshift range (e.g., lower luminosities at low redshift compared to high redshift).

\subsection{X-ray versus Optical AGN Clustering Properties}
\label{sec:disc_xopt}

We first compare the clustering properties of the X-ray and optical AGN samples.  
Table~\ref{CCF_MCMC_results} shows that the HOD parameters of the total X-ray and optical AGN samples agree  well within the uncertainties, including for the samples with H$\beta$-S/N$ > $5. 
Figure~\ref{fig:CMASSccfhod} shows that due to differences in the sample sizes, the optical AGN sample has much tighter constraints than the X-ray  AGN sample. The confidence ranges of these two samples have some overlap in the covered parameter spaces. An illustration of the overlap of the X-ray and optical AGN HODs is given in Fig.~\ref{fig:nm_nmphi}.

Although the HOD results are not highly constraining due to the limited statistics (resulting from the relatively small AGN sample sizes), we are able to determine some salient characteristics.
As shown in Fig.~\ref{contour_compare}, the X-ray AGN sample has on average higher-luminosity AGNs than the optical AGN sample, and the number density of the X-ray AGN sample is $\sim 1/5$ of that of the optical AGN sample. As the two-halo terms for the two samples are consistent with each other, the ``typical" DMH masses are also consistent. However, there are  differences in the one-halo term clustering properties, as seen in Fig.~\ref{fig:CMASSccf}. These differences are marginal, in that the 90\% confidence ranges in the HOD parameter spaces as well as in HODs themselves overlap (Figs.~\ref{fig:ccfhodconts}\ and \ref{fig:nm_nmphi}). However, for the X-ray sample, the preferred $\chi^2$-values are in the  $\alpha_{\rm s}<0$ region, while $\alpha_{\rm s}<0$ is almost excluded in the optical sample. As shown in Fig.~\ref{fig:CMASSccf}, the one-halo term of the X-ray selected sample has a higher amplitude on small scales, and the scale at which the two-halo term dominates is lower. 
This points to higher $M_{\rm min}$ and lower $\alpha_{\rm s}$ for X-ray AGN, where the AGN in satellites are more common in halos near $M_{\rm min}$ and the occupation decreases for more-massive halos. 

As discussed in \citet{miyaji_krumpe_2011}, possible mechanisms for low $\alpha_{\rm s}$ include the decrease of the cross-section of satellite-satellite mergers triggering AGN activity 
\citep{makino97,altamirano16} in high-velocity encounters in massive halos and
ram pressure stripping of cold gas in the intracluster medium. 

One caveat in interpreting $\alpha_{\rm s}$ is that an underlying assumption of our HOD model is that the radial distribution of satellite AGN follows the Navarro--Frenk--White (NFW) halo mass density profile \citep{nfw97} with the halo mass dependent concentration parameter of \citet{zheng_coil_2007}.  It is possible that this assumption may not be valid for all AGN samples. An HOD model fit to the same data that assumes a different satellite radial profile would change the resulting  $\alpha_{\rm s}$. However, such a model is beyond the scope of this paper and may be addressed when larger AGN samples such as those based on {\it eROSITA} become available in the future.

As mentioned above, our results indicate that X-ray selected AGN, in contrast to optically selected AGN, contain a large population of satellites at $10^{13}\,h^{-1} M_{\rm \odot}$. This is reflected in the $f_{\rm sat}$ distribution having higher values for the X-ray AGN sample. 
The marginal difference in the one-halo term behavior may be explained by X-ray selection identifying somewhat higher-luminosity AGN than optical selection. However, the difference is statistically marginal and therefore we conclude that X-ray and optically selected BLAGN samples show very similar clustering properties in the redshift range $0.44\leq z<0.64$.

\subsection{X-ray versus Optical AGN Clustering as a Function of Redshift}

In paper III we found no statistically significant difference in the clustering properties of X-ray and optical AGN at median redshifts of $z\sim 0.13$ (cross-correlation with SDSS main galaxies), $z\sim 0.27$ (cross-correlation with SDSS LRGs), and $z\sim 0.42$ (cross-correlation with the most-luminous SDSS LRGs). The amplitude and spatial extent of the one-halo term region are virtually identical in these lower-redshift ranges for X-ray and optically selected AGN (see Figure~8 in paper III).
Taking the results of all independent redshift measurements into  consideration, it appears that there are very similar bias values for  X-ray and optical BLAGN from very low redshift up to the redshift range studied of this work, with median $z\sim 0.53$.

\begin{figure}
  \centering
\resizebox{\hsize}{!}{  \includegraphics[]{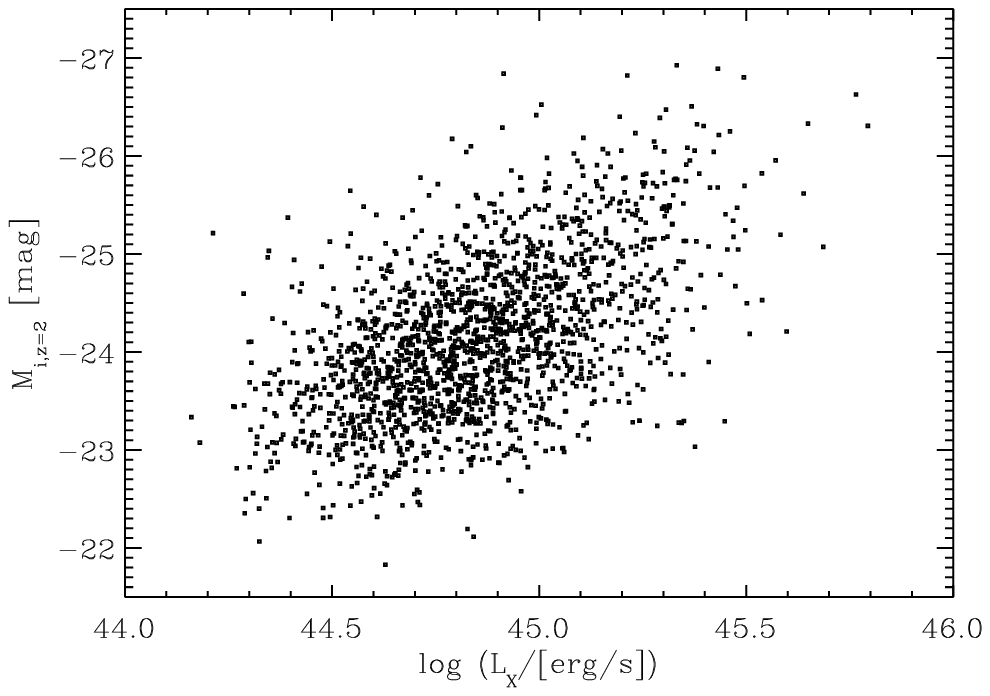}} 
      \caption{Comparison between the X-ray selected (RASS/SDSS) BLAGN sample and the optically selected (SDSS) BLAGN sample. 
      Absolute $i$-band magnitude ($k$-corrected to $z = 2$) is shown versus rest-frame X-ray luminosity (0.1--2.4 keV) for X-ray selected BLAGN with S/N at H$\beta$ $>5$.}
     \label{Xray_optical_comparison2}
\end{figure}

In each redshift range, the RASS X-ray selection and SDSS optical selection trace the full BLAGN population in different ways (see Fig.~\ref{Xray_optical_comparison}). 
Figure~\ref{Xray_optical_comparison2} shows that the X-ray and optical luminosities of AGN are correlated, though with substantial scatter. Thus, measuring clustering of optical AGN alone does not necessarily allow one to make conclusions about the clustering of X-ray selected AGN, and vice versa. 

In this work, we find statistically marginal differences in the HOD parameters
between the X-ray and optical AGN samples  due to differences in the one-halo
term clustering properties. This is reminiscent of similar differences seen
between  low- and high- $L_{\rm X}$ AGN as well as type I and type II AGN in our previous work with the {\it Swift}/ BAT+{\it INTEGRAL}/IBIS nearby AGN sample  \citep[][]{krumpe_miyaji_2018}.

Despite possible differences in the one-halo term for X-ray and optical AGN samples at low redshift, their two-halo term clustering (and therefore large-scale bias) agrees very well. 
This is in contrast with previous X-ray and optical AGN clustering studies at higher redshifts ($z>0.8$), which generally find that X-ray AGN samples are hosted by more-massive DMHs than optical AGN samples at the same redshift (see \citealt{porciani_magliocchetti_2004}; \citealt{coil_hennawi_2007}; \citealt{ross_shen_2009}; \citealt{gilli_daddi_2005,gilli_zamorani_2009}; \citealt{coil_georgakakis_2009}; \citealt{allevato_2011}; \citealt{eftekharzadeh_myers_2015}; \citealt{laurent_eftekharzadeh_2017}). 
The differences in these results are likely due to a combination of different methodologies used in the conversion from $b_{\rm DMH}$ to $M_{\rm DMH}^{typ}$ in different studies, as well as
differences in properties of the AGN samples (i.e., luminosity and BH mass) and
in the properties of the galaxies hosting those AGN.

\cite{aird_coil_2021} argued that AGN clustering results are most easily interpreted in terms of the relative bias of AGN to well-characterized galaxy samples and are more challenging to interpret in terms of the absolute bias value alone. Comparing the clustering properties of (inactive) galaxy samples with matched properties as observed in the AGN samples (e.g., in stellar mass, redshift range, and star formation rate) reveals whether AGN clustering can be  explained purely by the known galaxy-halo connection or whether the presence of AGN in the center of a galaxy alters the clustering signal. Current measurements using this relative bias approach do find that the AGN clustering signal can be produced by inactive galaxies closely matched to the AGN in spectral class, stellar mass, and redshift (e.g., \citealt{mendez_coil_2016}; \citealt{mountrichas_georgakakis_2019}; \citealt{krishnan_almaini_2020}). 
Measuring host galaxies properties such as stellar mass in the most-luminous BLAGN can be very challenging, making a relative bias measurement difficult. Alternatively, galaxy-galaxy lensing measurements can provide constraints on the $M_{\rm DMH}$ of the individual AGN host galaxies.
Semiempirical modeling suggests a halo mass distribution for X-ray selected AGN that peaks at $M_{\rm DMH} \sim 10^{12}\,h^{-1}\,M_\odot$ with a tail extending to higher halo masses (\citealt{leauthaud_benson_2015};  \citealt{georgakakis_comparat_2019}). This tail shifts the average halo mass to higher values than 
$M_{\rm DMH} \sim 10^{12}\,h^{-1}\,M_\odot$.
In addition, the HOD modeling of the one-halo term should be improved to allow for distributions beyond a single power law.
Equation~\ref{eq:agnhod} could be adjusted to allow for a more complex halo mass distribution of AGN. Ideally the model would allow for different slopes at different halo masses. However, the challenge for such models would be the limited S/N of current AGN clustering measurements, such that a more flexible model would result in  poor constraints on additional HOD parameters.

\begin{figure}
  \centering
 \resizebox{\hsize}{!}{ 
   \includegraphics[width=9.0cm,height=6.0cm]{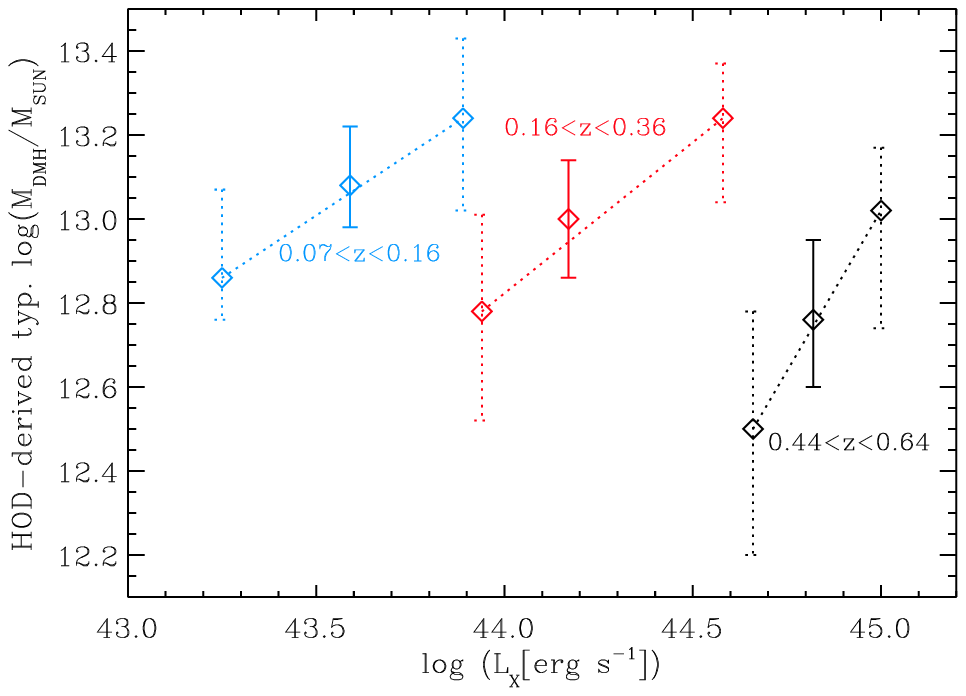}}
      \caption{{Typical DMH mass (derived from HOD modeling and grid fitting approach) as a function of 0.1--2.4 keV luminosity in independent redshift ranges (shown with different colors). Data points with the solid error bars show the full samples, while the data points with dashed error bars displays the respective subsamples split by X-ray luminosity.}}
      \label{fig:evolLX}
\end{figure}

\subsection{$L_{\rm X}$ Clustering Dependence of X-ray Selected AGN}
\label{sec:disc_lx}

We find a weak and marginal  clustering dependence on luminosity for X-ray selected BLAGN at a 2.1$\sigma$ confidence level ($0.44<z<0.64$), in that higher-luminosity X-ray AGN have a higher mean DMH mass than lower-luminosity AGN. Interestingly, we previously  discovered weak, positive, and marginally significant  clustering dependences with $L_{\rm X}$ in two lower-redshift ranges ($0.07<z<0.16$, $0.16<z<0.36$). 
Considering the detection of a weak and tentative $L_{\rm X}$ clustering dependence in three independent redshift ranges provides stronger evidence for the existence of this positive but weak correlation. 
We note that our X-ray selected AGN sample is drawn from a soft (0.1--2.4 keV) X-ray band, such that moderately X-ray absorbed AGN will be missed by our selection.
In paper IV, we show with cosmological simulations that a weak  
$L_{\rm X}$ dependence of the clustering is only expected for the most-luminous AGN while the more moderate or weak $L_{\rm X}$ AGN are not expected to exhibit a $L_{\rm X}$ dependence of the clustering properties.  \cite{aird_coil_2021} also presented constraints on DMH mass distribution as a function of $L_{\rm X}$ and have results for very luminous AGN that are consistent with our findings.

Figure~\ref{fig:evolLX} shows that there is not a simple continuous function of bias with $L_{\rm X}$ that is independent of redshift; what is observed is that AGN with the same $L_{\rm X}$ show different clustering strengths at different redshifts.
Within any redshift range, the most X-ray luminous AGN have a similar high typical DMH mass of $M_{\rm DMH} \sim 10^{13}\,h^{-1}\,M_\odot$.
Within each individual redshift range, the lower $L_{\rm X}$ AGN, being $\sim$5 times less luminous, have a typical $M_{\rm DMH} \sim 10^{12.7}\,h^{-1}\,M_\odot$. There may be a tentative trend of lower DMH mass with increasing redshift for AGN at a given X-ray luminosity.

To further investigate this observation, we compare the number density of AGN at a given luminosity with that of DMHs at the corresponding typical halo mass. The ratio of these densities represents a ``duty cycle", or the fraction of halos that contain an AGN. In principal, if a halo contains more than, e.g, two AGN at the same time, these count as two. However, as seen below, the duty cycle is much less than one, and such cases are rare. Here we compare the AGN duty cycles between the highest- and lowest-redshift bins shown in Fig. \ref{fig:evolLX}. We estimate the duty cycle by:
\begin{equation}
    \mathrm{duty\,cycle}\sim\frac{d\,\phi_{\rm X}/d\,\log L_{\rm X}}{d\,\phi_{\rm DMH}/d\,\log M_{\rm DMH}}\cdot\left|\frac{d\,\log L_{\rm X}}{d\log M_{\rm DMH}}\right|,
    \label{eq:duty}
\end{equation}
i.e., the ratio between the X-ray luminosity function (XLF) and the corresponding DMH mass function. The XLF $(d\,\phi_{\rm X}/d\,\log L_{\rm X})$ is obtained from the 0.5--2 keV XLF \citep{miyaji_2001,hasinger05} converted to our 0.1--2.4 keV band assuming a photon index of $\Gamma=2.4$. The DMH mass function ($d\,\phi_{\rm DMH}/d\,\log M_{\rm DMH}$) is from \citet{sheth_mo_2001} and is implemented in our HOD code.  The factor  $|d\,\log L_{\rm X}/d\log M_{\rm DMH}|$ is estimated from the slope for the given redshift bin in Fig. \ref{fig:evolLX}.

For our $z\sim 0.53$ sample with $M_{\rm DMH}\sim 10^{12.8}\,h^{-1}\,M_\odot$, the estimated duty cycle is $\sim 1\times 10^{-3}$, i.e.,  0.1\% of the DMHs at this halo mass could contain a BLAGN. Using the DMH mass growth rate of \citet{
fakhouri10}, the descendants of halos of mass $M_{\rm DMH}\sim 10^{12.8}\,h^{-1}\,M_\odot$ at $z\sim 0.53$ have $M_{\rm DMH}\sim 10^{13.0}\,h^{-1}\,M_\odot$ at $z\sim 0.13$. The duty cycle for these descendants (at $z\sim 0.13$) is estimated to be $\sim 2\%$. Thus, roughly speaking, the duty cycle represented by our high-$L_{\rm X}$ $z\sim 0.13$ sample at $M_{\rm DMH}\sim 10^{13.2}\,h^{-1}\,M_\odot$ DMHs has increased by an order of magnitude compared to their progenitor DMHs at $z\sim 0.53$. 

There are caveats in interpreting our results in this manner. 
We assume that the observed slope of the $\log M_{\rm DMH}$--$\log L_{\rm X}$ relation at each redshift range can be extended to unprobed luminosities. The error on the slope is large, and it is not guaranteed that the observed relation applies beyond the observed luminosity range. In particular, at $z\sim 0.53$, AGN with a luminosity of $L_{\rm 0.1-2\,keV}\la 10^{44}h_{\rm 70}^{-2}{\rm erg\,s^{-1}}$ are below the RASS flux limit and thus not included in our sample. Those AGN may well reside  in $\ga 10^{12.5}\,h^{-1}\,M_\odot$ DMHs. This would result in a smaller $|d\,\phi_{\rm DMH}/d\,\log M_{\rm DMH}|$ and thus a larger duty cycle. 
Future analysis of AGN samples collected with {\sl eROSITA} will allow us to explore the typical DMH masses of luminosity- and redshift-defined AGN samples to compare the differences in duty cycles in this space with unprecedented precision.

\subsection{$M_{\rm i}$ Clustering Dependence of Optically selected AGN}

We do not find a clear clustering dependence of the optically selected AGN
with $M_{\rm i}$ (see Table~\ref{tab:overview}). The subsample with
$-23.25<M_{\rm i}<-23.70$ deviates the most from other subsamples but at a
difference of only $\sim$1.4$\sigma$. Other than the $M_{\rm i} > -22.7$ mag
subsample, all samples have  similar redshift distributions. To remove
differences in the redshift distributions completely, we create
redshift-matched subsamples (except for the subsample $M_{\rm i} > -22.7$ mag) and analyze their clustering properties. The  bias values for these samples vary within their 1$\sigma$ uncertainties, such that no significant correlation is found. Considering the relatively small uncertainties of the optical sample bias measurements, there is little room for even a weak clustering dependence with $M_{\rm i}$.

\cite{shen_mcbride_2013} also explored the $M_{\rm i}$ clustering dependence of quasars using cross-correlation measurements between DR7 (optical) quasar catalog and DR10 CMASS galaxies. Their data are also consistent with no significant luminosity ($M_{\rm i}$) dependence. One difference is that they analyze a much wider redshift range from $0.3<z<0.9$, while we focus on a much narrower redshift range to exclude redshift evolution of the clustering properties as well as having better control of the systematics (e.g., constant stellar masses for CMASS galaxies). In addition, they also select their $M_{\rm i}$ subsamples differently. The advantage of these slightly different approaches, while finding the same (non)result, strengthens the robustness that there is no $M_{\rm i}$ clustering dependence for optical AGN.

We also did not find an $M_{\rm i}$ clustering dependence in our previous work
(paper III) in  the redshift ranges  $0.16<z<0.36$ and $0.36<z<0.50$. In the
lowest redshift range ($0.07<z<0.16$) we were not able to test for such a
dependence due to the low number of objects. Thus, we conclude that from low
to intermediate redshifts, there is no  $M_{\rm i}$ clustering dependence for
optically selected AGN. This is in contrast to the result that in all of these
redshift ranges we do find evidence for a weak $L_{\rm X}$ dependence in the
X-ray selected samples. In paper IV, we gave as a possible explanation the complex sample selection for the optical SDSS sample (e.g., host galaxy/AGN separation). However, in the redshift range studied here this effect should be minor. 
In the logical chain of correlations between $M_{\rm DMH}$ $\rightarrow$ $M_{\rm stellar}$ $\rightarrow$ $M_{\rm bulge}$ $\rightarrow$ $M_{\rm BH}$ $\rightarrow$ $M_{\rm i}$ there is a scatter at each stage, which sums across the chain. It is possible that the last step in the chain has a higher scatter than that of $M_{\rm BH}$ $\rightarrow$ $L_{\rm X}$, as for $L_{\rm X}$ we consistently detect a positive but weak clustering dependence and we find no dependence with $M_{\rm i}$.  Another potential explanation is that there are different dependencies of $M_{\rm stellar}$--$L_{\rm X}$ (e.g., \citealt{carraro_shankar_2022}) and $M_{\rm stellar}$--$M_{\rm i}$.


\begin{figure*}

\begin{center}
\hbox{\hspace*{-0.4cm}
 \includegraphics[width=9.0cm,height=6cm]{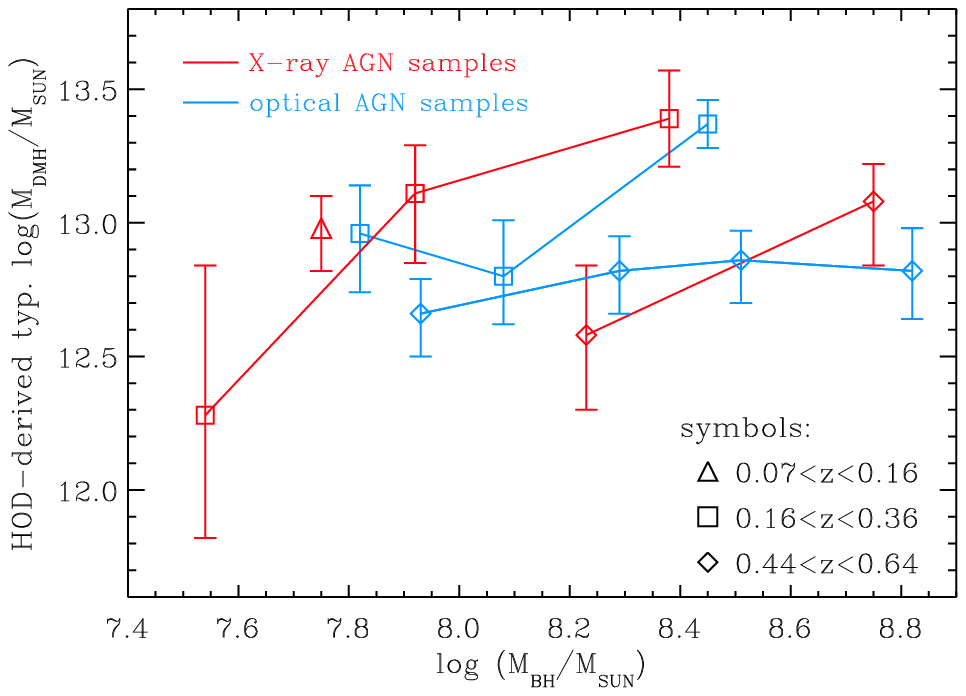}
\hspace{0.2cm}
 \includegraphics[width=9.0cm,height=6.0cm]{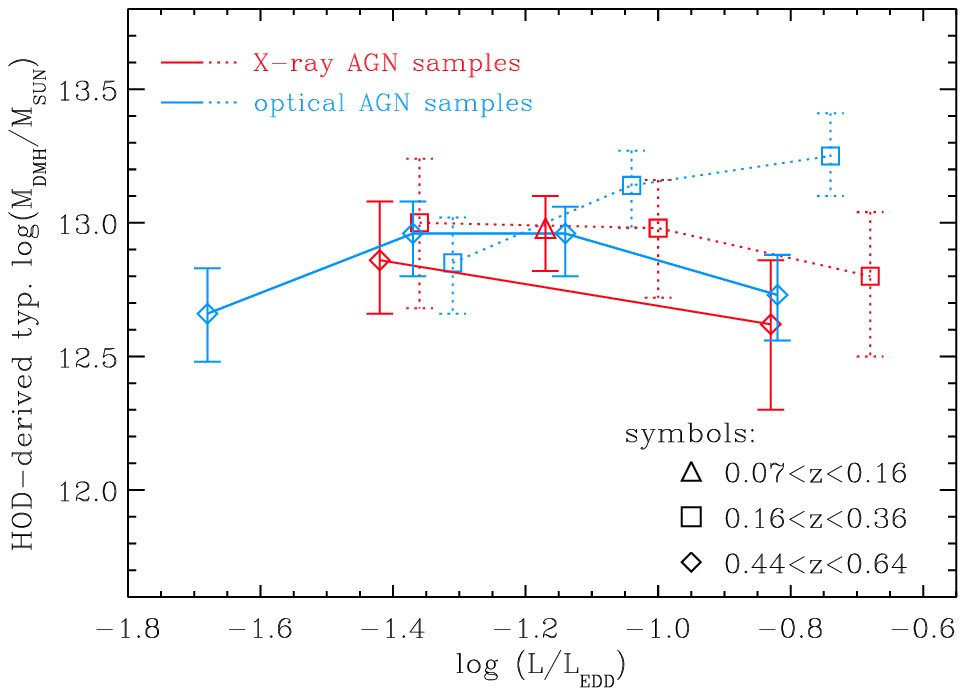}
}
\end{center}
\caption{{\it Left:} Typical DMH mass (derived from HOD modeling and grid fitting approach) as a function of $M_{\rm BH}$ at several independent redshifts (shown with different symbols) for X-ray (red symbols) and optical (blue symbols) AGN samples. 
{\it Right:} Similar to the left panel, here as a function of the Eddington ratio. We show the results in the redshift range $0.16<z<0.36$ with dotted lines to allow for a clear distinction of the different samples.}
\label{fig:evolMBHLLedd}
\vspace*{0.8cm}
\end{figure*}


\subsection{$M_{\rm BH}$ Clustering Dependence and Its Evolution with Redshift}
\label{sec:disc_mh}

In the redshift range of $0.44<z<0.64$ studied here, we  find only a tentative positive correlation of $M_{\rm BH}$ with the clustering strength for the X-ray AGN sample (1.3$\sigma$ when comparing $b$ values, 2.1$\sigma$ confidence when comparing mean DMH masses). Despite the much larger sample size of the optical AGN, the largest difference between the $M_{\rm BH}$ subsamples is only  1$\sigma$ when comparing their bias values and 1.4$\sigma$ when comparing mean DMH masses.  This result is not sensitive to the specific cuts used in creating these subsamples or to reducing the number of subsamples (e.g., two subsamples only).
Figure~\ref{contour_compare} (right) and Table~\ref{tab:overview} show that the X-ray and optical AGN samples span a similar parameter space in $M_{\rm BH}$. The samples differ only slightly in $L_{\rm H\beta}$ in that X-ray  AGN have on average higher $L_{\rm H\beta}$ than optical AGN (Fig.~\ref{contour_compare}, left). Thus if a very weak $M_{\rm BH}$ dependence of the clustering strength exists, it remains unclear why the optical AGN samples do not reveal this correlation as the X-ray selected AGN samples do.

In Fig.~\ref{fig:evolMBHLLedd} (left), we summarize the DMH mass dependence on $M_{\rm BH}$ measured in different redshift ranges for X-ray and optical BLAGN samples. At $0.07<z<0.16$, there are too few optical AGN to compute a robust clustering signal. In this redshift range, there is a clear signal for X-ray AGN; however, we are not able to create subsamples due to the relatively low number of objects. Thus in  Fig.~\ref{fig:evolMBHLLedd} we  show only the measurement for the total X-ray AGN sample in the redshift range $0.07<z<0.16$.

There is not a single $M_{\rm BH}$ versus DMH mass correlation across all redshifts; i.e., increasing $M_{\rm BH}$ does not necessarily result in higher DMH mass. Instead, we find that within each redshift range the DMH mass ranges are very similar across a range of $M_{\rm BH}$, consistent with a  typical $M_{\rm DMH} \sim 10^{12.7-13.3}\,h^{-1}\,M_\odot$.
For the redshift ranges in which $M_{\rm BH}$ subsamples can be created ($0.16<z<0.36$ and $0.44<z<0.64$), in both cases there is a weak positive correlation between $M_{\rm BH}$ and the typical DMH mass for X-ray selected AGN. For the optical AGN, no clear trend is found.
Within a given redshift range,   measurements between the X-ray and optical selections with similar $M_{\rm BH}$ agree well, within their uncertainties.

\subsection{$L/L_{\rm EDD}$ Clustering Dependence and Evolution with Redshift}

We do not find a correlation between $L/L_{\rm EDD}$ and typical DMH mass in
the redshift range $0.44<z<0.64$; all subsamples, for both X-ray and optical
AGN, agree in the typical and mean DMH masses within their combined
uncertainties at less than 1.3$\sigma$. In Fig.~\ref{fig:evolMBHLLedd} (right)
we show the measurements for all redshift ranges. The results show that across
all  redshift ranges studied, there is no clear clustering dependence with
$L/L_{\rm EDD}$ and that at a given $L/L_{\rm EDD}$ the measurements from
different redshifts agree well. Determining any $L/L_{\rm EDD}$ clustering
dependence and its evolution with redshift for a hard X-ray selected sample
would provide a more complete picture than the current sample, as it would mitigate many observational biases (e.g., soft X-ray selection for our AGN sample).

\subsection{$L_{\rm Bol}$ Clustering Dependence}
\label{Lbol_discussion}
We do not find a correlation between $L_{\rm Bol}$ and typical DMH mass in the redshift range $0.44<z<0.64$ for the two X-ray selected AGN subsamples. However, in the same redshift range we find a $>2\sigma$ $L_{\rm Bol}$ dependence of the clustering strength in three of the four optical AGN subsamples. 
However, the highest $L_{\rm Bol}$ subsample (median log ($L_{\rm Bol}/[\rm{erg \, s^{-1}}]) = 45.67$) has almost identical clustering properties as the lowest $L_{\rm Bol}$ subsample (median log ($L_{\rm Bol}/[\rm{erg \, s^{-1}}]) = 44.80$). 
Although the X-ray selected AGN sample has a smaller dynamic range in 
 $L_{\rm Bol}$, there is overlap with the $L_{\rm Bol}$ range of the optical selected AGN sample.
Within these individual subsamples we do not find a statistically significant discrepancy between the optical and X-ray selected AGN.  As the $L_{\rm Bol}$ results taken together do not present a consistent picture of a clear dependence, we refrain from interpreting a possible trend to the data.
 
Unfortunately no observational constraints on the $L_{\rm Bol}$ dependence of AGN clustering exists from cross-correlation measurements at other redshifts. In the future, larger samples will be needed to more fully study the $L_{\rm Bol}$ clustering dependence at different redshifts.

%
 \section{Conclusions}

In this paper, we extend our clustering studies of X-ray (\textit{ROSAT}/SDSS) and optically (SDSS) selected broad-line AGN to the redshift range of $z=0.44-0.64$. As in our previous work, we use a cross-correlation approach with a large set of tracer galaxies (here the CMASS galaxy sample) to substantially improve the S/N of the AGN clustering measurement. We define a CMASS galaxy sample to (i) have an identical average stellar mass (and thus clustering strength) over the entire redshift range used in this paper, and (ii) explore the AGN clustering properties down to scales of
$r_p = 0.1$ $h^{-1}$ Mpc. We estimate for the X-ray and optically selected broad-line AGN samples the supermassive black hole mass, accretion ratio relative to Eddington, and the bolometric luminosity based on spectral fits to the H$\beta$ line. We split the AGN samples by the physical properties of luminosity in the selection bandpass, $M_{\rm BH}$, $L/L_{\rm EDD}$, and bolometric luminosity. We apply HOD modeling directly to the cross-correlation measurements to infer the AGN distribution as a function of DMH mass. We also determine the HOD parameters for the CMASS galaxy sample. For the full X-ray and optical AGN samples, we apply an MCMC HOD parameter search, while for the AGN subsamples, we apply a grid search to estimate the large-scale bias and typical and mean DMH mass of each sample.

The HOD constraints on the optical BLAGN sample are 
tighter than those for the X-ray BLAGN sample, due to a larger sample size. 
X-ray and optical BLAGN have the same large-scale bias and thus extremely similar typical DMH masses (logarithmic DMH masses of $12.87^{+0.11}_{-0.19}\,h^{-1} M_{\odot}$ for the X-ray AGN sample and $12.72^{+0.06}_{-0.06}\,h^{-1} M_{\odot}$ for the optical AGN sample).
While this reflects the typical DHM mass, the full range of halo masses occupied by AGN is very broad (e.g., \citealt{georgakakis_comparat_2019}; \citealt{aird_coil_2021}). 
There is marginal statistical evidence that the one-halo clustering properties between the X-ray and optical AGN samples are different. In contrast to optically selected AGN, X-ray selected AGN may have  a larger population of satellite galaxies at 
$M_{\rm DMH} \sim 10^{13}\,h^{-1}\,M_\odot$, indicated by the higher satellite fraction and lower $\alpha_{\rm s}$ (satellite slope) found for the X-ray AGN HOD compared to the optical AGN HOD.

When we compare results as a function of AGN properties, we do not find statistically significant correlations with $L/L_{\rm EDD}$ and $M_i$. For the X-ray selected AGN sample, we find a positive correlation between mean DMH mass and $L_{\rm X}$ as well as with $M_{\rm BH}$ at a $>$ 2$\sigma$ confidence level. The optical AGN sample does not show a significant correlation between mean DMH mass and $M_{\rm BH}$, but the measured values for the different optical subsamples also do not contradict the $M_{\rm BH}$ versus mean and typical DMH mass correlations found for the X-ray selected AGN sample. 
Within the optical broad-line AGN sample, we find some dependencies at $>$ 2$\sigma$ confidence level in the clustering properties as a function of $L_{\rm Bol}$.
However, the lowest $L_{\rm Bol}$ and highest $L_{\rm Bol}$ samples have very similar clustering properties, so a clear trend is not present.

We compare our results with those obtained in our paper III, in which we
evaluate the clustering and HOD properties  of X-ray (RASS/SDSS) and optically
selected (SDSS) BLAGN in three lower-redshift ranges using different SDSS galaxy tracer sets. 
The full redshift range of $z=0.07-0.64$ covered by all of these studies spans a cosmic time interval of 5 Gyr.  In all redshift ranges we use the same cross-correlation technique to determine the AGN clustering properties, create samples from the same large area surveys (RASS and SDSS), and evaluate the results with similar HOD modeling.
We find that X-ray and optically selected BLAGN, despite their different optical and X-ray luminosities at different redshifts, occupy DMHs with $M_{\rm DMH}^{typ} \sim 10^{12.5-13.0}\,h^{-1}\,M_\odot$ across this full redshift range. In other words, we find no statistically significant difference in the typical DMH masses between X-ray and optically selected BLAGN samples. 
However, at higher redshift, the same DMH mass hosts more X-ray luminous AGN than at lower redshift.

Semianalytic cosmological simulations find that an $L_{\rm X}$ and $M_{\rm BH}$ dependence of the AGN clustering exists, but only for the most-luminous AGN and not for low- or moderate-luminosity AGN. Since our X-ray selected BLAGN sample contains the most-luminous objects in independent redshift bins, we can test this prediction.
From $z=0.07$ to $z=0.64$, we also find weak positive correlations between the typical DMH mass and $L_{\rm X}$ for the X-ray BLAGN samples. In the redshift ranges $z=0.16-0.36$ and $z=0.44-0.64$ we can estimate $M_{\rm BH}$, and we have enough objects to create subsamples in $M_{\rm BH}$ and $L/L_{\rm EDD}$.  We find that the weak positive $L_{\rm X}$ dependence of the AGN clustering appears to be due to a weak positive $M_{\rm BH}$ dependence of the clustering signal, in that more X-ray luminous (and larger $M_{\rm BH}$) BLAGN reside in more-massive DMHs than their lower X-ray luminosity (lower $M_{\rm BH}$) counterparts. The observed clustering strength across the full redshift range does not depend on $L/L_{\rm EDD}$. Consequently, higher accretion rate AGN do not reside in more dense environments.

The number of RASS-detected AGN decreases sharply below $z\lesssim 0.07$ and
above $z\gtrsim 0.6$, due to limited volumes at low redshift and ROSAT's flux
sensitivity. Consequently, the redshift range for high-S/N clustering studies
using RASS-selected AGN cannot be extended further. However, the X-ray
telescope eROSITA (\citealt{predehl_andritschke_2021}) is a game changer for
AGN studies. eROSITA is far more sensitive than ROSAT and scans the entire sky
several times at  X-ray wavelengths. The resulting stacked X-ray data will
reach a flux limit $\sim$30 times more sensitive than RASS. This will provide
unprecedented large AGN samples for clustering studies to improve constraints
on dependences with redshift, $L_{\rm X}$, and $M_{\rm BH}$, as well as to
test if the one-halo properties of X-ray and optically selected AGN differ. 
Recently the German eROSITA team published its first AGN clustering paper (\citealt{comparat_luo_2023}) using several thousand AGN detected in the $\sim$140 square degree eROSITA Final Equatorial Depth Survey field (\citealt{brunner_liu_2022}; \citealt{liu_buchner_2022}). The spectroscopic AGN samples from the eROSITA all-sky survey scans will contain hundreds of thousands of objects and will thus bring AGN clustering studies to a new level.


\acknowledgments
\section{acknowledgments}
We thank the referee for useful suggestions which helped to improve the manuscript.
The research leading to these results has received funding from the European
Community's Seventh Framework Programme (/FP7/2007-2013/) under grant agreement
No. 229517. 

M.K. acknowledges support by DFG grants KR 3338/3-1 and KR 3338/4-1.
T.M. and H.A. acknowledge support from UNAM-DGAPA (PAPIIT IN111319, IN114423 and PASPA) and 
CONACyT (grant Cient\'ifica B\'asica 252531). T.M. thanks Leibniz-Institut f\"ur 
Astrophysik Potsdam (AIP) for hospitality during his sabbatical leave from UNAM. 
A.L.C. acknowledges support from the Ingrid and Joseph W. Hibben Chair at UC San Diego.
We thank Xan Morice-Atkinson for his support on the photometric redshift estimates.  
The \textit{ROSAT} Project was supported by the Bundesministerium f{\"u}r Bildung 
und Forschung (BMBF/DLR) and the Max-Planck-Gesellschaft (MPG).

Funding for the Sloan Digital Sky Survey IV has been provided by the Alfred P. Sloan Foundation, the U.S. Department of Energy Office of Science, and the Participating Institutions. SDSS acknowledges support and resources from the Center for High-Performance Computing at the University of Utah. The SDSS website is www.sdss.org.

SDSS is managed by the Astrophysical Research Consortium for the Participating Institutions of the SDSS Collaboration including the Brazilian Participation Group, the Carnegie Institution for Science, Carnegie Mellon University, Center for Astrophysics/Harvard and Smithsonian (CfA), the Chilean Participation Group, the French Participation Group, Instituto de Astrofísica de Canarias, The Johns Hopkins University, Kavli Institute for the Physics and Mathematics of the Universe (IPMU) / University of Tokyo, the Korean Participation Group, Lawrence Berkeley National Laboratory, Leibniz-Institut für Astrophysik Potsdam (AIP), Max-Planck-Institut für Astronomie (MPIA Heidelberg), Max-Planck-Institut für Astrophysik (MPA Garching), Max-Planck-Institut für Extraterrestrische Physik (MPE), National Astronomical Observatories of China, New Mexico State University, New York University, University of Notre Dame, Observatório Nacional / MCTI, The Ohio State University, Pennsylvania State University, Shanghai Astronomical Observatory, United Kingdom Participation Group, Universidad Nacional Autónoma de México, University of Arizona, University of Colorado Boulder, University of Oxford, University of Portsmouth, University of Utah, University of Virginia, University of Washington, University of Wisconsin, Vanderbilt University, and Yale University.
\\



\appendix

\section{A --- Recovering the Small-scale Clustering Signal} \label{smallscales}
The full CMASS sample is designed to facilitate the primary science goal of
detection of the baryon acoustic oscillation signal at $z \sim 0.6$
(\citealt{cuesta_vargas_2016}). Since two fibers cannot be placed closer than
62 arcsecs for approximately 5.5\% of all photometrically selected CMASS candidates, no optical 
spectrum could be
obtained for these sources (\citealt{guo_zehavi_2013}). This strongly affects 
small-scale clustering measurements. 
The CMASS survey accounts for these restrictions by
introducing weights to observed galaxies
(\citealt{reid_ho_2016}; \citealt{ross_bautista_2020}). 
If a spectroscopically observed CMASS galaxy has, e.g., two close neighbors that fulfill  the CMASS selection criteria but are spectroscopically missed due to fiber collisions, the CMASS galaxy which was observed is given a weight of three, to account for the two missing galaxies. Other
observational biases are also considered and folded into the weights. 
These weights ensure that large-scale galaxy-galaxy pair counts are correctly recovered.
The total weight for each object is calculated by $w_{\rm tot}=(w_{\rm CP} +
w_{\rm NOZ} -1) \times w_{\rm sys}$, where $w_{\rm CP}$ is the weight based on
fiber-collision pairs, $w_{\rm NOZ}$ accounts for spectroscopic objects for which no redshift
information could be obtained, and $w_{\rm sys}$ accounts for systematic effects due to
the varying star density across the SDSS footprint, which affects the observed density of galaxies. 
More details on the weights can be found in  \cite{ross_percival_2012}
and \cite{anderson_aubourg_2012}.

\begin{figure*}[t]
\begin{center}
\hbox{\hspace*{-0.3cm}
 \includegraphics[width=9.4cm]{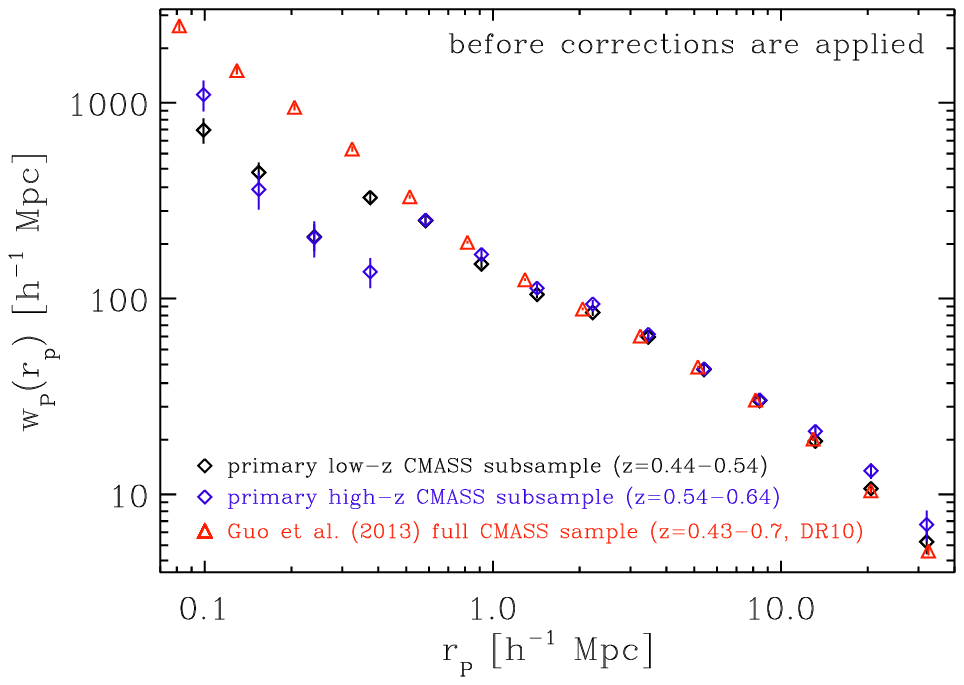}
\hspace{-0.4cm}
 \includegraphics[width=9.4cm]{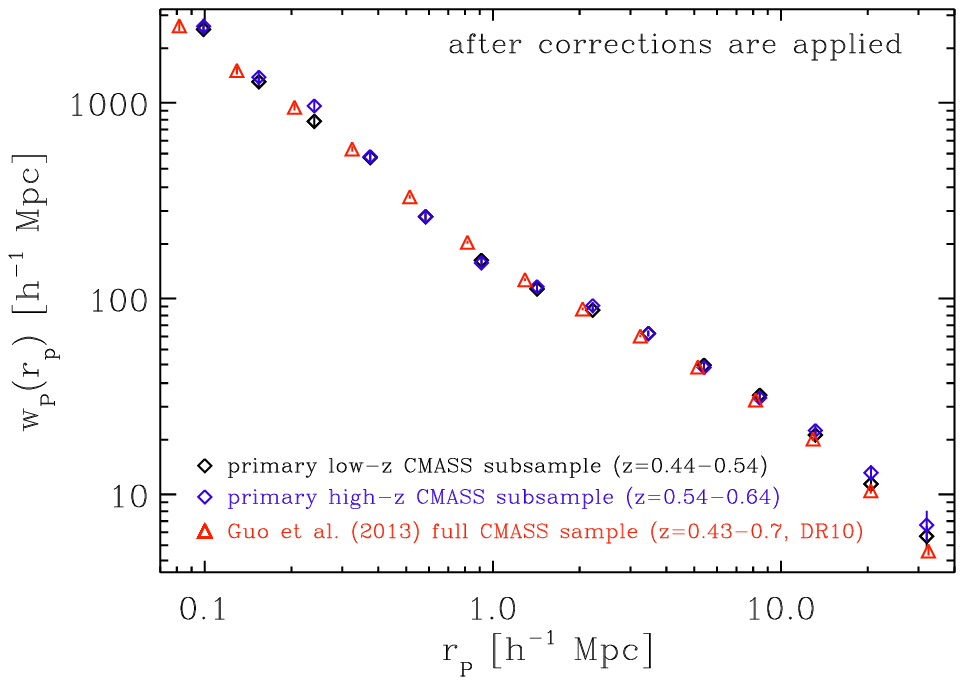}
\hspace{-0.5cm}
}
\end{center}
\vspace*{-0.5cm}
\caption{{\it Left}: comparison between the ACFs of a CMASS sample 
with $0.44<z<0.64$ and $11.25 < \rm{log} (M_{stellar}/M_{\odot})<11.43$ when
split into low- and high-redshift subsamples, 
  \textit{before} applying our correction algorithm. In addition we show the full \cite{guo_zehavi_2013} CMASS
  sample which utilizes a different approach to correct the small-scale clustering.
{\it Right}: similar to the left panel showing the low- and high-redshift CMASS subsamples, here  \textit{after} our correction algorithm is applied.}
\label{fig:wprp_split_compare}
\end{figure*}

While such weighting recovers
missed galaxy pair counts on large scales, this approach does not allow us to correctly recover the correlation function on scales less than  
$\sim$0.5 $h^{-1}$ Mpc (corresponding to 62 arcsecs at at $z=0.55$, the median redshift of our sample), as shown in Figure~\ref{fig:wprp_split_compare} (left). 
On these small scales, it is necessary to either use information from the photometric catalogs about pairs of galaxies that could not both be spectroscopically observed, or other correction methods developed in the literature. 
To recover the CMASS small-scale clustering signal, different approaches have been used (e.g., \citealt{guo_zehavi_2013}; \citealt{mohammad_2020}).
\cite{guo_zehavi_2013} presented a
method that divides the CMASS sample into two distinct populations. The first
is drawn from SDSS regions covered by multiple masks where fiber collisions are minimized. The second population originates from SDSS regions that are
affected by fiber collisions because they are only observed by one mask. The
clustering signal on small scales can be estimated using the first population only, 
within the overlapping regions,  and on larger scales is measured using the full SDSS footprint.  Recently, \citet{mohammad_2020} introduced a ``pairwise-inverse probability and angular upweighting" method for this correction, which recovers clustering measurements down to $\sim 0.1\,h^{-1}$ Mpc scales. As shown below, our alternative approach, which is developed independently and explained in the rest of this subsection, also allows us to correct for the fiber collision at similar small scales.

To begin, we use the DR12 target catalog, which is based solely on SDSS photometric data. 
Using the different selection criteria, e.g.,
color cuts, this catalog contains information on which photometric objects
should receive a fiber and why this target was
selected (i.e., which object class). As described above, not all of these targets can be spectroscopically observed 
due to fiber collisions.  
The target catalog encodes information about why an object was targeted for spectroscopic 
follow-up observation using target 
IDs\footnote{\url{ http://www.sdss3.org/dr10/algorithms/bitmask\_boss\_target1.php}}.
These targets IDs are bitmask values.
We identify all photometric target IDs in the North Cap that lead to spectroscopically
confirmed CMASS objects. Of those, we consider only IDs that lead to at least 100
spectroscopically confirmed CMASS galaxies.  For the resulting 
target IDs (001, 134, 135), we compute the fraction $c$ of spectroscopic CMASS
targets with each of these IDs over photometric targets with these IDs ($c_{001}=0.253$,
$c_{134}=0.872$, $c_{135}=0.008$). The meaning of these number is, e.g., that
25.3\% of photometric targets with ID 001 led to
spectroscopically confirmed CMASS objects.   


\begin{figure*}
\begin{center}
\hbox{\hspace*{0.0cm}
 \includegraphics[width=9.1cm]{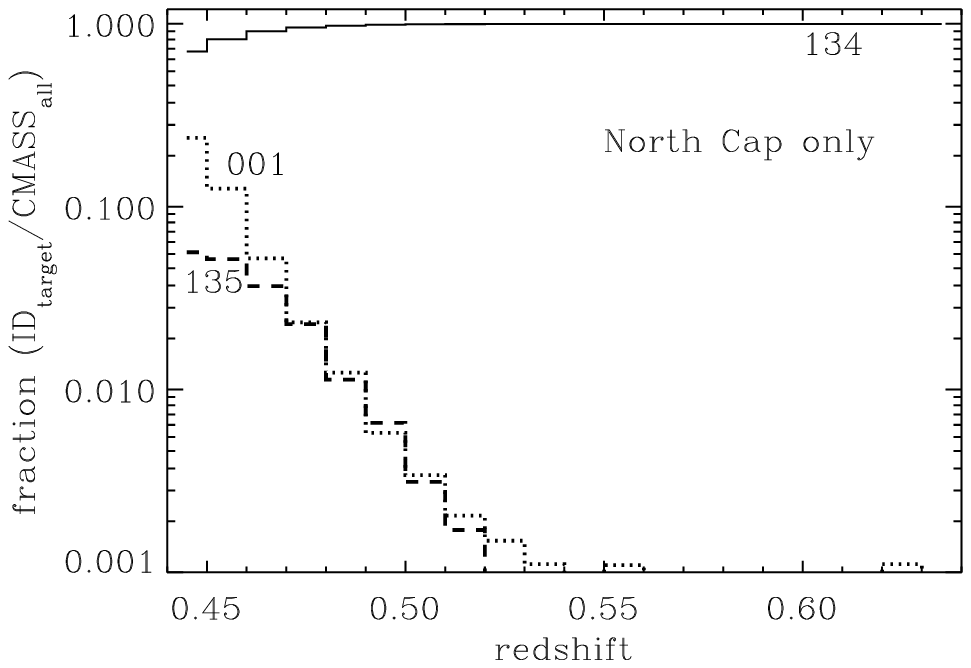}
\hspace{-0.4cm}
 \includegraphics[width=9.1cm]{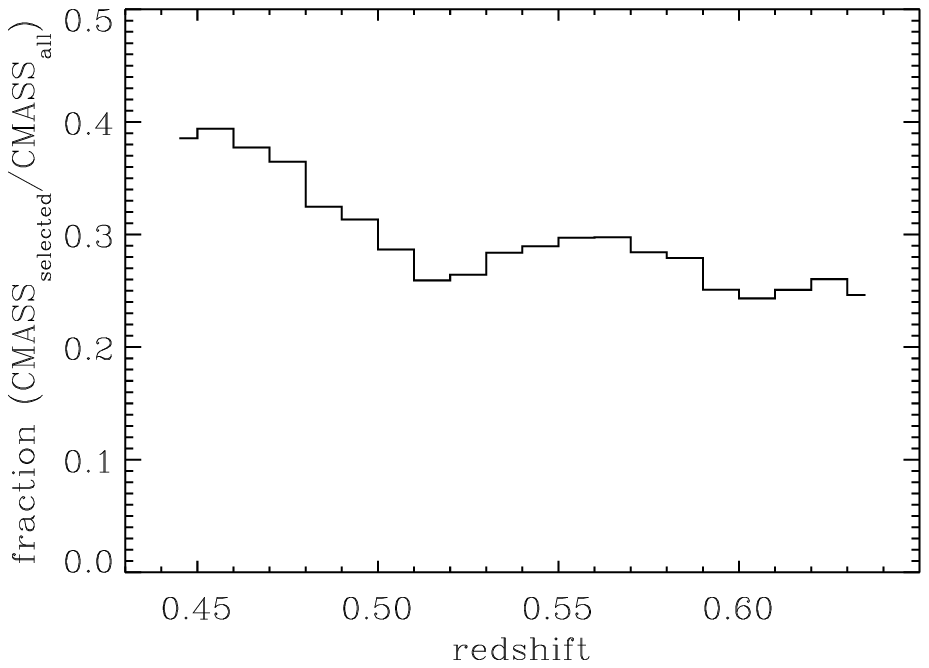}
\hspace{-0.5cm}
}
\end{center}
\vspace*{-0.5cm}
\caption{{\it Left}: fraction of different target IDs (bitmask values) to the full
  spectroscopic CMASS sample as a function of redshift (SDSS North Cap). Each line represents a
  different target ID whose tag is stated directly next to the
  corresponding line. 
{\it Right}: fraction of spectroscopic CMASS sources that fulfill our final selection
criteria to the full spectroscopic CMASS sample as a function of redshift. 
}
\label{fig:collided}
\vspace*{+0.5cm}
\end{figure*}

Then we calculate the redshift distribution ($f_{\rm ID\_target}(z)$) of spectroscopic CMASS galaxies with these IDs in the range $0.44<z<0.64$ (bin size $\Delta z=0.01$). 
Figure~\ref{fig:collided} (left) shows that spectroscopic CMASS galaxies with ID 134 are by far the dominant population. In each redshift bin, 
the sum over all three target IDs is 1. Thus this provides the normalized relative 
contribution of each target ID as a function of redshift.

Next we determine from Fig.~\ref{stellarM_z_CMASS} the fraction $f_{\rm {selected/all}}(z)$ of spectroscopic CMASS objects that fulfill our final selection criteria of 
$0.44 \le z \le 0.64$ and stellar mass selection (red data points) relative to the
full spectroscopic CMASS sample (red+black data points) in as a function of
redshift. The result is shown in Fig.~\ref{fig:collided} (right).
With this information in hand, we are able to perform a statistical correction for
the fiber collisions based on the photometric target catalog.  
To recover the small-scale clustering signal, we 
compute, for each photometric galaxy that has a fiber collision with a spectroscopic CMASS object, the probability that 
the photometric source (based on its photometric target ID and the redshift) 
should be assigned the redshift of the spectroscopic CMASS object and include it in the final sample as follows:

\begin{equation}
P(ID,z)= c_{\rm ID\_target} \times f_{\rm ID\_target}(z) \times f_{\rm
  {selected/all}}(z) -P_{\rm spur}
\label{eq:collision}
\end{equation}

In practice, we search for all photometric targets that have IDs with the values 001, 134,
or 135 within a radius of 62 arcsecs of each selected spectroscopic CMASS galaxy.
Photometric targets within the search radius that already have spectroscopic
data are rejected. If the spectroscopic CMASS galaxy has $w_{\rm CP} > 1$, we
use the redshift of the spectroscopic CMASS object that the
object collides with and calculate the probability $P(ID,z)$ according
Eq.~\ref{eq:collision}.  We then draw a random value $n_{\rm random}$
between 0 and 1 for each colliding target. If $n_{\rm random} \le P(ID,z)$, we consider this object for
our collision correction further; otherwise, we reject it. The term 
$P_{\rm spur}$ corrects for spurious matches. The number density per sky area
for the photometric targets with the CMASS IDs is $8.78\times 10^{-6}$ arcsec$^{-2}$. 
We calculate the distance between the spectroscopically confirmed CMASS object
and the collided object. Using a simple circular area calculation (distance = radius) 
and considering the number density provides the chance that the collided object
will spuriously fall at its observed position.   

If $n_{\rm random} \le P(ID,z)$, we adopt the R.A. and decl. of the photometric object and assign
it the same redshift as the nearby spectroscopic CMASS galaxy. Finally, we recompute $w_{\rm CP}$ for the spectroscopic
CMASS galaxy by $w_{\rm CP,NEW} = w_{\rm CP,OLD} -1$ if the collided object
ends up being in our sample. This approach is
repeated for the same spectroscopic CMASS galaxy if it collides with more
than one object until $w_{\rm CP} > 1$ is not fulfilled anymore. If more than one
object collides with a spectroscopic CMASS galaxy, we sort all objects by
distance from the spectroscopic object and start by calculating (thus 
potentially correcting) the closest photometric object first. This
approach accounts for the fact that close neighbors have a substantially higher probability of having similar  redshifts.   

Our approach for correcting the fiber-collision issue in the CMASS sample has
several advantages: (i) it does not affect the large-scale clustering above 
$\sim$0.5 $h^{-1}$ Mpc, because for each collision-corrected object we include in our
final sample, we reduce the weight ($w_{\rm CP}$) by 1, (ii) the resulting clustering signal is very similar in shape and amplitude to that measured for
the full CMASS sample from SDSS DR10 by \cite{guo_zehavi_2013} ($0.43<z<0.7$), 
and most importantly for our cross-correlation method (iii) when
split into lower- and higher-redshift subsamples ($0.44<z<0.54$ and $0.54<z<0.64$,
respectively), the clustering signal has minimal differences,  well within the
expected statistical uncertainties (Fig.~\ref{fig:wprp_split_compare}, right). 
We additionally tested various approaches using photometric redshifts to correct the small-scale clustering signal. However, these approaches resulted in strong discrepancies between the low- and high-redshift CMASS samples at small scales.

\section{B --- Reliability of the H$\beta$ Broad-line Fit} 
\label{reliability}

In order to assess the reliability of our broad-line fits, we carry out Monte
Carlo simulations using 21 objects in the X-ray catalog with S/N$>50$,
stepwise degrading their S/N by adding random Gaussian noise, fitting the
degraded spectra with our fitting routine, and measuring the FWHM from the best
fit to the degraded spectrum. For each AGN, we perform 100 simulations per S/N
bin. Following paper~IV, we require an uncertainty on the H$\beta$ FWHM
measurement of less than 40\% as above that value the uncertainties
exceed the commonly assumed systematic uncertainties for viral \mbh estimates
of $\sim$0.3 dex. As shown in the upper panel in
Fig.~\ref{fig:SNsim} at S/N=5, roughly 90\% of our targets fulfill
this requirement on the recovered FWHM. We therefore use S/N$=5$ as our
default threshold for spectra for which we determine $M_{\rm BH}$, and we verify that our results do not
depend on the precise S/N value chosen. Our results agree with previous work,
suggesting that at an S/N below $\sim5$ the uncertainties on FWHM
dominate the \mbh measurement (e.g., \citealt{denney_peterson_2009,shen_richards_2011}). 
The typical FWHM uncertainty at S/N$=5$ is $\sim0.06$~dex, as
shown in the bottom panel of Fig.~\ref{fig:SNsim}. After applying the S/N$>5$ 
threshold, we have a sample size of 1632 for the X-ray AGN sample and 8889 
for the optical AGN sample.


\section{C --- Comparing our H$\beta$ Fit Results to Literature Measurements}
\label{comparing}

\begin{figure*}
\begin{center}
\hbox{\hspace*{-0.2cm}
 \includegraphics[width=9.1cm,height=8cm]{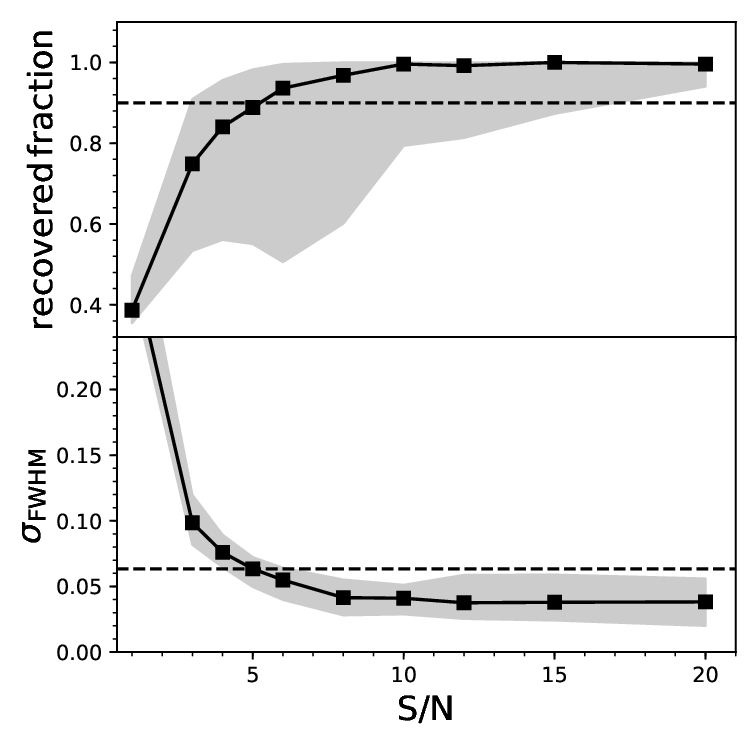}
\hspace{-0.2cm}
 \includegraphics[width=9.1cm,height=8cm]{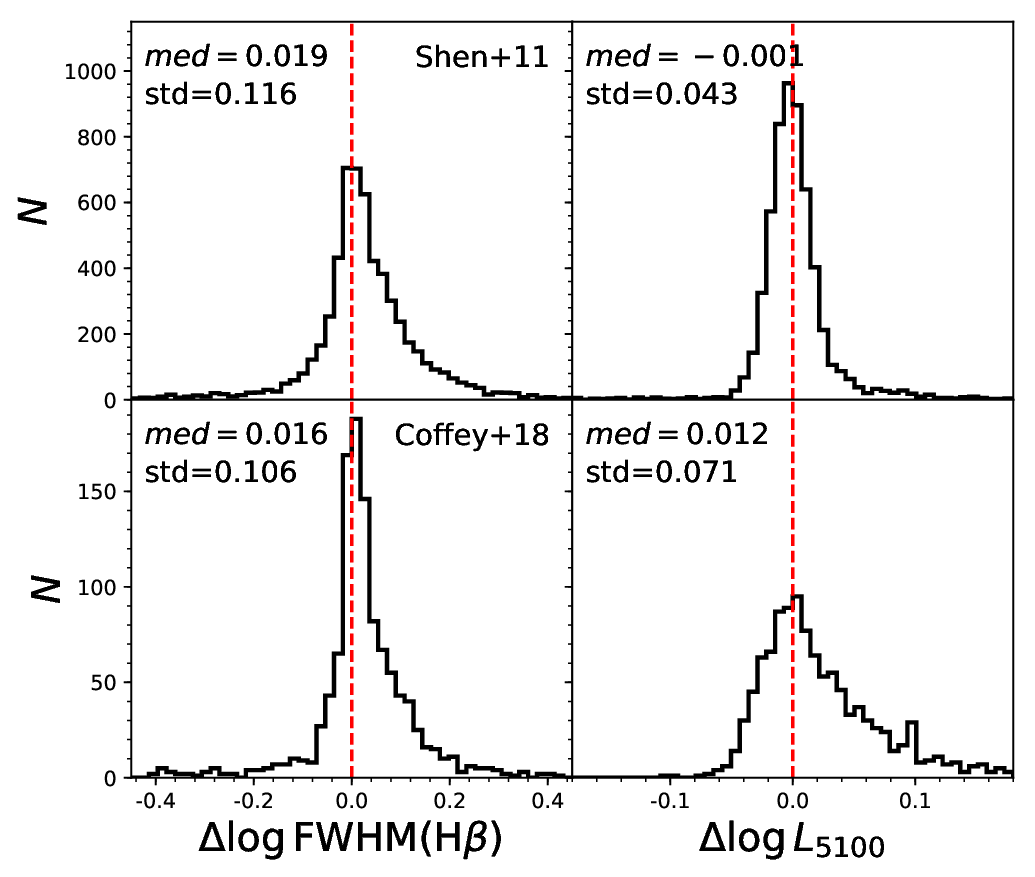}
\hspace{-0.5cm}
}
\end{center}
\vspace*{-0.5cm}
\caption{{\it Left}: Upper panel: Simulation of the recovery rate of the intrinsic FWHM
        of H$\beta$ within a 40\% uncertainty as a function of spectral S/N, based on
        Monte Carlo simulations for 21 AGN spectra. Lower
        panel: Dependence of the mean uncertainty on the FWHM measurement as a
        function of S/N. The gray areas show the 68\% ($1\sigma$) uncertainty
        range based on the simulated  spectra with degraded S/N properties.
     \label{fig:SNsim}
{\it Right}: Comparison of the FWHM (left panels) and $L_{5100}$
        measurements (right panels) between our samples and those from the
        literature. Upper panels: Comparison between our optical DR14 AGN
        catalog and the SDSS DR7 quasar catalog by \citet{shen_richards_2011}. Lower
        panels: Comparison of our X-ray sample with the SDSS/SPIDERS (2RXS)
        catalog by \citet{coffey_salvato_2019}.}
     \label{fig:spec_comparison}
     \vspace*{+0.5cm}
\end{figure*}

Our sample overlaps with previous studies of the spectral properties of AGN in
the SDSS.  We compare our  H$\beta$ measurements with other studies, to investigate the consistency of our results with previous
work. In particular we cross-match our optical AGN sample (DR14) with the SDSS DR7
quasar catalog of \citet{shen_richards_2011} and our SPIDERS X-ray selected
AGN sample with the recent
study by \citet{coffey_salvato_2019}. The SDSS DR7 quasar catalog is a subset of the
DR14 quasar catalog by \citet{paris_petitjean_2018} with 5764 objects in
common with our optical DR14 sample. In the upper panels of
Figure~\ref{fig:spec_comparison} we compare the FWHM and $L_{5100}$
measurements of  \citet{shen_richards_2011} to ours. We find an excellent agreement overall, 
with median differences in $\log {\rm FWHM}=0.02$ and $\log L_{5100}=0.00$
and a standard deviation of 0.12~dex and 0.04~dex, respectively.  
\citet{coffey_salvato_2019} present an $M_{\rm BH}$ catalog for SPIDER AGN from the 2RXS
catalog with 1124 AGN in common with our X-ray 
sample. We show this comparison in the lower panels of
Figure~\ref{fig:spec_comparison}. For the FWHM measurement, we find a similar
agreement as in the comparison with \citet{shen_richards_2011}, providing an estimate
of the systematic uncertainties of the line width measurements. For the
continuum luminosity, $L_{5100}$, we find a consistent mode, but there is a tail with
lower $L_{5100}$ in the \citet{coffey_salvato_2019} catalog. This is caused by the
inclusion of a host galaxy template in the fitting procedure of the continuum
in \citet{coffey_salvato_2019}. We do not explicitly account for the host galaxy
contribution in our fitting routine, since for our sample it is often
difficult to break the degeneracy between host galaxy spectrum and AGN
continuum emission. We account for the effect of host galaxy
contamination to $L_{5100}$ in a statistical fashion, as detailed in
Section~\ref{sec:mbh}.

The differences between our method and those of \cite{shen_richards_2011} and \cite{coffey_salvato_2019} lead to uncertainties in the \mbh estimates that are lower than the systematic uncertainties of $\sim$0.3 dex; thus, we conclude that
our line fitting routine provides robust measurements of the
H$\beta$ region spectral properties and enables the estimation of
black hole masses based on these measurements.

\begin{thebibliography}{}

\bibitem[Abolfathi et al.(2018)]{abolfathi_aguado_2018} Abolfathi, B., Aguado, D.S., Aguilar, G., et al., 2018, ApJS, 235, 42

\bibitem[Adelman-McCarthy et al.(2006)]{adelman_mccarthy_agueeros_2006} Adelman-McCarthy, J., Ag\"ueros, M.A., Allam, S.S., et al. 2006, ApJS, 162, 38
 
\bibitem[Aird \& Coil(2021)]{aird_coil_2021} Aird, J. \& Coil, A.L. 2021, MNRAS, 502, 5962

\bibitem[Alam et al.(2015)]{alam_albareti_2015} Alam, S., Albareti, F.D., Allende Prieto, C., et al. 2015, ApJS, 219, 12

\bibitem[Allevato et al.(2011)]{allevato_2011} Allevato, V., Finoguenov, A., Cappelluti, N., et al. 2011, ApJ, 736, 99

\bibitem[Altamirano-D{\'e}vora et al.(2016)]{altamirano16} Altamirano-D{\'e}vora, L., Miyaji, T., Aceves, H., et al.\ 2016, RMxAA, 52, 11

\bibitem[Anderson et al.(2012)]{anderson_aubourg_2012} Anderson, L., Aubourg, E., Bailey, S., et al. 2012, MNRAS, 427, 3435

\bibitem[Bentz et al.(2009)]{bentz_peterson_2009} Bentz, M.C., et al. 2009, ApJ, 697, 160

\bibitem[Blanton et al.(2017)]{blanton_bershady_2017} Blanton, M.R., Bershady, M.A., Abolfathi, B., et al. 2017, AJ, 154, 28

\bibitem[Boller et al.(2016)]{boller_freyberg_2016} Boller, Th., Freyberg,   M.J., Tr\"umper, J., et al. 2016 A\&A, 588, 103

\bibitem[Bolton et al.(2012)]{bolton_schlegel_2012} Bolton, A.S., Schlegel, D.J., Aubourg, {\'E}., et al. 2012, \aj, 144, 144

\bibitem[Booth \& Schaye(2010)]{booth_schaye_2010} Booth, C.M. \& Schaye, J. 2010, MNRAS, 405, L1

\bibitem[Boroson \& Green(1992)]{boroson_green_1992} Boroson, T.A., \& Green, R.F. 1992, \apjs, 80, 109

\bibitem[Brunner et al.(2022)]{brunner_liu_2022} Brunner, H., Liu, T., Lamer, G., et al. 2022, A\&A, 661, 1 

\bibitem[Cappelluti et al.(2010)]{cappelluti_ajello_2010} Cappelluti, N., Ajello, M., Burlon, D., et al. 2010, ApJL, 716, L209 

\bibitem[Cardelli et al.(1989)]{cardelli_clayton_1989} Cardelli, J.A., Clayton, G.C., \& Mathis, J.S. 1989, \apj, 345, 245

\bibitem[Carraro et al.(2022)]{carraro_shankar_2022} Carraro, R., Shankar, F., Viola, A., et al. 2022, \mnras, 512, 1185

\bibitem[Coffey et al.(2019)]{coffey_salvato_2019} Coffey, D., Salvato, M., Merloni, M., et al. 2019, A\&A, 625, 123

\bibitem[Coil et al.(2009)]{coil_georgakakis_2009} Coil, A. L., Georgakakis, A., Newman, J. A., et al.  2009, ApJ, 701, 1484

\bibitem[Coil et al.(2007)]{coil_hennawi_2007} Coil, A. L.,  Hennawi, J.F., Newman, J., et al. 2007, ApJ, 654, 115

\bibitem[Comparat et al.(2023)]{comparat_luo_2023} Comparat, J., Luo, W., Merloni, A.,  et al. 2023, A\&A, 673, 122
  
\bibitem[Comparat et al.(2017)]{comparat_maraston_2017} Comparat, J., Maraston, C., Goddard, D., et al. 2017, arXiv:1711.06575

\bibitem[Comparat et al.(2020)]{comparat_merloni_2020} Comparat, J., Merloni, A., Dwelly, T., et al. 2020, A\&A, 636, 22

\bibitem[Comparat et al.(2019)]{comparat_merloni_2019} Comparat, J., Merloni, A., Salvato, M., et al. 2019, MNRAS, 487, 2005
  
\bibitem[Cooray \& Sheth(2002)]{cooray_sheth_2002} Cooray, A. \& Sheth, R., 2002, PhR, 372, 1

\bibitem[Cuesta et al.(2016)]{cuesta_vargas_2016} Cuesta, A.J., Vargas-Magnana, M., Beutler, F. et al., 2016, MNRAS, 457, 1770

\bibitem[Davis \& Peebles(1983)]{davis_peebles_1983} Davis, M. \& Peebles, P.J.E. 1983, ApJ, 267, 465

\bibitem[Dawson et al.(2013)]{dawson_schlegel_2013} Dawson, K. S., Schlegel, D. J., Ahn, C. P., et al. 2013, AJ, 145, 10

\bibitem[Denney et al.(2009)]{denney_peterson_2009} Denney, K. D., Peterson, B. M., Dietrich, M., Vestergaard, M., \& Bentz, M. C. 2009, ApJ, 692, 246

\bibitem[Eftekharzadeh et al.(2015)]{eftekharzadeh_myers_2015}  Eftekharzadeh, S., Myers, A.D., White, M., et al. 2015, MNRAS, 453, 2779

\bibitem[Eisenstein et al.(2011)]{eisenstein_weinberg_2011} Eisenstein, D. J., Weinberg, D.H., Agol, E., et al. 2011, AJ, 142, 72

\bibitem[\protect\citeauthoryear{Fakhouri, Ma, \& Boylan-Kolchin}{2010}]{fakhouri10} Fakhouri O., Ma C.-P., Boylan-Kolchin M., 2010, MNRAS, 406, 2267

\bibitem[Georgakakis et al.(2019)]{georgakakis_comparat_2019} Georgakakis, A., Comparat, J., Merloni< A., et al. 2019, MNRAS, 487, 275

\bibitem[Gilli et al.(2005)]{gilli_daddi_2005} Gilli, R., Daddi, E., Zamorani, G., et al. 2005, A\&A, 430, 811

\bibitem[Gilli et al.(2009)]{gilli_zamorani_2009} Gilli, R., Zamorani, G., Miyaji, T., et al. 2009, A\&A, 494, 33

\bibitem[Gunn et al.(2006)]{gunn_siegmund_2006} Gunn, J.E., Siegmund, W.A., Mannery, E.J., et al. 2006, AJ, 131, 2332

\bibitem[Guo et al.(2013)]{guo_zehavi_2013} Guo, H., Zehavi, I., Zheng, Z., et al. 2013, ApJ, 767, 122

\bibitem[Guo et al.(2015)]{guo_zheng_2015} Guo, H., Zheng, Z., Zehavi, I., et al. 2015, \mnras, 446, 578

\bibitem[\protect\citeauthoryear{Hasinger, Miyaji, \& Schmidt}{2005}]{hasinger05} Hasinger G., Miyaji T., Schmidt M., 2005, A\&A, 441, 417

\bibitem[Kaiser(1987)]{kaiser87} Kaiser, N.\ 1987, \mnras, 227, 1

\bibitem[Kaspi et al.(2000)]{kaspi_smith_2000} Kaspi, S., Smith, P. S., Netzer, H., et al. 2000, ApJ, 533, 631

\bibitem[Kelly \& Shen(2013)]{kelly_shen_2013} Kelly, B.C., \& Shen, Y. 2013, \apj, 764, 45

\bibitem[Koutoulidis et al.(2013)]{koutoulidis_plionis_2013} Koutoulidis, L., Plionis, M., Georgantopoulos, I., Fanidakis, N. 2013, MNRAS, 428, 1382 

\bibitem[Krishnan et al.(2020)]{krishnan_almaini_2020} Krishnan, C., Almaini, O., Hatch, N.A., et al. 2020, MNRAS, 494, 1693

\bibitem[Krumpe et al.(2010)]{krumpe_miyaji_2010} Krumpe, M., Miyaji, T., \& Coil, A. L. 2010a, ApJ, 713, 558 (paper I)

\bibitem[Krumpe et al.(2014)]{krumpe_miyaji_2014} Krumpe, M., Miyaji, T., \& Coil, A. L., 2014, AcPPP, 71
  
\bibitem[Krumpe et al.(2012)]{krumpe_miyaji_2012} Krumpe, M., Miyaji, T., Coil, A. L., \& Aceves, H. 2012, ApJ, 746, 1 (paper III)

\bibitem[Krumpe et al.(2015)]{krumpe_miyaji_2015} Krumpe, M., Miyaji, T., Husemann, B., et al. 2015, ApJ, 815, 21 (paper IV)

\bibitem[Krumpe et al.(2018)]{krumpe_miyaji_2018} Krumpe, M., Miyaji, T., Coil, A. L., \& Aceves, H. 2018, MNRAS, 474 ,1773 

\bibitem[Landy \& Szalay(1993)]{landy_szalay_1993} Landy, S.D. \& Szalay, A.S. 1993, ApJ, 412, 64

\bibitem[Larson et al.(2011)]{larson_dunkley_2011} Larson, D., Dunkley, J., Hinshaw, G., et al. 2011, ApJS, 192, 16

\bibitem[Laurent et al.(2017)]{laurent_eftekharzadeh_2017} 
 Laurent, P., Eftekharzadeh, S., Le Goff, J.-M. et al. 2017, JCAP, 07, 017 

\bibitem[Law-Smith \& Eisenstein(2017)]{lawsmith_eisenstein_2017} Law-Smith, J., Eisenstein, D.J., 2017, ApJ accepted, arxiv:1702.03933

\bibitem[Leauthaud et al.(2015)]{leauthaud_benson_2015} Leauthaud, A., Benson, A., Civano, F., et al. 2015, MNRAS, 446, 1874

\bibitem[Leauthaud et al.(2016)]{leauthaud_bundy_2016} Leauthaud, A., Bundy, K., Saito, S., et al. 2016, MNRAS, 457, 4021

\bibitem[Liu et al.(2022)]{liu_buchner_2022} Liu, T., Buchner, J., Nandra, K., et al. 2022, A\&A, 661, 5 

\bibitem[Makino \& Hut(1997)]{makino97} Makino, J., \& Hut, P.\ 1997, \apj, 481, 83 

\bibitem[Maraston \& Str\"omb\"ack(2011)]{maraston_2011} Maraston, C. \&  Str\"omb\"ack, G., 2011 MNRAS, 418, 2785

\bibitem[Maraston et al.(2013)]{maraston_2013} Maraston, C., Pforr, J., Henriques, B.M., et al., 2013, MNRAS, 435, 2764

\bibitem[Marconi et al.(2004)]{marconi_risaliti_2004} Marconi, A., Risaliti, G., Gilli, R., et al. 2004 MNRAS, 351, 169

\bibitem[Markwardt(2009)]{markwardt_2009} Markwardt, C.B., 2009, in Astronomical Society of the Pacific Conference Series, Vol. 411, Astronomical Data Analysis Software and Systems XVIII, ed.  D.~A. {Bohlender}, D.~{Durand}, \& P.~{Dowler}, 251

\bibitem[McLure \& Dunlop(2004)]{mclure_dunlop_2004} McLure, R.J., Dunlop, J.S. 2004, MNRAS, 352, 1390

\bibitem[McLure \& Jarvis(2002)]{mclure_jarvis_2002} McLure, R.J., Jarvis, M.J. 2002, \mnras, 337, 109

\bibitem[Mej{\'{\i}}a-Restrepo et al.(2016)]{mejia_trakhtenbrot_2016} Mej{\'{\i}}a-Restrepo, J.E., Trakhtenbrot, B., Lira, P., Netzer, H., \& Capellupo, D.M. 2016, \mnras, 460, 187

\bibitem[Melnyk et al.(2018)]{melnyk_elyiv_2018} Melnyk, O., Elyiv, A., Smolcic, V., et al. 2018, A\&A, 620, A6

\bibitem[Mendez et al.(2016)]{mendez_coil_2016} Mendez, A.J., Coil, A.L, Aird, James, et al. 2016, ApJ, 821, 55

\bibitem[Meneux et al.(2009)]{meneux_guzzo_2009} Meneux, B., Guzzo, L., de la Torre, S., et al. 2009, A\&A, 505, 463

\bibitem[Miyaji et al.(2001)]{miyaji_2001} Miyaji, T., Hasinger, G., \& Schmidt, M.\ 2001, \aap, 369, 49

\bibitem[Miyaji et al.(2011)]{miyaji_krumpe_2011} Miyaji, T., Krumpe, M., Coil, A. L., \& Aceves, H. 2011, ApJ, 726, 83 (paper II)

\bibitem[Mohammad et al.(2020)]{mohammad_2020} Mohammad, F.G., Percival, W.J., Seo, H.-J., et al.\ 2020, \mnras, 498, 128 

\bibitem[Mountrichas \& Georgakakis(2012)]{mountrichas_georgakakis_2012} Mountrichas, G., Georgakakis, A. 2012, MNRAS, 420, 514

\bibitem[Mountrichas et al.(2019)]{mountrichas_georgakakis_2019} Mountrichas, G., Georgakakis, A., Georgantopoulos, I. 2019, MNRAS, 483, 1374

\bibitem[Mullaney et al.(2013)]{mullaney_alexander_2013} Mullaney, J.R., Alexander, D.M., Fine, S., et al. 2013, \mnras, 433, 622

\bibitem[Myers et al.(2015)]{myers_palanque-delabrouille_2015} Myers, A.D., Palanque-Delabrouille, N., Prakash, A., et al. 2015, ApJS, 221, 27

\bibitem[Navarro et al.(1997)]{nfw97} Navarro, J.~F., Frenk, C.~S., \& White, S.~D.~M.\ 1997, ApJ, 490, 493, MNRAS, 332, 827

\bibitem[Netzer \& Trakhtenbrot(2007)]{netzer_trakhtenbrot_2007} Netzer, H., \& Trakhtenbrot, B. 2007, \apj, 654, 754

\bibitem[Nuza et al.(2013)]{nuza_sanchez_2013} Nuza, S., S\'anchez, A.G., Prada, F., et al. 2013, \mnras, 432, 743

\bibitem[P\^aris et al.(2018)]{paris_petitjean_2018} P\^aris, J., Petitjean, P., Aubourg, E., et al. 2018, A\&A, 613, 51

\bibitem[Peebles(1980)]{peebles_1980} Peebles, P. J. E. 1980, The Large-Scale Structure of the Universe, (Princeton, NJ: Princeton Univ. Press)

\bibitem[Peterson \& Wandel(2000)]{peterson_wandel_2000} Peterson, B.M., \& Wandel, A. 2000, ApJL, 540, L13
	
\bibitem[Plionis et al.(2018)]{plionis_koutoulidis_2018} Plionis, M., Koutoulidis, L, Koulouridis, E., et al. 2018, A\&A, 620, A17 

\bibitem[Porciani et al.(2004)]{porciani_magliocchetti_2004} Porciani, C., Magliocchetti, M., \& Norberg, P. 2004, MNRAS, 355, 1010 

\bibitem[Powell et al.(2018)]{powell_cappelluti_2018} Powell, M.C.,Cappelluti, N., Urry, C.M. et al. 2018, ApJ, 858, 110

\bibitem[Powell et al.(2020)]{powell_urry_2020} Powell, M.C., Urry, C.M., Cappelluti, N., et al. 2020, ApJ, 891, 41

\bibitem[Predehl et al.(2021)]{predehl_andritschke_2021} Predehl, P., Andritschke, R., Arefiev, V., et al. 2021, A\&A, 647, 1

\bibitem[Prugniel et al.(2007)]{prugniel_soubiran_2007} Prugniel, P., Soubiran, C., Koleva, M., \& Le Borgne, D. 2007, arXiv astro-ph/0703658

\bibitem[Reid et al.(2016)]{reid_ho_2016} Reid, B., Ho, S., Padmanabhan, N., et al., 2016, MNRAS, 455, 1553

\bibitem[Richards et al.(2006b)]{richards_strauss_2006} Richards, G. T., Strauss, M.A., Fan, X., et al. 2006b, AJ, 131, 2766

\bibitem[Rodr\'iguez-Torres et al.(2016)]{rodriguez_chuang_2016} Rodr\'iguez-Torres, S.A., Chuang, C.-H., J., Prada, F., et al., 2016, MNRAS, 460, 1173

\bibitem[Rodr\'iguez-Torres et al.(2017)]{rodriguez_comparat_2017} Rodr\'iguez-Torres, S.A., Comparat, J., Prada, F., et al., 2017, MNRAS, 468, 728

\bibitem[Ross et al.(2020)]{ross_bautista_2020} Ross, A.J., Bautista, J., Tojeiro, R., et al., 2020, MNRAS, 498, 2354

\bibitem[Ross et al.(2017)]{ross_beutler_2017} Ross, A.J., Beutler, F., Chuang, C.-H., et al. 2017, MNRAS, 464, 1168

\bibitem[Ross et al.(2012)]{ross_percival_2012} Ross, A.J., Percival, W.J., Sanchez, A.G., et al. 2012, MNRAS, 424, 564

\bibitem[Ross et al.(2009)]{ross_shen_2009} Ross, N. P., Shen, Y., Strauss, M. A., et al. 2009, ApJ, 697, 1634

\bibitem[Salpeter et al.(1955)]{salpeter_1955} Salpeter, E. E. 1955, ApJ, 121, 161

\bibitem[Salvato et al.(2018)]{salvato_buchner_2018} Salvato, M., Buchner, J., Budav\'ari, T., et al. 2018, MNRAS, 473, 4937

\bibitem[Savitzky \& Golay(1964)]{savitzky_golay_1964} Savitzky, A., \& Golay, M.J.E. 1964, AnaCh, 36, 1627

\bibitem[Schlegel et al.(1998)]{schlegel_finkbeiner_1998} Schlegel, D.J., Finkbeiner, D.P., \& Davis, M. 1998, \apj, 500, 525

\bibitem[Schulze et al.(2015)]{schulze_bongiorno_2015} Schulze, A., Bongiorno, A., Gavignaud, I., et al. 2015, \mnras, 447, 2085

\bibitem[Schulze et al.(2017)]{schulze_schramm_2017} Schulze, A., Schramm, M., Zuo, W., et al. 2017, \apj, 848, 104 

\bibitem[Schulze et al.(2018)]{schulze_silverman_2018} Schulze, A., Silverman, J.D., Kashino, D., et al. 2018, \apjs, 239, 22

\bibitem[Shankar et al.(2020)]{shankar_allevato_2020} Shankar, F., Allevato, V., Bernardi, M., et al. 2020, NatAs, 4, 282

\bibitem[Shen et al.(2013)]{shen_mcbride_2013} Shen, Y., McBride, C.K., White, M., et al. 2013, \apj, 778, 98

\bibitem[Shen et al.(2011)]{shen_richards_2011} Shen, Y., Richards, G.T., Strauss, M.A., et al. 2011, \apjs, 194, 45

\bibitem[Sheth et al.(2001)]{sheth_mo_2001} Sheth, R.K., Mo, H.J., \& Tormen, G. 2001, MNRAS, 323, 1

\bibitem[Smee et al.(2013)]{smee_gunn_2013} Smee, S.A., Gunn, J.E, Uomoto, A., et al. 2013, AJ, 146, 32

\bibitem[Springel(2005)]{springel_2005} Springel V. 2005, MNRAS, 364, 1105

\bibitem[Sunyaev et al.(2021)]{sunyaev_arefiev_2021} Sunyaev, R., Arefiev, V., Babyshkin, V. et al. 2021, A\&A submitted

\bibitem[Tinker et al.(2005)]{tinker_weinberg_2005} Tinker, J.L., Weinberg, D.H., Zheng, Z., \& Zehavi, I. 2005, ApJ, 631, 41

\bibitem[Trakhtenbrot \& Netzer(2012)]{trakhtenbrot_netzer_2012} Trakhtenbrot, B., \& Netzer, H. 2012, \mnras, 427, 3081

\bibitem[Tr\"umper et al.(1982)]{truemper_1982} Tr\"umper, J. 1982, Advances in Space Research, 2, 241

\bibitem[\protect\citeauthoryear{van den Bosch et al.}{2013}]{vandenbosch13} van den Bosch F.~C., More S., Cacciato M., Mo H., Yang X., 2013, MNRAS, 430, 725

\bibitem[Vestergaard \& Peterson(2006)]{vestergaard_peterson_2006} Vestergaard, M., \& Peterson, B.M. 2006, \apj, 641, 689

\bibitem[Viitanen et al.(2019)]{viitanen_allevato_2019} Viitanen, A., Allevato, V., Finoguenov, A., et al. 2019, A\&A, 629, A14

\bibitem[Voges et al.(1999)]{voges_aschenbach_1999} Voges, W., Aschenbach, B., Boller, T., et al. 1999, A\&A, 349, 389

\bibitem[Voges et al.(2000)]{voges_aschenbach_2000} Voges, W., Aschenbach, B., Boller, T., et al. 2000, VizieR Online Data Catalog, 9029, 0

\bibitem[White et al.(2011)]{white_blanton_2011} White, M., Blanton, M., Bolton, A., et al. 2011, \apj, 728, 126

\bibitem[White \& Frenk(1991)]{white_frenk_1991} White, S.D.M. \& Frenk, C.S., 1991, \apj, 379, 52

\bibitem[Wilkinson et al.(2017)]{wilkinson_2017} Wilkinson, D.M., Maraston, C., Goddard, D. , Thomas, D., Parikh, T., 2017, MNRAS, 472, 4297

\bibitem[Yang et al.(2006)]{yang_mushotzky_2006} Yang, Y., Mushotzky, R. F., Barger, A. J., \& Cowie, L. L. 2006, ApJ, 645, 68

\bibitem[Zehavi et al.(2011)]{zehavi_zheng_2011} Zehavi, I., Zheng, Z., Weinberg, D. H., et al. 2011, ApJ, 736, 59

\bibitem[Zheng et al.(2007)]{zheng_coil_2007} Zheng, Z., Coil, A., \& Zehavi, I. 2007, ApJ, 667, 760  

\bibitem[Zheng et al.(2009)]{zheng_zehavi_2009} Zheng, Z., Zehavi, I.,  Eisenstein, D.~J., Weinberg, D.~H., \& Jing, Y.~P.\ 2009, ApJ, 707, 554  


\end{thebibliography}
\end{document}